\def\ts     {\thinspace}
\def\kms    {\ifmmode{{\rm \ts km\ts s}^{-1}}\else{\ts km\ts s$^{-1}$}\fi}
\def\msol   {\ifmmode{{\rm M}_{\odot} }\else{M$_{\odot}$}\fi}
\def\lsol   {\ifmmode{L_{\odot}}\else{$L_{\odot}$}\fi}
\def\lfir   {\ifmmode{L_{\rm FIR}}\else{$L_{\rm FIR}$}\fi}
\def\lir   {\ifmmode{L_{\rm IR}}\else{$L_{\rm IR}$}\fi}
\def\zsol   {\ifmmode{{\rm Z}_{\odot}}\else{Z$_{\odot}$}\fi}

\def\aco    {\ifmmode{{\rm CO}(J\!=\!1\! \to \!0)}\else{{\rm CO}($J$=1$\to$0)}\fi}
\def\bco    {\ifmmode{{\rm CO}(J\!=\!2\! \to \!1)}\else{{\rm CO}($J$=2$\to$1)}\fi}
\def\cco    {\ifmmode{{\rm CO}(J\!=\!3\! \to \!2)}\else{{\rm CO}($J$=3$\to$2)}\fi}
\def\dco    {\ifmmode{{\rm CO}(J\!=\!4\! \to \!3)}\else{{\rm CO}($J$=4$\to$3)}\fi}
\def\eco    {\ifmmode{{\rm CO}(J\!=\!5\! \to \!4)}\else{{\rm CO}($J$=5$\to$4)}\fi}
\def\fco    {\ifmmode{{\rm CO}(J\!=\!6\! \to \!5)}\else{{\rm CO}($J$=6$\to$5)}\fi}
\def\gco    {\ifmmode{{\rm CO}(J\!=\!7\! \to \!6)}\else{{\rm CO}($J$=7$\to$6)}\fi}
\def\hco    {\ifmmode{{\rm CO}(J\!=\!8\! \to \!7)}\else{{\rm CO}($J$=8$\to$7)}\fi}
\def\ico    {\ifmmode{{\rm CO}(J\!=\!9\! \to \!8)}\else{{\rm CO}($J$=9$\to$8)}\fi}
\def\jco    {\ifmmode{{\rm CO}(J\!=\!10\! \to \!9)}\else{{\rm CO}($J$=10$\to$9)}\fi}
\def\kco    {\ifmmode{{\rm CO}(J\!=\!11\! \to \!10)}\else{{\rm CO}($J$=11$\to$10)}\fi}
\def\ci     {\ifmmode{{\rm C}{\rm \small I}}\else{C\ts {\scriptsize I}}\fi}
\def\hi     {\ifmmode{{\rm H}{\rm \small I}}\else{H\ts {\scriptsize I}}\fi}
\def\hh     {\ifmmode{{\rm H}_2}\else{H$_2$}\fi}
\def\cone {\ifmmode{{\rm C}{\rm \small I}(^3\!P_1\!\to^3\!P_0)}
     \else{C\ts {\scriptsize I}{\small$(^3\!P_1\!\to^3\!\!\!P_0)$}}\fi}
\def\ctwo {\ifmmode{{\rm C}{\rm \small I}(^3\!P_2\!\to^3\!P_1)}
     \else{C\ts {\scriptsize I}{\small$(^3\!P_2\!\to^3\!\!\!P_1)$}}\fi}
\def \cline{\ifmmode{{\rm C}{\rm \small I}(2-1)}\else{C\ts {\scriptsize I}(2--1)}\fi}
\def\cij {\ifmmode{{\rm C}{\rm \small I}\,(^3P_i\to^3P_j)}\else{C\ts {\scriptsize I}\,{\small$(^3P_i\to^3P_j)$}}\fi}
\def\cii    {\ifmmode{{\rm C}{\rm \small II}}\else{C\ts {\scriptsize II}}\fi}
\def\nii    {\ifmmode{{\rm N}{\rm \small II}}\else{N\ts {\scriptsize II}}\fi}
\def\tex {\ifmmode{{T}_{\rm ex}}\else{$T_{\rm ex}$}\fi}
\def\tmb {\ifmmode{{T}_{\rm mb}}\else{$T_{\rm mb}$}\fi}
\def\tkin {\ifmmode{{T}_{\rm kin}}\else{$T_{\rm kin}$}\fi}
\def\microns {\ifmmode{\mu{\rm m}}\else{$\mu$m}\fi}
\def\um{\ifmmode{\mu{\rm m}}\else{$\mu$m}\fi}
\def\nhh   {\ifmmode{n({\rm H}_2)}\else{$n$(H$_2$)}\fi}
\def\gradv {\ifmmode{{\rm dv/dr}}\else{dv/dr}\fi}
\def\rxj   {\ifmmode{{\rm RXJ0911.4+0551}}\else{RXJ0911.4+0551}\fi}
\def\dfof  {\ifmmode{{\Delta F/F}}\else{$\Delta F/F$}\fi}
\def\adfof  {\ifmmode{{\mid\!\Delta F/F\!\mid}}\else{$\mid\!\Delta F/F\!\mid$}\fi}
\def\daoa  {\ifmmode{{\Delta \alpha/\alpha}}\else{$\Delta \alpha/\alpha$}\fi}
\def\adaoa  {\ifmmode{{\mid\!\Delta \alpha/\alpha\!\mid}}\else{$\mid\!\Delta \alpha/\alpha\!\mid$}\fi}
\def\aadot  {\ifmmode{{\mid\!\dot{\alpha}/\alpha\!\mid}}\else{$\mid\!\dot{\alpha}/\alpha\!\mid$}\fi}
\def\duou  {\ifmmode{{\Delta \mu/\mu}}\else{$\Delta \mu/\mu$}\fi}
\def\aduou  {\ifmmode{{\mid\!\Delta \mu/\mu\!\mid}}\else{$\mid\!\Delta \mu/\mu\!\mid$}\fi}
\def\water    {\ifmmode{{\rm H}_2{\rm O}}\else{H$_2$O}\fi}
\def\tdust    {\ifmmode{{\rm T}_{\rm dust}}\else{T$_{\rm dust}$}\fi}
\documentclass{emulateapj}

\usepackage{natbib}
\usepackage{here}
\usepackage{epsf}
\usepackage{epstopdf}

\usepackage{color}

\slugcomment{ApJ draft}

\shorttitle{SPT+ALMA redshift survey}
\shortauthors{Wei\ss, A.\ et al.}

\begin{document}

\title{ALMA REDSHIFTS OF MILLIMETER-SELECTED GALAXIES FROM THE SPT SURVEY: \\ THE REDSHIFT DISTRIBUTION OF DUSTY STAR-FORMING GALAXIES}
\def\MPIfR{1}
\def\ESO{2}
\def\Arizona{3}
\def\Caltech{4}
\def\UPenn{5}
\def\UChicago{6}
\def\CfA{7}
\def\Harvard{8}
\def\KICPChicago{9}
\def\EFIChicago{10}
\def\Saclay{11}
\def\PhysicsUChicago{12}
\def\JPL{13}
\def\Miss{14}
\def\AAUChicago{15}
\def\ANL{16}
\def\Dal{17}
\def\Cambridge{18}
\def\McGill{19}
\def\Davis{20}
\def\Berkeley{21}
\def\UFlorida{22}
\def\UCL{23}
\def\Colorado{24}
\def\LBNL{25}
\def\UCLA{26}
\def\ATNF{27}
\def\Michigan{28}
\def\Carnegie{29}
\def\STScI{30}
\def\CaseWestern{31}
\def\SAIC{32}
\def\IAS{33}

\author{
% top tier
 A.~Wei\ss\altaffilmark{\MPIfR},
C.~De~Breuck\altaffilmark{\ESO},
D.~P.~Marrone\altaffilmark{\Arizona},   
J.~D.~Vieira\altaffilmark{\Caltech},
% alphabetical
 J.~E.~Aguirre\altaffilmark{\UPenn},
 K.~A.~Aird\altaffilmark{\UChicago},
 M.~Aravena\altaffilmark{\ESO},
 M.~L.~N.~Ashby\altaffilmark{\CfA},
% M.~Banerji\altaffilmark{\Cambridge},
 M.~Bayliss\altaffilmark{\CfA,\Harvard}, 
B.~A.~Benson\altaffilmark{\KICPChicago,\EFIChicago}, 
M.~B\'ethermin\altaffilmark{\Saclay},
 A.~D.~Biggs\altaffilmark{\ESO},
L.~E.~Bleem\altaffilmark{\KICPChicago,\PhysicsUChicago}, 
 J.~J.~Bock\altaffilmark{\Caltech,\JPL},
 M.~Bothwell\altaffilmark{\Arizona},
C.~M.~Bradford\altaffilmark{\JPL},
M.~Brodwin\altaffilmark{\Miss},
J.~E.~Carlstrom\altaffilmark{\KICPChicago,\EFIChicago,\PhysicsUChicago,\AAUChicago,\ANL}, 
C.~L.~Chang\altaffilmark{\KICPChicago,\EFIChicago,\ANL}, 
S.~C.~Chapman\altaffilmark{\Dal,\Cambridge},
T.~M.~Crawford\altaffilmark{\KICPChicago,\AAUChicago}, 
A.~T.~Crites\altaffilmark{\KICPChicago,\AAUChicago},
T.~de~Haan\altaffilmark{\McGill}, 
M.~A.~Dobbs\altaffilmark{\McGill}, 
T.~P.~Downes\altaffilmark{\Caltech}, 
 C.~D.~Fassnacht\altaffilmark{\Davis},
%E.~B.~Fomalont\altaffilmark{\NRAO},
E.~M.~George\altaffilmark{\Berkeley}, 
M.~D.~Gladders\altaffilmark{\KICPChicago,\AAUChicago}, 
A.~H.~Gonzalez\altaffilmark{\UFlorida}, 
T.~R.~Greve\altaffilmark{\UCL},	
N.~W.~Halverson\altaffilmark{\Colorado}, 
Y.~D.~Hezaveh\altaffilmark{\McGill},
F.~W.~High\altaffilmark{\KICPChicago,\AAUChicago}, 
G.~P.~Holder\altaffilmark{\McGill}, 
W.~L.~Holzapfel\altaffilmark{\Berkeley}, 
S.~Hoover\altaffilmark{\KICPChicago,\EFIChicago}, 
J.~D.~Hrubes\altaffilmark{\UChicago}, 
K.~Husband\altaffilmark{\Cambridge},
%T.~R.~Hunter\altaffilmark{\NRAO},
% M.~Johnson\altaffilmark{\UCLA},
R.~Keisler\altaffilmark{\KICPChicago,\PhysicsUChicago}, 
%L.~Knox\altaffilmark{\Davis}, 
A.~T.~Lee\altaffilmark{\Berkeley,\LBNL}, 
E.~M.~Leitch\altaffilmark{\KICPChicago,\AAUChicago}, 
M.~Lueker\altaffilmark{\Caltech}, 
D.~Luong-Van\altaffilmark{\UChicago}, 
 M.~Malkan\altaffilmark{\UCLA},
 V.~McIntyre\altaffilmark{\ATNF},
J.~J.~McMahon\altaffilmark{\KICPChicago,\EFIChicago,\Michigan}, 
J.~Mehl\altaffilmark{\KICPChicago,\AAUChicago}, 
 K.~M.~Menten\altaffilmark{\MPIfR},
S.~S.~Meyer\altaffilmark{\KICPChicago,\EFIChicago,\PhysicsUChicago,\AAUChicago}, 
%L.~M.~Mocanu\altaffilmark{\KICPChicago,\AAUChicago},
%T.~E.~Montroy\altaffilmark{\CaseWestern}, 
 E.~J.~Murphy\altaffilmark{\Carnegie},
%T.~Natoli,\altaffilmark{\KICPChicago,\PhysicsUChicago}, 
S.~Padin\altaffilmark{\Caltech,\KICPChicago,\AAUChicago}, 
T.~Plagge\altaffilmark{\KICPChicago,\AAUChicago}, 
%C.~Pryke\altaffilmark{\KICPChicago,\AAUChicago,\Minnesota}, 
C.~L.~Reichardt\altaffilmark{\Berkeley}, 
A.~Rest\altaffilmark{\STScI}, 
 M.~Rosenman\altaffilmark{\UPenn},
J.~Ruel\altaffilmark{\Harvard}, 
J.~E.~Ruhl\altaffilmark{\CaseWestern}, 
% K.~Sharon\altaffilmark{\KICPChicago,\AAUChicago,\AstroMichigan},
K.~K.~Schaffer\altaffilmark{\KICPChicago,\SAIC}, 
%L.~Shaw\altaffilmark{\McGill,\Yale}, 
E.~Shirokoff\altaffilmark{\Caltech}, 
J.~S.~Spilker\altaffilmark{\Arizona},
B.~Stalder\altaffilmark{\CfA}, 
% S.~A.~Stanford
Z.~Staniszewski\altaffilmark{\Caltech,\CaseWestern}, 
A.~A.~Stark\altaffilmark{\CfA}, 
K.~Story\altaffilmark{\KICPChicago,\PhysicsUChicago}, 
K.~Vanderlinde\altaffilmark{\McGill}, 
% W.~Walsh
N.~Welikala\altaffilmark{\IAS},
R.~Williamson\altaffilmark{\KICPChicago,\AAUChicago}
}

\altaffiltext{\MPIfR}{Max-Planck-Institut f\"{u}r Radioastronomie, Auf dem H\"{u}gel 69 D-53121 Bonn, Germany}
\altaffiltext{\ESO}{European Southern Observatory, Karl-Schwarzschild Stra\ss e, D-85748 Garching bei M\"unchen, Germany}
\altaffiltext{\Arizona}{Steward Observatory, University of Arizona, 933 North Cherry Avenue, Tucson, AZ 85721, USA}
\altaffiltext{\Caltech}{California Institute of Technology, 1200 E. California Blvd., Pasadena, CA 91125, USA}
\altaffiltext{\UPenn}{University of Pennsylvania, 209 South 33rd Street, Philadelphia, PA 19104, USA}
\altaffiltext{\UChicago}{University of Chicago, 5640 South Ellis Avenue, Chicago, IL 60637, USA}
\altaffiltext{\CfA}{Harvard-Smithsonian Center for Astrophysics, 60 Garden Street, Cambridge, MA 02138, USA}
\altaffiltext{\Harvard}{Department of Physics, Harvard University, 17 Oxford Street, Cambridge, MA 02138, USA}
\altaffiltext{\KICPChicago}{Kavli Institute for Cosmological Physics, University of Chicago, 5640 South Ellis Avenue, Chicago, IL 60637, USA}
\altaffiltext{\EFIChicago}{Enrico Fermi Institute, University of Chicago, 5640 South Ellis Avenue, Chicago, IL 60637, USA}
\altaffiltext{\Saclay}{Laboratoire AIM-Paris-Saclay, CEA/DSM/Irfu - CNRS - Universit\'e Paris Diderot, CEA-Saclay, Orme des Merisiers, F-91191 Gif-sur-Yvette, France}
\altaffiltext{\PhysicsUChicago}{Department of Physics, University of Chicago, 5640 South Ellis Avenue, Chicago, IL 60637, USA}
\altaffiltext{\JPL}{Jet Propulsion Laboratory, 4800 Oak Grove Drive, Pasadena, CA 91109, USA}
\altaffiltext{\Miss}{Department of Physics and Astronomy, University of Missouri, 5110 Rockhill Road, Kansas City, MO 64110, USA}
\altaffiltext{\AAUChicago}{Department of Astronomy and Astrophysics, University of Chicago, 5640 South Ellis Avenue, Chicago, IL 60637, USA}
\altaffiltext{\ANL}{Argonne National Laboratory, 9700 S. Cass Avenue, Argonne, IL, USA 60439, USA}
\altaffiltext{\Dal}{Department of Physics and Atmospheric Science, Dalhousie University, Halifax, NS B3H 3J5 Canada}
\altaffiltext{\Cambridge}{Institute of Astronomy, University of Cambridge, Madingley Road, Cambridge CB3 0HA, UK}
%\altaffiltext{\NRAO}{National Radio Astronomy Observatory, 520 Edgemont Road, Charlottesville, VA 22903, USA}
\altaffiltext{\McGill}{Department of Physics, McGill University, 3600 Rue University, Montreal, Quebec H3A 2T8, Canada}
\altaffiltext{\Davis}{Department of Physics,  University of California, One Shields Avenue, Davis, CA 95616, USA}
\altaffiltext{\Berkeley}{Department of Physics, University of California, Berkeley, CA 94720, USA}
\altaffiltext{\UFlorida}{Department of Astronomy, University of Florida, Gainesville, FL 32611, USA}
\altaffiltext{\UCL}{Department of Physics and Astronomy, University College London, Gower Street, London WC1E 6BT, UK}
\altaffiltext{\Colorado}{Department of Astrophysical and Planetary Sciences and Department of Physics, University of Colorado, Boulder, CO 80309, USA}
\altaffiltext{\LBNL}{Physics Division, Lawrence Berkeley National Laboratory, Berkeley, CA 94720, USA}
\altaffiltext{\UCLA}{Department of Physics and Astronomy, University of California, Los Angeles, CA 90095-1547, USA}
\altaffiltext{\ATNF}{Australia Telescope National Facility, CSIRO, Epping, NSW 1710, Australia}
\altaffiltext{\Michigan}{Department of Physics, University of Michigan, 450 Church Street, Ann Arbor, MI, 48109, USA}
\altaffiltext{\Carnegie}{Observatories of the Carnegie Institution for Science, 813 Santa Barbara Street, Pasadena, CA 91101, USA}
%\altaffiltext{\Minnesota}{Physics Department, University of Minnesota, 116 Church Street S.E., Minneapolis, MN 55455, USA}
\altaffiltext{\STScI}{Space Telescope Science Institute, 3700 San Martin Dr., Baltimore, MD 21218, USA}
\altaffiltext{\CaseWestern}{Physics Department, Center for Education and Research in Cosmology  and Astrophysics,  Case Western Reserve University, Cleveland, OH 44106, USA}
%\altaffiltext{\AstroMichigan}{Department of Astronomy, University of Michigan, 500 Church Street, Ann Arbor, MI, 48109, USA}
\altaffiltext{\SAIC}{Liberal Arts Department, School of the Art Institute of Chicago,  112 S Michigan Ave, Chicago, IL 60603, USA}
%\altaffiltext{\Yale}{Department of Physics, Yale University, P.O. Box 208210, New Haven, CT 06520-8120, USA}
\altaffiltext{\IAS}{Institut d'Astrophysique Spatiale, B\^atiment 121, Universit\'e Paris-Sud XI \& CNRS, 91405 Orsay Cedex, France}

\begin{abstract}
  Using the Atacama Large Millimeter/submillimeter Array (ALMA), we
  have conducted a blind redshift survey in the 3\,mm atmospheric
  transmission window for 26 strongly lensed dusty star-forming
  galaxies (DSFGs) selected with the South Pole Telescope (SPT).  The
  sources were selected to have S$_{1.4{\rm mm}}$$>$$20$ mJy and a
  dust-like spectrum and, to remove low-z sources, not have bright radio (S$_{\rm 843\,MHz}<6$\,mJy) 
  or far-infrared counterparts (S$_{100\um}<1$\,Jy, S$_{60\um}<200$\,mJy).  
  We robustly detect 44 line features in our survey,
  which we identify as redshifted emission lines of $^{12}$CO,
  $^{13}$CO, \ci, H$_2$O, and H$_2$O$^+$. We find one or more spectral
  features in 23 sources yielding a $\sim$$90\%$ detection rate for
  this survey; in 12 of these sources we detect multiple lines, while
  in 11 sources we detect only a single line.  For the sources with
  only one detected line, we break the redshift degeneracy with
  additional spectroscopic observations if available, or infer the
  most likely line identification based on photometric data.  
  This yields secure redshifts for $\sim$70\% of the sample. The
  three sources with no lines detected are tentatively placed in the
  redshift desert between 1.7$<$$z$$<$2.0.  The resulting mean
  redshift of our sample is $\bar z$=3.5.  This finding is in contrast
  to the redshift distribution of radio-identified DSFGs, which have a
  significantly lower mean redshift of $\bar z$=2.3 and for which only
  10-15\% of the population is expected to be at $z$$>$3.  We discuss
  the effect of gravitational lensing on the redshift distribution and
  compare our measured redshift distribution to that of models in the
  literature.
\end{abstract}


%%%%%%%%%%%%%%%%%%%%%%%%%%%%%%%%%%%%%
% 1. Intro
\section{Introduction} 
%%%%%%%%%%%%%%%%%%%%%%%%%%%%%%%%%%%%%

In the last decade, impressive progress has been made in our
understanding of galaxy formation and evolution based on
multi-wavelength deep field studies.  Millimeter (mm) and
submillimeter (submm) continuum observations demonstrated that
luminous, dusty galaxies were a thousand times more abundant in the
early Universe than they are at present day
\citep[e.g.,][]{smail97,blain99b,chapman05}.  The first surveys of the
redshift distribution of dusty star-forming galaxies (DSFGs) suggested
that the DSFG population peaks at redshift $\sim2$
\citep[e.g.,][]{chapman03, chapman05}, coeval with the peak of black
hole accretion and the peak of the star formation rate density as
measured in the optical/UV \citep[e.g.,][]{hopkins06}. These studies
suggested that the bulk of star formation activity in the universe at
$z=2-3$ could be taking place in DSFGs, hidden from the view of
optical/UV observations due to the high dust obscuration
\citep[e.g.,][]{hughes98,blain99b}.

Optical surveys now allow estimates of the history of star formation
\citep[the `Madau-Lilly' plot; ][]{madau96, lilly96, hopkins06} out to
$z\sim8$ \citep[e.g.,][]{bouwens10,bouwens11}, but have uncertain dust
extinction corrections.  Submm observations can provide a more
complete picture of the amount of highly obscured star formation over
a large range of look-back times. However, such studies have been
hampered by the difficulty of obtaining robust redshifts for DSFGs.
This difficulty increases strongly as a function of redshift, and
mainly arises from the coarse spatial resolution ($\sim$$20''$) of
single-dish submm observations and the dust-obscured nature of the
sources, which often prohibits identification of counterparts at other
wavelengths.
% for which redshifts could be determined.
%This difficulty increases strongly as a function of redshift, and
%mainly arises from the coarse spatial resolution of submm deep blank
%field observations ($\sim20''$), which often prohibits identification
%of counterparts at other wavelengths for which redshifts could be
%determined. Furthermore the large dust content of DSFGs means that
%they often have only weak (if any) counterparts in the rest-frame
%ultraviolet and optical due to dust obscuration. 
The solution has been to obtain higher spatial resolution data,
usually at radio and/or mid-infrared wavelengths, in which the most
likely counterpart to the submm emission could be identified
\citep[e.g.,][]{ivison02,ashby06,pope06,wardlow11,yun12}.  The slope
of the spectral energy distributions (SEDs) of galaxies in the radio
or mid-infrared (MIR), however, is such that the K-correction is positive,
and galaxies become more difficult to detect at high redshifts.  By
contrast, the steeply rising spectrum of dusty sources leads to a
negative K-correction for DSFGs at submm wavelengths, resulting in
fluxes roughly constant with redshift \citep{blain93}.  Therefore,
while DSFGs may be discoverable at submm wavelengths at almost any
redshift, their emission may be hidden at other other wavelengths.
%As a result, it can be difficult to identify the multi-wavelength counterparts to a DSFGs, particularly at high redshift.
Indeed, in submm surveys typically 50\% of DSFGs lack robust 
counterparts \citep[e.g.,][]{biggs11} albeit the fraction depends on 
the depth of the radio/MIR observations. This mismatch in the wavelength 
sensitivity could bias the DSFG redshift distribution, particularly at $z>3$.

%The spectral energy distributions (SEDs) of galaxies result in positive 
%$K$-corrections and cosmic dimming for measurments in the radio or 
%mid-infrared; however, measurements at mm-wavelengths experience a 
%negative $K$-correction resulting in fluxes roughly constant with redshift 
%\citep{blain93}. Therefore, while DSFGs exhibit the remarkable feature of 
%being detectable at any redshift, the multi-wavelength counterparts for 
%DSFGs may thus remain undetected. Indeed, in submm surveys
%typically 30-50\% of DSFGs lack robust counterparts
%\citep[e.g.,][]{biggs11}. This mismatch in the wavelength sensitivity
%could bias the DSFG redshift distribution, particularly at $z>3$.

A more reliable and complete method to obtain secure multi-wavelength
identifications of DSFGs is to follow the single-dish detections up
with mm interferometry. Prior to ALMA this method has proven to be
time-intensive, requiring entire nights of time detect a single
source; the first sample detected blindly in the continuum with mm
interferometry was published by \citet{younger07}.  A larger sample
was published recently by \citet{smolcic12}, which included optical
spectroscopic redshifts for roughly half the sample and photometric
redshift estimates the remaining sources in the sample which suggested
that the previous spectroscopically determined redshift distributions
\citep[e.g.,][]{chapman05} were biased low.

A more direct and unbiased way to derive redshifts of DSFGs is via
observations of molecular emission lines at millimeter wavelengths
which can be related unambiguously to the (sub)mm continuum source.
This method has only become competitive over the past years with the
increased bandwidth of mm/submm facilities. Its power to measure
reliable redshifts has been demonstrated in the case of SMMJ14009+0252
and HDF850.1 \citep{weiss09a,walter12}, two of the first DSFGs
detected by SCUBA, for which other methods failed to deliver redshifts
for more than a decade. While CO redshift surveys of a representative
sample of DSFGs will remain observing time expensive till the
operation of full ALMA, CO line redshifts for strongly lensed systems
can be obtained easily \citep[e.g.][]{swinbank10,cox11,frayer11}.

In the past studies of strongly lensed sources have been limited to a
handful of targets due to their rareness and the lack of large scale
mm/submm surveys. This has changed dramatically over the past years
with the advent of large area surveys from {\it Herschel}
\citep[specifically H-ATLAS and HerMES][]{eales10,oliver10} and the
South Pole Telescope \citep[SPT-SZ,][]{carlstrom11}. These surveys
have detected hundreds of strongly lensed high-redshift DSFGs
\citep[][]{vieira10,negrello10}.  First CO redshift measurements at mm
\citep{lupu12} and centimeter \citep{harris12} wavebands of H-ATLAS
sources suggested that the lensed DSFGs lie within the same redshift
range as unlensed, radio-identified sources \citep{chapman05}.
Although a large overlap between the SPT and {\it Herschel}
populations is expected, SPT's longer selection wavelength of 1.4\,mm
predicts a broader redshift distribution than {\it Herschel} detected
sources and indeed photometric redshifts of DSFGs discovered by the
SPT confirm this expectation \citep{greve12}.

In this paper, we present the results from an ALMA CO redshift survey of
a sample of 26 strongly lensed DSFGs selected from 1300\,deg$^2$ of
SPT-SZ survey data \citep[][]{carlstrom11}.  The depth of the SPT-SZ
survey data, which is sufficient to detect $S_{1.4\rm mm}$$\sim$20 mJy
sources at 5$\sigma$, combined with the flat redshift selection
function of DSFGs at this wavelength \citep[e.g.,][]{blain93}, has
produced an optimal sample for mm molecular line redshift searches in
strongly lensed DSFGs.  In an accompanying paper, \citet{vieira13}
show that these sources are virtually all strongly lensed, while
\citet{hezaveh13} report the associated lens modeling procedure.

In Section \ref{sect:observations}, we describe the target selection
and observations.  The biases on the observed redshift distribution,
resulting from the source selection and the effect of gravitational
lensing are discussed in Section \ref{sect:discussion}.  Our results
are summarized in Section \ref{sect:summary}.  Throughout this paper
we have adopted a flat WMAP7 cosmology, with $H_{\rm 0} = 71$
\kms\,Mpc$^{-1}$, $\Omega_\Lambda=0.73$ and $\Omega_m=0.27$
\citep[][]{komatsu11}.

%%%%%%%%%%%%%%%%%%%%%%%%%%%%%%%%%%%%%
% 2. Observations
\section{Observations}
\label{sect:observations}
%%%%%%%%%%%%%%%%%%%%%%%%%%%%%%%%%%%%%

We observed a sample of 26 bright (S$_{1.4\rm{mm}}> 20$\,mJy), 1.4\,mm
selected SPT sources with ALMA. Sources were selected from the first
1300 deg$^2$ of the now complete 2500\,deg$^2$ SPT-SZ survey
\citep[for more details on the survey, see][]{williamson11,story12}.
The flux density cut is done on the initial
raw flux density, and not the final de-boosted flux density, the
details of which can be found in \citet[][]{vieira10} and
\citet[][]{crawford10}. To remove synchrotron dominated systems
we required dust-like spectra between 1.4 and 2~mm 
\citep[S$_{1.4{\rm mm}}/$S$_{2.0{\rm mm}}$$>$$2$; ][]{vieira10}.
In addition, we used far-infrared (FIR) and/or radio criteria to
remove low-redshift contaminants (see Section \ref{Sect:lowzbias}).
In order to refine the relatively coarse SPT source positions
  (the SPT's beam size is $1'.05$ at 1.4\,mm) we further required
  follow up observations at higher spatial resolution (typically
  870\micron\ images from the Large Apex BOlometer CAmera (LABOCA) or
  1\,mm data from the Submillimeter Array). Based on 1.4~mm flux
densities, our Cycle 0 targets comprise a representative sample of the
SPT sources meeting these selection criteria.  This is shown in
  Appendix \ref{sect:supplementary_photometry} where we present the
  SPT 1.4\,mm, LABOCA 870\micron, and {\it Herschel}-SPIRE 350\micron\
  flux density properties of this subsample compared to all SPT
  sources which have been observed with {\it Herschel} and LABOCA.

In order to optimize the ALMA
observing efficiency, we assembled 5 groups of targets that lie within
15$^\circ$ of each other on the sky -- this restriction precluded a
complete flux-limited sample. 
We excluded two sources with
redshifts previously determined by Z-Spec \citep[a wide-band, low
resolution spectrometer operating between 190-310\,GHz,
see][]{bradford04} on the Atacama Pathfinder Experiment (APEX)
telescope and XSHOOTER \citep{Vernet11} or the the FOcal Reducer and
Spectrograph \citep[FORS2;][]{appenzeller98} on the ESO Very Large
Telescope (VLT).

\begin{figure}[htb]
\centering
\includegraphics[width=8.5cm,angle=0]{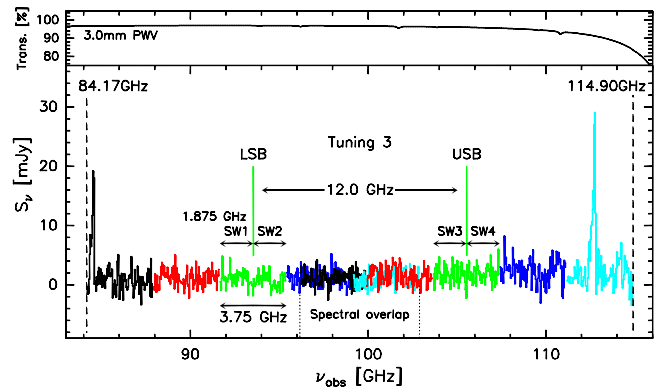}
\caption{Spectral setup and frequency coverage of our 5 tunings
  (shown in different colors) in ALMA band 3 for the source
  SPT0103-45.  In each tuning, four spectral windows covering 1.875\,GHz each
  were placed in contiguous pairs in the lower and upper sidebands
  (LSB/USB). Note that the frequency range 96.2--102.8\,GHz (delimited
  by dotted vertical lines) is covered twice. The total spectral range
  is indicated by the dashed vertical lines. The top panel shows the
  atmospheric transmission across band 3 at Chajnantor for 3\,mm
  precipitable water vapor (PWV). }
\label{tunings}
\end{figure}

The ALMA observations were carried out in 2011 November and 2012
January in the Cycle 0 early science compact array configuration.  We
performed a spectral scan in the 3\,mm atmospheric transmission window
with five tunings in dual polarization mode.  Each tuning covers
7.5~GHz in two 3.75~GHz wide sidebands, each of which is covered by
two 1.875~GHz spectral windows in the ALMA correlator.  This setup
spans 84.2 to 114.9\,GHz (with 96.2 to 102.8~GHz covered twice; see
Figure~\ref{tunings}), nearly the entire bandwidth of the Band~3
(84$-$116~GHz) receiver. Over this frequency range ALMA's primary
  beam is $61''-45''$.
%much larger than the $870\um$ source position
%accuracy ($<6''$) which have been used as pointing centers.  
The observations employed between 14 and 17 antennas in different
sessions, and resulted in typical synthesized beams of
7$''\times$5$''$ to 5$''\times$3$''$ (FWHM) from the low- to
high-frequency ends of the band.  Each target was observed for $\sim
120$ seconds in each tuning, or roughly 10 minutes per source in
total, not including overheads.

Typical system temperatures for the observations were T$_{\rm
  sys}=60$\,K. Flux calibration was performed on planets (Mars,
Uranus, or Neptune) or Jupiter's moons (Callisto or Ganymede), with
passband and phase calibration determined from nearby quasars. The
data were processed using the Common Astronomy Software Application
package \citep[CASA,][]{mcmullin07,petry12}.  Calibrated data cubes
were constructed with a channel width of 19.5\,MHz ($\sim50-65$\,\kms\
for the highest and lowest observing frequency). The typical noise per
channel is 2\,mJy\,beam$^{-1}$ across the band and
1.4\,mJy\,beam$^{-1}$ between 96.0 and 102.8~GHz where two tunings
overlap.  Continuum images generated from the full bandwidth have
typical noise levels of 70\,$\mu$Jy\,beam$^{-1}$.

The spectral coverage of this experiment includes CO(1--0) for
0.003$<$$z$$<$0.36 and one or more CO lines, between the (2--1) and
(7-6) transitions, between 1.0$<$$z$$<$8.6, with the exception of a
small redshift ``desert'' between 1.74$<$$z$$<$2.00 (see
Figure~\ref{tuningZ}).  An additional redshift desert at
0.36$<$$z$$<$1.0 is also present, but our high 1.4~mm flux density
threshold effectively requires that our sources be gravitationally
lensed (\S\ref{Sect:lowzbias}) and it is highly unlikely that sources
at this redshift will be lensed (\S\ref{Sect:lensbias}).

%---------------------------------------------
% Fig 1 Spectral tunings
%---------------------------------------------

%---------------------------------------------
% Fig 2 Redshift coverage
%---------------------------------------------
\begin{figure}[htb]
\centering
\includegraphics[width=8.5cm,angle=0]{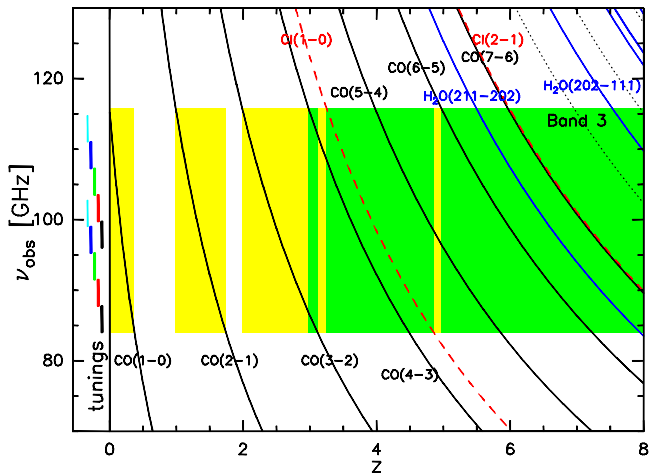}
\caption{Spectral coverage of the CO, [CI], and H$_2$O emission lines
  as a function of redshift.  The \emph{green} shaded region marks the
  redshifts where two or more strong lines lines provide an
  unambiguous redshift, while the \emph{yellow} region marks redshift
  range where only a single line is detectable.  The five
  frequency tunings are shown in the left panel (see also
  Figure~\ref{tunings}).  }
\label{tuningZ}
\end{figure}

%%%%%%%%%%%%%%%%%%%%%%%%%%%%%%%%%%%%%
% 3. Results
\section{Results} 
\label{sect:results}
%%%%%%%%%%%%%%%%%%%%%%%%%%%%%%%%%%%%%

We detect 3\,mm continuum emission with a high signal-to-noise ratio
(SNR 8--30) in all 26 sources; all sources remain spatially
unresolved in these compact configuration data. Within the primary beam
of ALMA we do not detect any other source at the sensitivity limit of 
our observations. Table
\ref{Tab:contflux} lists the ALMA 3\,mm continuum positions, while the
3\,mm continuum flux densities are given in
Appendix~\ref{sect:supplementary_photometry} together with other
photometric measurements.

Figure\,\ref{Fig:spectra} presents the spectra. In total, we detect 44
line features with line integrated ${\rm SNR}>5$ in our survey, which we identify as emission lines of
$^{12}$CO, $^{13}$CO, \ci, H$_2$O, and H$_2$O$^+$. Our spectra can be
grouped into three categories:
\begin{itemize}
\item Spectra with no line features (3 sources). 
\item Spectra with a single line feature (11 sources). For these spectra we cannot determine the redshift 
unambiguously and use other spectroscopic and photometric measurements  to constrain the redshift.
\item Spectra with multiple line features (12 sources). In this case, a unique redshift solution can be 
derived from the ALMA 3\,mm spectral scans alone.
\end{itemize}
Table \ref{Tab:lineIDs} summarizes the detected line features and the
derived redshifts. Uncertainties for the redshifts are based on
Gaussian fits to the line profiles.  The identification of the
ambiguous features is discussed in Section \ref{Sect:lineIDs}.

%---------------------------------------------
% Fig 3 Spectra
%---------------------------------------------

\begin{figure*}[htb]
\centering
\includegraphics[width=17.5cm,angle=0]{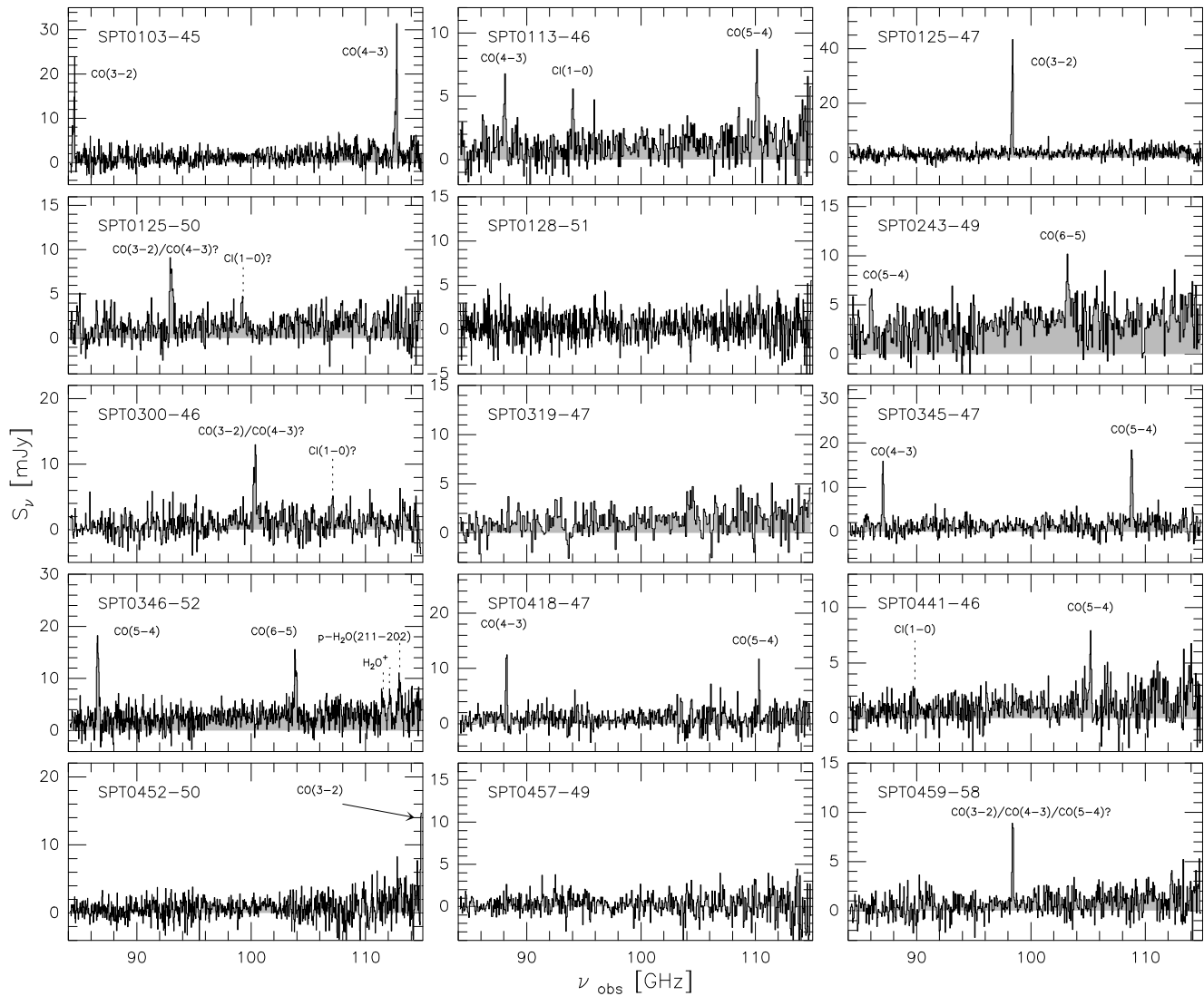}
\caption{Spectra for target galaxies in the ALMA the 
3\,mm band. Spectra are shown at a resolution of 
40--70\,MHz ($\sim 100-250$\,\kms) depending on the line width and the signal to noise ratio.}
\label{Fig:spectra} 
\addtocounter{figure}{-1}
\end{figure*}

\begin{figure*}[htb]
%\ContinuedFloat
\centering
\includegraphics[width=17.5cm,angle=0]{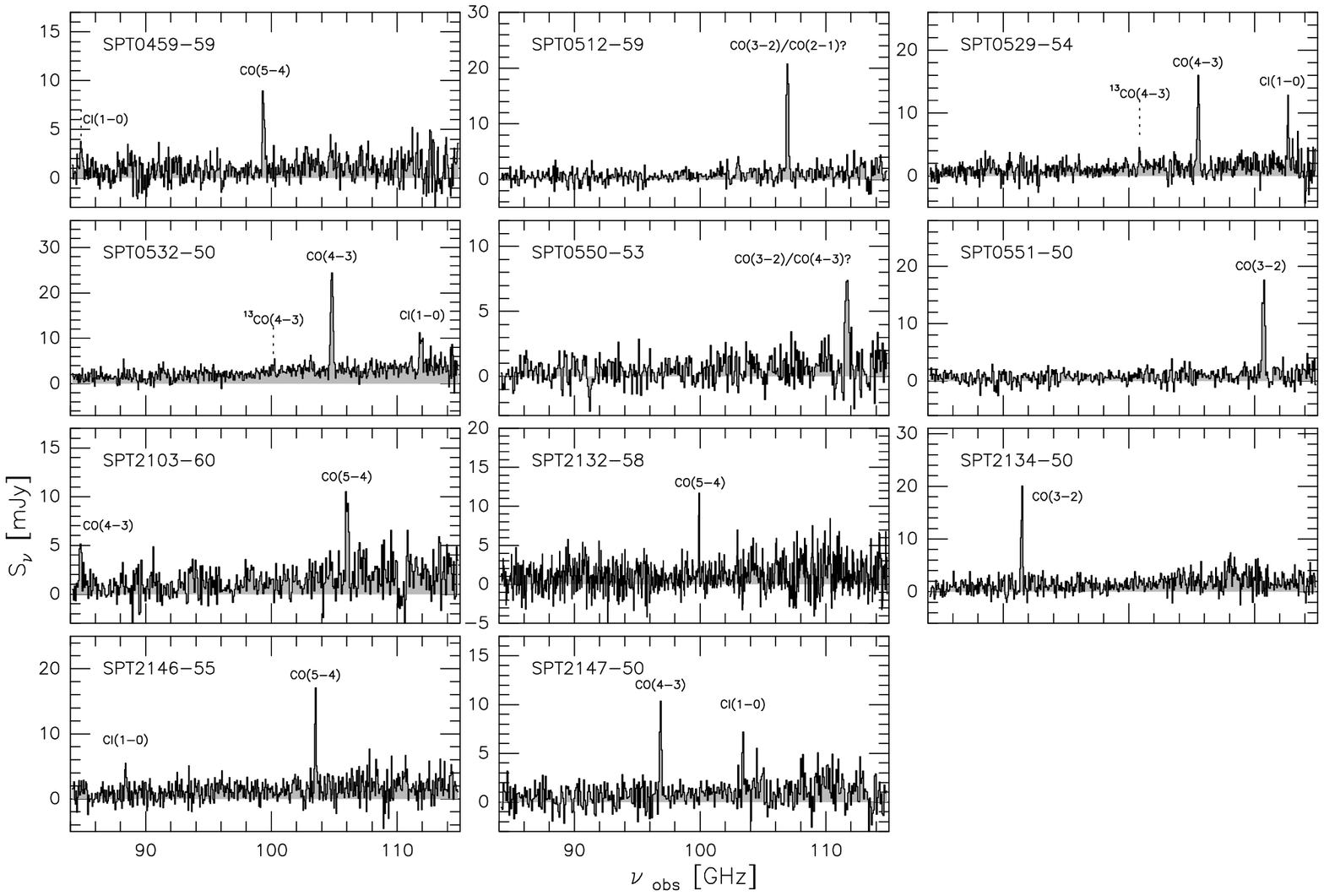}
\caption{Figure\,\ref{Fig:spectra}, continued.}
\label{Fig:spectra2} 
\end{figure*}

%---------------------------------------------
% Table 1 positions and fluxes
%---------------------------------------------

\begin{deluxetable}{cccc}
\tablecaption{ALMA source positions\label{Tab:contflux}}
\startdata
\tableline
\tableline
\\
Short name & Source& R.A. & Dec.\\
           &       &   \multicolumn{2}{c}{J2000} \\
\tableline
\\
SPT0103-45 &SPT-S J010312-4538.8& 01:03:11.50 & -45:38:53.9 \\ 
SPT0113-46 &SPT-S J011308-4617.7& 01:13:09.01 & -46:17:56.3 \\ 
SPT0125-47 &SPT-S J012506-4723.7& 01:25:07.08 & -47:23:56.0 \\ 
SPT0125-50 &SPT-S J012549-5038.2& 01:25:48.45 & -50:38:20.9 \\ 
SPT0128-51 &SPT-S J012809-5129.8& 01:28:10.19 & -51:29:42.4 \\ 
SPT0243-49 &SPT-S J024307-4915.5& 02:43:08.81 & -49:15:35.0 \\ 
SPT0300-46 &SPT-S J030003-4621.3& 03:00:04.37 & -46:21:24.3 \\ 
SPT0319-47 &SPT-S J031931-4724.6& 03:19:31.88 & -47:24:33.7 \\ 
SPT0345-47 &SPT-S J034510-4725.6& 03:45:10.77 & -47:25:39.5 \\ 
SPT0346-52 &SPT-S J034640-5204.9& 03:46:41.13 & -52:05:02.1 \\ 
SPT0418-47 &SPT-S J041839-4751.8& 04:18:39.67 & -47:51:52.7 \\ 
SPT0441-46 &SPT-S J044143-4605.3& 04:41:44.08 & -46:05:25.5 \\ 
SPT0452-50 &SPT-S J045247-5018.6& 04:52:45.83 & -50:18:42.2 \\ 
SPT0457-49 &SPT-S J045719-4932.0& 04:57:17.52 & -49:31:51.3 \\ 
SPT0459-58 &SPT-S J045859-5805.1& 04:58:59.80 & -58:05:14.0 \\ 
SPT0459-59 &SPT-S J045912-5942.4& 04:59:12.34 & -59:42:20.2 \\ 
SPT0512-59 &SPT-S J051258-5935.6& 05:12:57.98 & -59:35:41.9 \\ 
SPT0529-54 &SPT-S J052902-5436.5& 05:29:03.09 & -54:36:40.0 \\ 
SPT0532-50 &SPT-S J053250-5047.1& 05:32:51.04 & -50:47:07.5 \\ 
SPT0550-53 &SPT-S J055001-5356.5& 05:50:00.56 & -53:56:41.7 \\ 
SPT0551-50 &SPT-S J055138-5058.0& 05:51:39.42 & -50:58:02.1 \\ 
SPT2103-60 &SPT-S J210328-6032.6& 21:03:30.90 & -60:32:40.3 \\ 
SPT2132-58 &SPT-S J213242-5802.9& 21:32:43.23 & -58:02:46.2 \\ 
SPT2134-50 &SPT-S J213404-5013.2& 21:34:03.34 & -50:13:25.1 \\ 
SPT2146-55 &SPT-S J214654-5507.8& 21:46:54.02 & -55:07:54.3 \\ 
SPT2147-50 &SPT-S J214720-5035.9& 21:47:19.05 & -50:35:54.0 \\ 
\enddata
\tablenotetext{}{Source names are based on positions measured with the SPT. Source positions are
based on the ALMA 3\,mm continuum data.}
\end{deluxetable}

%---------------------------------------------
% Table 2 Redshifts
%---------------------------------------------

\begin{deluxetable*}{lccll}

\tablecaption{Redshifts and line identification \label{Tab:lineIDs}}
\startdata
\tableline
\tableline
\\
Source &z & T$_{\rm dust}  [K]$ & lines & comment\\
\\
%       &  & [K]            &        &  \\
\tableline 
\multicolumn{5}{c}{secure redshifts}\\
\hline
SPT0103-45&  {\bf 3.0917(3)}$^a$   & 33.3$\pm$2.5   & CO(3-2) \& CO(4-3)           & \\
SPT0113-46&  {\bf 4.2328(5) }   & 31.8$\pm$3.1  & CO(4-3), CI(1-0) \& CO(5-4)  & \\
SPT0125-47&  {\bf 2.51480(7)}   & 40.7$\pm$4.2       & CO(3-2)                      & CO(1-0) from the ATCA\\
SPT0243-49&  {\bf 5.699(1)  }   & 30.1$\pm$4.9       & CO(5-4) \& CO(6-5)           & \\
SPT0345-47&  {\bf 4.2958(2) }   & 52.1$\pm$7.8       & CO(4-3) \& CO(5-4)           & \\
SPT0346-52&  {\bf 5.6559(4) }   & 52.9$\pm$5.3       & CO(5-4), CO(6-5), H$_2$O \& H$_2$O$^+$  & \\
SPT0418-47&  {\bf 4.2248(7) }   & 52.9$\pm$7.5       & CO(4-3) \& CO(5-4)           & \\
SPT0441-46&  {\bf 4.4771(6) }   & 39.3$\pm$3.9       & CI(1-0) \& CO(5-4)           & CI(1-0) feature low SNR, [\cii] confirmation with APEX\\
SPT0452-50&  {\bf 2.0104(2) }   &20.9$\pm$1.8        & CO(3-2)                      & alternative redshifts excluded due to lack of higher J CO lines \\
SPT0459-59&  {\bf 4.7993(5) }   &36.0$\pm$3.7        & CI(1-0) \& CO(5-4)           & \\
SPT0529-54&  {\bf 3.3689(1) }   &31.9$\pm$2.4        & CO(4-3), CI(1-0) \& $^{13}$CO(4-3) &\\
SPT0532-50&  {\bf 3.3988(1) }   &35.1$\pm$3.0        & CO(4-3), CI(1-0) \& $^{13}$CO(4-3) & \\
SPT0551-50&  {\bf 2.1232(2) }   &26.3$\pm$2.0        & CO(3-2)                      & VLT C{\tiny IV}\,${\rm 1550\,\AA}$ detection\\
SPT2103-60&  {\bf 4.4357(6) }   &38.6$\pm$3.5        & CO(4-3) \& CO(5-4)           & \\   
SPT2132-58&  {\bf 4.7677(2)}     &37.8$\pm$4.5        & CO(5-4)                      & [\cii] from APEX \\ 
SPT2134-50&  {\bf 2.7799(2) }   &40.5$\pm$4.6        & CO(3-2)                      & CO(7-6) \& CO(8-7) detections from Z-Spec \& the SMA\\
SPT2146-55&  {\bf 4.5672(2) }   &38.7$\pm$5.1        & CI(1-0) \& CO(5-4)           & \\
SPT2147-50&  {\bf 3.7602(3) }   &41.8$\pm$4.1        & CO(4-3) \& CI(1-0)           & \\
\\
SPT0538-50& {\bf 2.783}      &31.2$\pm$7.1        & CO(7-6), CO(8-7), Si\,{\tiny IV}\,${\rm 1400\,\AA}$  & ZSpec/VLT from \citet[][]{greve12}; no ALMA data\\
SPT2332-53& {\bf 2.738}      &32.9$\pm$3.6        & CO(7-6), Ly$\alpha$, C{\tiny IV}\,${\rm 1549\,\AA}$  & ZSpec/VLT from \citet[][]{greve12}; no ALMA data\\
\hline
\multicolumn{5}{c}{ambiguous redshifts}\\
\hline

SPT0125-50             & {\bf 3.9592(5)} & 43.3$\pm$5.2 & CO(4-3) \& CI(1-0) & CI(1-0) feature low SNR\\
\multicolumn{1}{r}{...}&  2.7174(6)      & 29.5$\pm$3.2 & CO(3-2)            & alternative ID if CI(1-0) is not real\\
SPT0300-46             & {\bf 3.5956(3)} &38.6$\pm$3.6  & CO(4-3) \& CI(1-0) & CI(1-0) feature low SNR\\
\multicolumn{1}{r}{...}& 2.4474(3)       &26.7$\pm$2.2  & CO(3-2)            & alternative ID if CI(1-0) is not real\\
SPT0459-58             & {\bf 3.6854(2)} &32.0$\pm$4.5  & CO(4-3)            & \\
\multicolumn{1}{r}{...}& 4.8565(2)       &40.8$\pm$6.0  & CO(5-4)            & similarly likely ID\\
\multicolumn{1}{r}{...}& 2.5142(1)       &22.4$\pm$2.9  & CO(3-2)            &\\
SPT0512-59             & {\bf 2.2335(2)} &33.2$\pm$3.0  & CO(3-2)            & \\
\multicolumn{1}{r}{...}& 1.1557(1)       &20.4$\pm$1.6  & CO(2-1)            &\\
SPT0550-53             & {\bf 3.1286(5)} &30.6$\pm$4.6  & CO(4-3)            & \\
\multicolumn{1}{r}{...}& 2.0966(4)       &21.6$\pm$2.9  & CO(3-2)            &\\
\\
\hline
\multicolumn{5}{c}{no CO line detections}\\
\hline\\
SPT0128-51             & $-$             & $-$          & no lines           & $z$=1.74--2.00 ? ; $z_{\rm photo}=3.8\pm0.5$ for \tdust=37.2\,K\\
SPT0319-47             & $-$             & $-$          & no lines           & $z$=1.74--2.00 ? ; $z_{\rm photo}=4.2\pm0.2$ for \tdust=37.2\,K\\
SPT0457-49             & $-$             & $-$          & no lines           & $z$=1.74--2.00 ? ; $z_{\rm photo}=3.3\pm0.2$ for \tdust=37.2\,K
\enddata
\tablecomments{In case of ambiguous redshifts,  preferred solutions are shown in bold.
}
\tablenotetext{a}{The number in brackets is the redshift uncertainty in the last decimal derived from Gaussian fits to the line profiles.}
\end{deluxetable*}

\subsection{Additional spectroscopic observations}
\label{sec:moarSpec}
For five of the sources in our sample for which we have detected only a
single line in our 3\,mm scan, we determine the redshift using
additional mm/submm or optical spectroscopy. We describe the observations 
and show the spectra in Appendix~\ref{sect:supplementary_redshift_info}.

\subsection{Ambiguous cases \label{Sect:lineIDs}}

The most likely candidates for a single line feature in the 3\,mm band
are redshifted transitions of CO up to J=3--2 (see
Figure~\ref{tuningZ}). The CO(4--3) and CO(5--4) lines may also 
appear as single lines across the band in cases where the \cone\ line 
falls out of the covered frequency range or may be too faint to be 
detected (the lowest flux density ratio between \cone\ and CO(4--3) or 
CO(5--4) that we observe in our survey is $<$0.15 (3$\sigma$)).
Single-line spectra cannot result from CO transitions of J=6--5 or
higher or molecular lines that can appear at flux densities comparable
to CO \citep[such as \water,][2011]{vanderWerf10}, because these lines
would be accompanied by another line within the observing band (see
Figure~\ref{tuningZ}). The detection of FIR fine 
structure lines, such as 122\,$\mu$m and 205\,$\mu$m [\nii]  and 
158\,$\mu$m [\cii] would require extreme redshifts ($z$$>$11) which are 
inconsistent with mm/submm continuum measurements.

Photometric measurements allow us to discriminate between the possible line 
assignments in our single-line sources. 
The thermal dust emission of our sources is sampled by 3\,mm ALMA, 2
\& 1.4 mm SPT, and 870\,$\mu$m LABOCA as well as 500, 350, and
250 \,$\mu$m {\it Herschel}--SPIRE observations. The photometry is given
in Appendix~\ref{sect:supplementary_photometry} .

%All sources 
%are also part of an ongoing {\it Herschel}--SPIRE 500, 350, and
%250\,$\mu$m photometry project; data for the 24 sources available to
%us at the time of writing are presented in
%Appendix~\ref{sect:supplementary_photometry}.  
%For SPT0103-45, which
%has not yet been observed with SPIRE, we have 350\,$\mu$m fluxes from
%the Submillimetre APEX Bolometer Camera (SABOCA) at APEX.  Both
%sources without SPIRE photometry show multiple lines in our ALMA
%spectra and have unambiguous redshifts.

For the fitting of the thermal dust continuum we have used the method
described in \citet{greve12} which uses a greybody fit with a spectral
slope of $\beta$=2 and an optically thin/thick transition wavelength of
$100$\micron, where the only free parameters are the dust
luminosity and the dust temperature, \tdust.  As in \citet{greve12},  we exclude
data points shortward of $\lambda_{rest}=50\um$ from the fit 
because a single-temperature SED model typically cannot match both 
sides of the SED peak simultaneously due to the presence of dust at multiple temperatures. 
Both the spectral slope
and transition wavelength affect the derived dust temperatures.  For
the present purpose, we seek only a consistent measure of the location
of the SED peak in each source; the ``temperatures'' should not be
interpreted as physical temperatures.  The dust temperature is better
derived using the source structural information that will be available
with lens models based on high spatial resolution ALMA observations
\citep{hezaveh13}, which will help constrain the dust opacity
\citep[e.g.,][]{weiss07}.

\begin{figure*}[ht]
\centering
\includegraphics[width=18cm,angle=0]{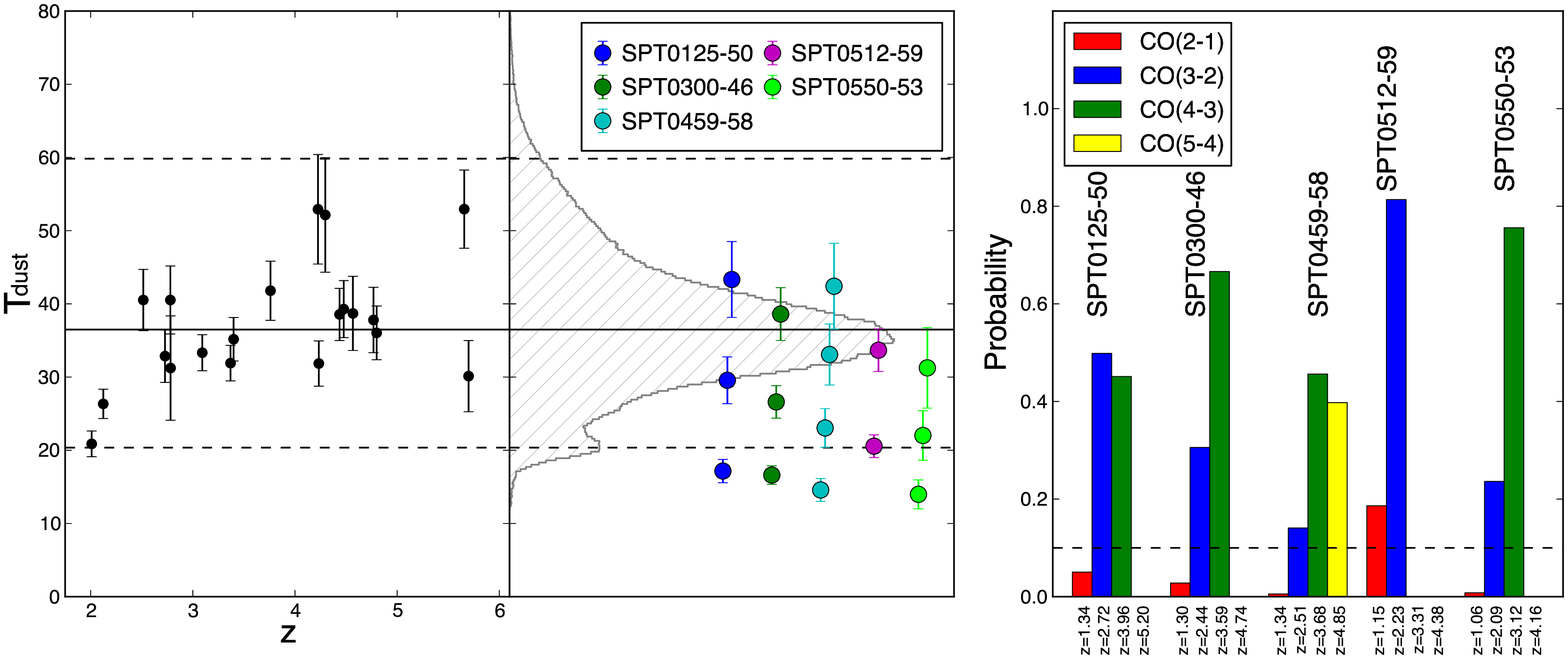}
\caption{
  {\it Left:} Dust temperatures for the sources in our sample with
  unambiguous redshifts.  {\it Center:} Combined histogram of dust
  temperatures derived from the posterior likelihood distributions for
  the sources with unambiguous redshifts.  Overplotted are the dust
  temperatures determined for each redshift option for those sources
  with uncertain redshift; horizontal spacing is arbitrary.  The solid
  and dashed lines show the median and 95\% confidence interval dust
  temperatures for those sources with unambiguous redshifts.  {\it
    Right:} Probability for the single line detected in our ALMA
  spectrum to be identified as one of the four possible CO transitions
  for the five sources with ambiguous redshifts. The probabilities were
  calculated by comparing the dust temperature associated with each
  line identification to the dust temperature distribution of our
  sources with known redshifts. The horizontal dashed line shows a probability of 10\%, the cut off above 
which we  consider the line identification to be plausible.}
\label{Fig:tdust} 
\end{figure*}

Given the fundamental degeneracy between \tdust\ and redshift due to
Wien's displacement law, it is not possible to solve for $z$ and
\tdust\ simultaneously. We therefore determine \tdust\ for each of the
possible redshifts and compare these to the dust temperature
distribution for targets with unambiguous redshifts (see Table
\ref{Tab:lineIDs}), including the two SPT sources with previously
known redshifts from \citet[][]{greve12} which share the same selection criteria than
the sample discussed here.  For these sources we find
a mean of T$_{\rm dust}$=37.2$\pm$8.2\,K and no apparent trend with redshift
(see Figure\,\ref{Fig:tdust}, left).  Based on the distribution of the
temperatures in this sub-sample (19 sources) we have calculated the
probability for each of the four dust temperature/redshift options for
the six sources with a single detected line and ambiguous redshifts.
This analysis strongly prefers a single redshift for one additional
source (SPT0452-50, see Appendix~\ref{sect:supplementary_singleline}).  
We have included SPT0452-50 in the
sample of sources with known redshifts, bringing the total to 20.

Figure\,\ref{Fig:tdust} shows the five remaining sources with ambiguous
redshifts.  We retain only dust temperatures with probabilities
$>$10\%, and find two plausible line identifications/redshifts for
four sources.  In one case this threshold only rules out CO(2--1),
leaving three plausible redshifts.
Table\,\ref{Tab:lineIDs} lists the possible redshifts together with
the implied dust temperatures.  Entries in bold face show the most
likely redshift solution.

In the case of the three sources without line features, we derive
photometric redshifts based on the FIR data using the mean dust temperature 
of the objects with unambiguous redshifts. This places these three sources 
between $z$=3.3-4.2 (see right column of Table~\ref{Tab:lineIDs}). Of the two
redshift ranges for which we cannot observe a CO line, the
0.36$<$$z$$<$1 range can be excluded because the SED would then imply
T$_{\rm dust}$ lower than the dust temperatures of the Milky Way and
other spiral galaxies ($\tdust$$<$15\,K for all sources) and due to
the small lensing probability (\S\ref{Sect:lensbias}).  The galaxies
may then be in the redshift desert at $z$=1.74--2.00 or at higher
redshift with CO line intensities below our detection threshold.  Our
redshift survey sensitivity was intended to detect CO lines out to
$z$$\sim$6, based on molecular gas estimates from the dust continuum,
and strong detections of emission lines in 90\% of the targets out to
$z=5.7$ lends credibility to the sensitivity target. However, two of
our non-detections are among the 1.4mm-faintest sources which leaves
open the possibility that the line sensitivity is inadequate in these
cases, albeit we do detect CO lines at similar 1.4\,mm flux density in
the survey.  Yet, estimates of the CO (and \ci) line intensities based
on the dust continuum observations alone require several strong
assumptions (e.g., on the gas-to-dust mass ratio and the molecular gas
excitation). Thus we cannot rule out that these systems represent a
class of galaxies with lower than expected line to continuum ratio,
with the lines falling below our detection limit. If we place these
three galaxies at $z$=1.74--2.00, we obtain low
($\tdust$$\approx$20\,K), but still plausible dust temperatures given
the \tdust\ distribution in our sub-sample with known redshifts.  We
note that this redshift identification is by no means secure, but
represents the lowest plausible redshift range given the estimates
based on the photometric data discussed above.

A discussion of the 9 individual cases which have zero or one CO line detected with ALMA 
and no additional spectroscopic observations is presented in Appendix\,\ref{sect:supplementary_photometry}.\\

%---------------------------------------------
% Fig 4 Dust Temperatures
%---------------------------------------------

%%%%%%%%%%%%%%%%%%%%%%%%%%%%%%%%%%%%%
% 4. Discussion
\section{Discussion}
\label{sect:discussion}
%%%%%%%%%%%%%%%%%%%%%%%%%%%%%%%%%%%%%

%---------------------------------------------
% Fig 5  Veto and SEDs
%---------------------------------------------
\begin{figure}[htb]
\centering
\includegraphics[width=8.4cm,angle=0]{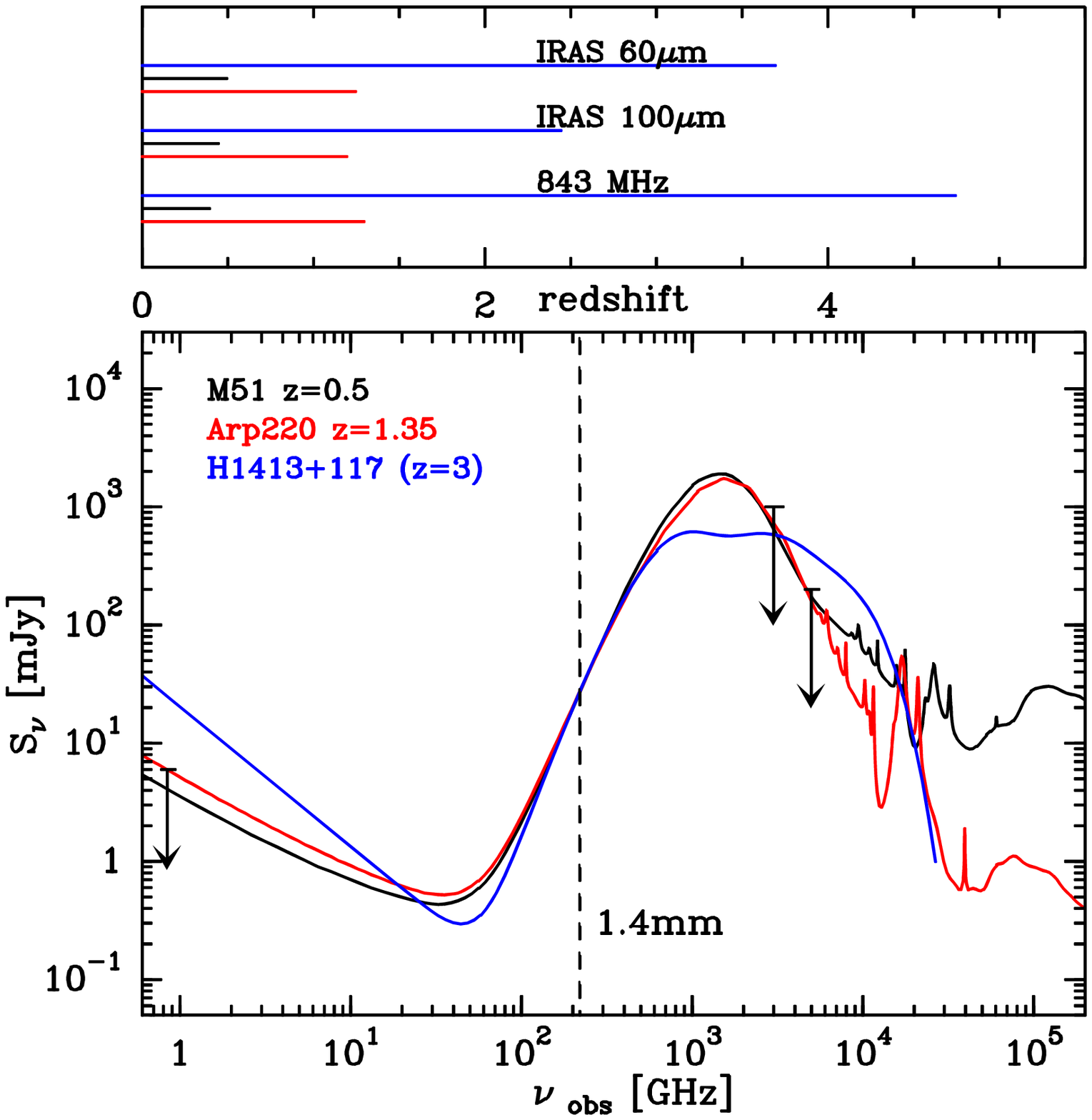}
\caption{{\it Top:} Redshift bias due to our IRAS 60 and 100\microns, and 843\,MHz 
radio flux vetos. The bars show the redshift range for which specific radio-to-IR SEDs 
are excluded from our sample. The color coding of the bars corresponds to galaxies shown 
in the bottom part of the figure. {\it Bottom:} Radio to optical SEDs of M51 (the Whirlpool Galaxy), 
Arp\,220 (the nearest ultraluminous infrared galaxy) and H1413+117 (the Cloverleaf QSO). 
These galaxies represent a range of possible SED types and are normalized to 
S$_{\rm 1.4mm}=28$\,mJy (the mean 1.4mm flux density of our sample). The dashed horizontal line shows our
selection wavelength of 1.4mm. The arrows show the
843\,MHz, 100\microns\ and 60\microns\ upper limits used for our source selection. 
The SEDs are shown for the lowest redshift (value indicated in the figure) for which each source 
matches our selection criteria, except for H1413+117  which is shown at $z=3.0$. 
}\label{Fig:lowzcut}
\end{figure}

\subsection{Redshift biases due to source selection criteria \label{Sect:lowzbias}}

From our line identifications in Table\,\ref{Tab:lineIDs}, it is
apparent that the lowest secure redshift detected in our survey is at
$z$=2.010. Only five sources are possibly at $z$$\le$2 (assuming that sources 
without a line detection fall into the redshift desert $z$=1.74--2.00). This 
is in contrast to the expectation from radio-identified DSFG redshift surveys, where
typically $\sim$50\% of all sources fall into the redshift range
0.5$<$$z$$<$2.0 \citep[e.g.,][]{chapman05,wardlow11}.

Part of this discrepancy arises from our source selection criteria.
In order to select strongly lensed, dusty high-redshift sources from
the SPT 1.4\,mm maps efficiently, additional criteria are used to
distinguish the high-$z$ population from the low-$z$ and
synchrotron-dominated sources that dominate the number counts of
S$_{1.4{\rm mm}}$$>$20\,mJy sources. 
\citet[][]{vieira10} present a discussion of the classification of
these populations and the details on how to distinguish them. Below,
we provide a summary of the selection criteria and discuss their
impact.

We first select sources whose mm flux is dominated by thermal dust
emission.  This step is based on the ratio of 1.4 to 2.0 mm flux
density and is efficient at removing any synchrotron-dominated source
from the sample, the majority of which are flat-spectrum radio quasars
(FSRQs) and have previously been cataloged at radio wavelengths.  We
impose a flux density cut on the sample of dust-dominated sources of
$S_{1.4{\rm mm}}>20$ mJy based on the raw fluxes determined on
the 1.4\,mm maps.
%This flux density cut is done on the initial
%raw flux density, and not the final de-boosted flux density, the
%details of which can be found in \citet[][]{vieira10} and
%\citet[][]{crawford10}.

The second step is to use external FIR catalogs to
remove (`veto') low-redshift sources from the sample of dusty sources.
Any source detected in the IRAS Faint Source Catalog
\citep[IRAS-FSC,][]{moshir92} at 60 or 100\micron\ (which implies
S$_{60\micron}$$<$200\,mJy and S$_{100\micron}$$<$1\,Jy over the entire SPT field) is omitted
from our source sample.  This removes $\sim70\%$ of the dusty
sources from our sample.
%and should effectively remove any dusty source
%at a redshift of $z$$\lesssim$1.4, depending on the characteristic
%dust temperature of the source.  
Every dusty source with a counterpart
in both the SPT and IRAS-FSC catalogs has a published spectroscopic
redshift at $z$$<$0.03 and is not strongly lensed.

The third step is to use external radio catalogs to remove
low-redshift and radio-loud sources from the sample of dusty sources.
Any source detected in the 843\,MHz Sydney University Molonglo Sky
Survey \citep[SUMSS,][]{bock98} (with a $\sim$$6$\,mJy 5$\sigma$ flux density
threshold over the entire SPT field) is omitted from our source sample.  
The SUMSS veto removes an additional $\sim15$\% of the dusty sources which passed the IRAS veto.  
%Eight dusty
%sources identified in the 1300~deg$^2$ SPT-SZ survey were not in the
%IRAS FSC, and were found in the SUMSS catalog, and were therefore
%excluded from this sample.  
This step is intended to ensure that no FSRQs were allowed into the sample. 
%and to 
%exclude very low-redshift galaxies that lie in gaps in the IRAS coverage within the SPT survey. 
The mean radio flux density reported in
the SUMSS catalog for these sources is 
$\langle$$S_{843{\rm MHz}}\rangle$=52\,mJy, well above the catalog threshold.  

The effect of these selections on the redshift distribution of the
1.4\,mm sources targeted in this study depends on the intrinsic
radio-IR SEDs of the DSFGs.  Figure\,\ref{Fig:lowzcut} shows the
redshift limits beyond which different radio-IR SEDs pass our
source veto criteria.  We show here well-studied examples of quiescent and
star-forming local galaxies, as well as an example for a
high-redshift, radio-loud active galactic nucleus (AGN) host galaxy.
The figure demonstrates that galaxies which follow the local radio-FIR
correlation 
%and have moderate far-infrared (FIR) luminosities 
and have relatively cold dust temperatures (\tdust$\lesssim$$30$ K, e.g., M51)
would pass our source selection criteria at relatively low redshift
($z$$\gtrsim$0.5). 
%However, in order for such a galaxy to be luminous enough to 
%be detected by SPT, it must be very strongly magnified, as the 1.4~mm emission from an unlensed 
%M51 at $z=0.01$ is just a few milli-Jansky. 

Sources with Arp\,220-like SEDs would pass our selection criteria at
higher redshifts ($z$$\gtrsim$1.4). 
%Again, such sources must be lensed, as Arp~220 
%itself falls below the SPT detection threshold at $z\sim0.06$, which is consistent with the redshifts 
%found for IRAS-vetoed sources. 
Other local and high-$z$ IR luminous sources, including M82,
SMM\,J2135-0102 (`The Eyelash' -- \citealt{swinbank10}), and HR10
\citep[][not shown]{stern06}, are allowed at redshifts similar to
Arp~220.  Sources with FIR SEDs dominated by hotter dust (due to AGN
heating, as in H1413+117, also known as `The Cloverleaf';
\citealt{benford99}) than is typical for star-forming systems would be
found in IRAS and excluded from the sample out to $z\sim3$.  
%The SED of ARP~220 is observed to be the closest match to the measured SEDs
%of the SPT sources, while the M51 and Cloverleaf SEDs may be
%considered the extreme ends of possible SEDs.

The SUMSS veto may exclude a few source classes from our sample.
Figure~\ref{Fig:lowzcut} shows that systems with much higher radio
power than implied by the radio-IR correlation, such as lensed
radio-loud AGN with significant dust emission (e.g., the Cloverleaf),
are excluded from our sample over a large redshift range.  This veto
may also exclude lensed DSFGs at $z$$\lesssim$$1.5$ (coincidentally
close to the IRAS redshift veto limit), where the radio-FIR
correlation predicts the radio emission will exceed the SUMSS limit.
Finally, DSFGs lensed by foreground galaxies with radio-active AGN and
residual FSRQs will be excluded in a redshift-unbiased way by this
veto.
%AGN population \cite[e.g.,][]{kellermann89} and our sample therefore
%should not have an appreciable bias towards star-forming galaxies
%without strong AGN activity.

%---------------------------------------------
% Fig 6 Lensing probability
%---------------------------------------------

\begin{figure}[h]
\centering
\includegraphics[width=8.2cm,angle=0]{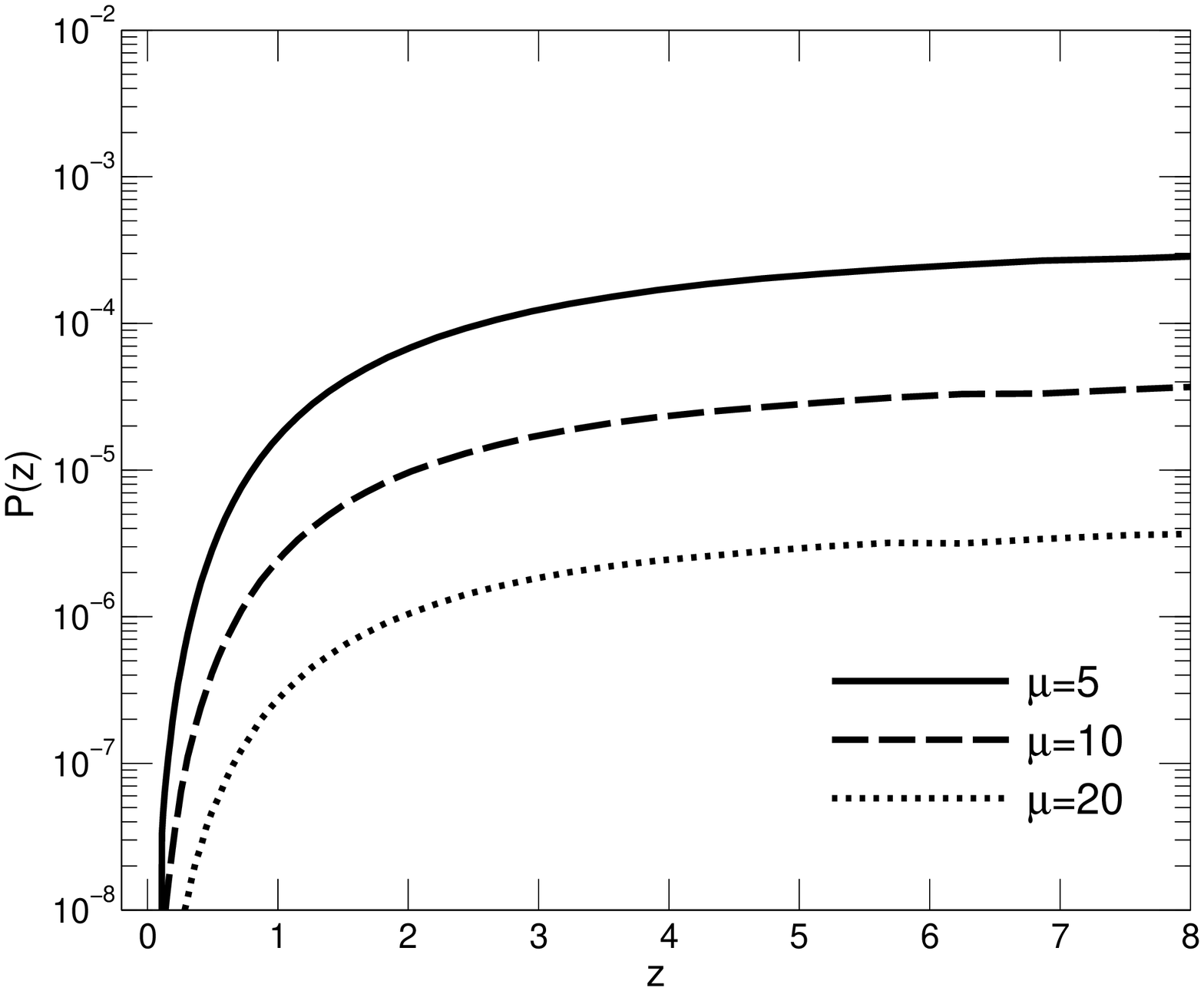}
\caption{Probability of strong gravitational lensing as a function of redshift for different source magnifications  ($\mu$) calculated 
from the models of \citet{hezaveh11} . The model assumes no size evolution for the underlying DSFG population. 
The figure demonstrates the strong decline of the lensing probability for $z$$\lesssim$$1.5$, independent of the magnification.}
\label{Fig:lensProb}
\end{figure}
 
\subsection{Redshift biases due to gravitational lensing}\label{Sect:lensbias}
%As discussed in the previous section, 
The high 1.4\,mm flux density cut of our target selection implies that
even the most infrared-luminous galaxies are too faint to be included
in the SPT dusty-source sample at $z\gtrsim0.5$ without assistance
from gravitational lensing ($\lir>3\cdot10^{13}\,\lsol$ for a
  Arp220 like SED).  This expectation is confirmed by our ALMA 
high-angular resolution imaging that resolves our sources into
arcs or Einstein rings -- hallmarks of gravitational lensing  \citep{vieira13}.
The redshift-dependent probability of strong
gravitational lensing therefore has important effects on our redshift
distribution. In Figure~\ref{Fig:lensProb}, we show the differential
probability of strong lensing versus redshift, calculated from the
models of \citet{hezaveh11} and \citet{hezaveh12b}, which use
gravitational lensing by a realistic population of intervening halos
to match the observed number counts of bright DSFGs. The strong
evolution in the lensing probability (the fractional volume at each
redshift subject to high magnification), a factor of 20 between
$z$$\sim$$2$ and $z\sim0.5$, demonstrates that the requirement that we
find lensed sources
%(observationally verified in \citealt{vieira13}) 
strongly suppresses sources at $z$$\lesssim$$1.5$. For $z$$\gtrsim$$2$
the lensing probability varies much more slowly, implying weaker
effects on the lensed source counts.

At higher redshifts, other lensing effects can more significantly
alter the normalized redshift distribution, d$n$/d$z$, especially
changes in source sizes.  To evaluate such effects, we compare an
assumed intrinsic redshift distribution to the model distribution of
strongly lensed sources (S$_{\rm 1.4mm}$$>$15\,mJy, consistent with
the deboosted 1.4\,mm flux densities of our sources, see Table
  \ref{tab:photometry}).  As discussed in \citet{hezaveh12b},the
selection of a sample of millimeter-bright DSFGs, lensed by
intervening galaxies, will preferentially identify those with more
compact emission regions. This implies that the observed redshift
distribution could be biased if DSFGs undergo a size evolution with
redshift.

Observationally, it is well established that high-redshift DSFGs are
significantly larger than local ULIRGs. In the high-redshift
($z$$\gtrsim$2) sources the star-forming regions extend over
$\sim$5\,kpc diameter, while lower-redshift ($z$$\lesssim$1) ULIRGs
typically form stars in kpc-sized regions \citep[see, e.g.,][and
references therein]{tacconi06,engel10}. Whether DSFGs undergo a size
evolution in the redshift range $z$=1.5--6, the relevant redshift
range for our study, is, however, largely unknown due to the small
number of high-redshift objects for which spatially resolved
observations of the submm emission region exist and the large
diversity of morphologies.  Evidence for extended molecular gas
reservoirs ($>$10\,kpc diameter) has been found in some DSFGs out to
redshift $z$$\approx$4 \citep[e.g.,][]{genzel03,ivison10b,
  younger10,carilli11,ivison11,riechers11} while the molecular gas
distribution in IR luminous AGN host galaxies, which have been
measured out to redshift $z$=6.4, are typically more compact
\citep[$\sim$2--3\,kpc diameter, e.g., ][]{walter04,walter09}. These
differences, however, mainly reflect the diversity of submm-detected
objects and possibly an evolutionary link between DSFGs and AGN host
galaxies \citep{riechers11b} rather than an overall size evolution of
submm-selected high-$z$ galaxies.

\begin{figure}[htb]
\centering
\includegraphics[width=7.0cm,angle=0]{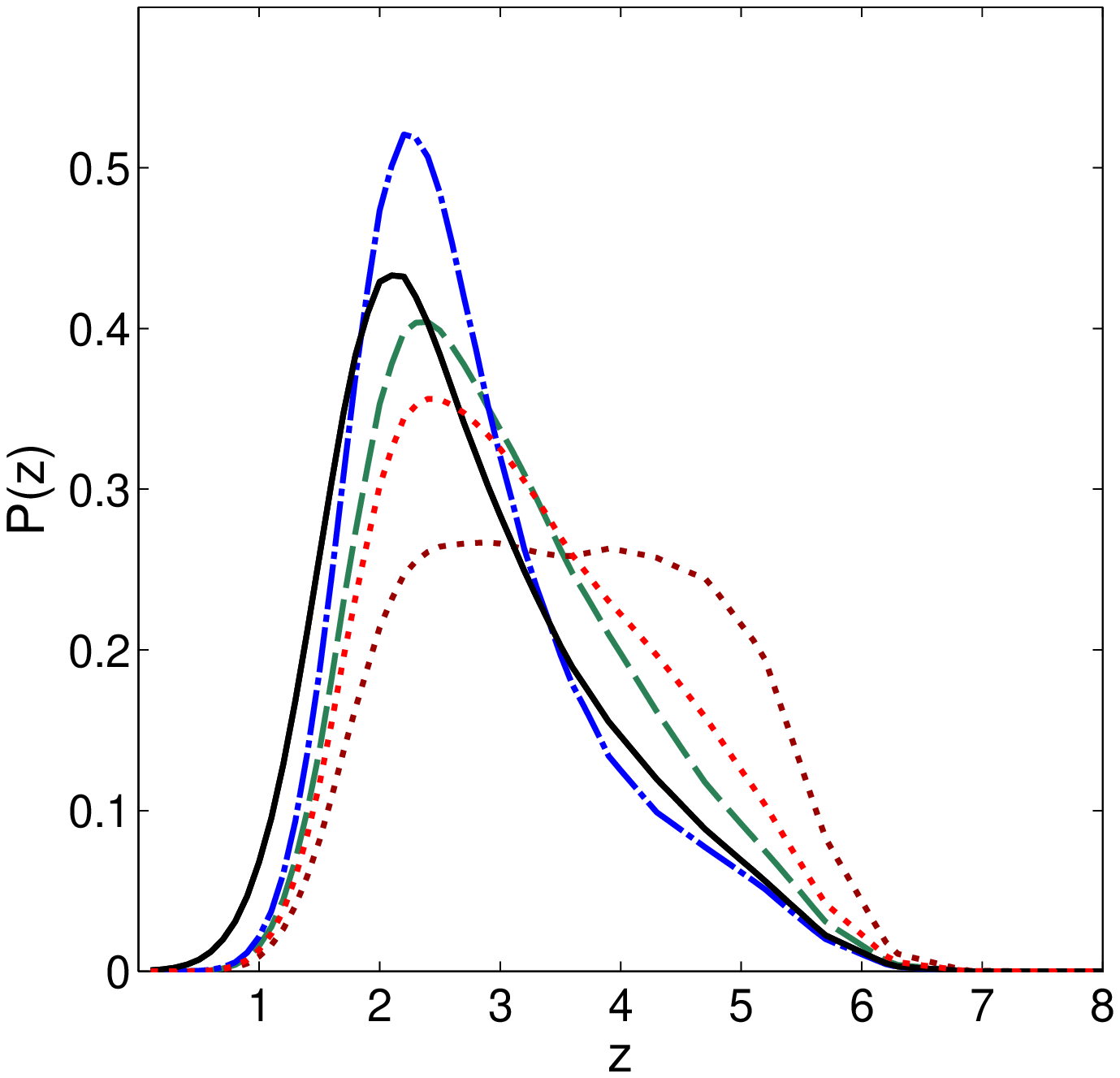}
\caption{
Comparison between an assumed intrinsic redshift distribution 
(d$n$/d$z$, \textit{solid black}) consistent with spectroscopic observations (see text for references)
and distributions modified by gravitational lensing (using the models described in \citet{hezaveh11}) under 
different size evolution scenarios. The \textit{green dashed} line shows the bias to the 
redshift distribution due to gravitational lensing assuming no size evolution versus redshift.  
The \textit{blue dot-dashed} line show the bias to the redshift distribution due to gravitational 
lensing if the size of DSFGs increases with redshift, from r=1\,kpc at $z$=2 to 3\,kpc at $z$=5.  
The \textit{red dotted} line shows the bias of the redshift distribution 
due to gravitational lensing if the size of DSFGs decreases moderately 
with redshift (r=2.5\,kpc at $z$=2 as measured for DSFGs \citep{engel10}, to 
r=1.5\,kpc at $z$=6 using the measured submm QSO host sizes \citep{walter09} 
as lower bound to the size of DSFGs).
The \textit{maroon dotted}  line exemplifies 
the extreme size evolution which would be required to bring the redshift distribution of \citet[][]{chapman05} 
into agreement with our observations (r=2.5\,kpc at $z$=2 to r=0.2\,kpc at $z$=6).}
\label{Fig:lensbias}
\end{figure}

In Figure~\ref{Fig:lensbias}, we compare different size-evolution
scenarios, where the intrinsic distribution was prescribed to be
consistent with the observed redshift distribution from
radio-identified DSFGs including recent spectroscopic data from the literature 
\citep{chapman05, capak08,coppin09,daddi09b, daddi09a, riechers10,banerji11,walter12}.  
The figure demonstrates
that the effect of gravitational lensing on the observed redshift
distribution is relatively small when there is no size evolution or
increasing source sizes with redshift.  For example, in the redshift
range $z$=2--4 the difference between d$n$/d$z$ derived from the
unlensed and lensed sources is smaller than $\sim20\%$ in both cases.
In the case of no size evolution the observed redshifted distribution
is displaced by $\Delta z$$\sim$0.3 towards higher redshifts
compared to the unlensed case.  Given the steep increase of d$n$/d$z$
between $z$=1--2 of the redshift distribution \citep{chapman05,banerji11}, this
shift causes an underestimate of the source counts in this redshift
interval by roughly a factor of two which may explain the low number
of $z$$<$2 objects detected in our survey.  For decreasing source
sizes with redshift \citep[as suggested by optical observations,][]{fathi12}
the difference between the observed and intrinsic
redshift distribution can become significant also for $z$$>$3, with
the counts of the high-redshift galaxies increased compared to the
intrinsic distribution.

A compilation of the effective source radii for $z$=1--6 derived from
an analysis of the dust SEDs of unlensed submm detected DSFGs and
quasi-stellar object (QSO) host galaxies has been published in
\citet{greve12}. Their Figure\,5 shows the submm source radii as a function
of redshift. The size of the highest redshift sources ($z$=5--6) in
this diagram tend to fall below the average size of $z$=1--3 objects,
but as mentioned above, these high-redshift sources are all QSO host
galaxies and as such cannot be taken as evidence for a size evolution
of the whole DSFG population.  The sample of source radii in the
literature \citep{tacconi08, engel10, rujopakarn11}, which were
directly measured from high-resolution imaging, show no clear evidence
for size evolution above $z$$>$0.4.  In the absence of conclusive
observational constraints, it is difficult to quantify the redshift
bias due to gravitational lensing. We note, however, that making our
observed redshift distribution consistent with an intrinsic
distribution like the one from \citet[][]{chapman05} would require an
extreme growth of DSFGs between $z$=6 and $z$=2 (r=0.2\,kpc to
2.5\,kpc in 2.3\,Gyr, see Figure\,\ref{Fig:lensbias}).  Likewise, a
modest evolution %consistent with observations 
(r=1.5\,kpc at $z$=6 to
2.5\,kpc at $z$=2, using the QSO size measurements as lower limits to the size
of DSFGs at $z=6$, see above) results in a steeper redshift
distribution than that implied by our most likely redshifts. Both
suggest that gravitational lensing is unlikely to be the dominant
source for the differences in d$n$/d$z$ between the present sample and
the radio-identified samples.

%---------------------------------------------
% Fig 6  redshift bias
%---------------------------------------------

\subsection{The redshift distribution}\label{Sect:redshiftDist}

Even with the conservative choice of taking all ambiguous sources to
be at their lowest redshift option (see Table \ref{Tab:lineIDs}), at
least 50\% of the SPT sample is at $z$$>$3. Only five sources are
possibly at $z$$\le$2 (assuming that sources without a line detection
fall into the redshift desert $z$=1.74--2.00), consistent with the
expectations for a sample of strongly lensed objects.  Our sample mean
redshift is $\bar z$=3.5.  This redshift distribution is in contrast
to that of radio identified DSFGs which have a significantly lower
mean redshift of $\bar z$=2.3 and for which only 10-15\% of the
population is expected to be at $z$$>$3
\citep[e.g.,][]{chapman05,wardlow11}.

\begin{figure}[htb]
\centering
\includegraphics[width=8.5cm,angle=0]{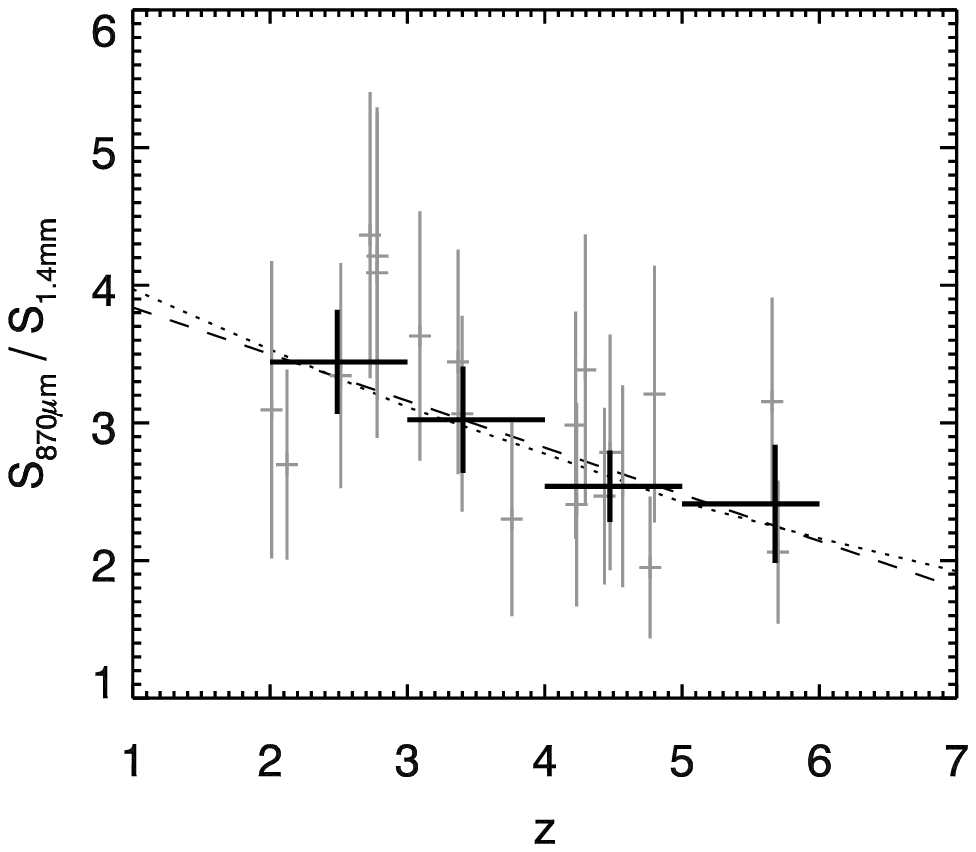}
\caption{
Observed 870\,$\mu$m to 1.4\,mm flux density ratio as a function of redshift 
for our sample of 20 sources with unambiguous spectroscopic redshifts. The grey 
points show the individual measurements and their error bars taking absolute 
calibration uncertainties into account. The black crosses show the mean flux 
density ratio in redshift bins of  $\Delta z$=1 centered at the weighted mean $z$. 
The dashed line is a linear fit to the data ($S_{870\micron}/S_{1.4mm}=4.18-0.34\,z$ for $z=2-6$). 
The dotted line shows the expectation for a Arp220 like dust SED.}
\label{Fig:1mm870mu}
\end{figure}

A potential difference between our redshift distribution and the
850-\um-selected samples in the literature arises from the interaction
of the SED of the typical DSFG and the selection wavelength.  This has
been discussed in several papers, including \citet{greve08} and
\citet{smolcic12}. 
%It has been argued that 850\,\um\ selection results
%in lower redshift samples because the redshifting of the FIR peak of
%the SED creates 850\,\um\ dropouts.  
 It has been argued that 850~\um\ selection results in lower redshift samples than 1.4~mm selection because 
the negative $K$-correction ceases once the SED peak is redshifted into the detection band, which occurs at 
lower redshift for shorter wavelength observations.
 Because our sources have been
selected at 1.4\,mm (SPT) and also observed at 870\,$\mu$m (LABOCA),
we can examine the effect that 850~\um\ selection would have on our
sample. The flux ratio as a function of redshift is shown in
Figure\,\ref{Fig:1mm870mu}, it reveals a modest decrease of the
870\,$\mu$m/1.4\,mm flux ratio for increasing redshift.  Our
observations therefore support the notion that 850~\um\ selection will
preferentially remove sources at the highest redshifts.  We caution,
however, that this effect will operate only on the fainter population
of high-redshift sources, those near to the detection limit where the
850\,$\mu$m may fall below the detection threshold while the 1.4\,mm
signal remains detectable.

Some studies of submm selected galaxies from blank field surveys 
presented evidence for a correlation between observed submm flux density 
and the source redshift \citep[e.g.][]{ivison02,pope05,ivison07,biggs11}. If confirmed,
this could imply a possible bias towards higher redshift for our study if 
the intrinsic IR luminosity of our sample is on average higher than that
of unlensed mm/submm selected samples. So far, lens models based on spatially
resolved images of the 870\um\ continuum are only available for four SPT sources 
\citep{hezaveh13}. These have magnifications of $\mu=5-21$ with a mean of 
$\bar \mu$=14. The gravitational flux amplification of the SPT sources has also
been discussed in \citet[][]{greve12}. They derive $\bar \mu=11-22$ based
on an analysis of the FIR properties of 11 SPT sources compared to unlensed
samples, in reasonable agreement with the lens models. Adopting an average
magnification of $\bar \mu=15$ for the sources studied here, our sample is expected
to cover intrinsic flux densities of ${\rm S}_{\rm 1.4mm}=1.0-3.0$\,mJy and  
${\rm S}_{870\um}=1.7-9.5$\,mJy with means of $\bar {\rm S}_{\rm 1.4mm}=1.8$\,mJy and 
$\bar {\rm S}_{\rm 870\um}=5.4$\,mJy. These intensities ranges are well match with
unlensed source flux densities observed in mm/submm blank fields surveys 
\citep[e.g.][]{borys03,coppin06,pope06,austermann09,weiss09} which implies that our
sample should be representative for the submm selected galaxy population at $z>1.5$.
We further note that the claimed correlation between observed submm flux density 
and source redshift has recently been questioned \citep{wardlow11,karim12}.

An additional difference between this sample and earlier spectroscopic
measurements of the DSFG redshift distribution is the radio selection.
As noted above, previous DSFG redshift searches have primarily relied
upon radio counterpart identification to provide optical spectroscopy
targets and therefore have a radio detection requirement. Here we have
excluded sources with bright radio counterparts, which might be
expected to oppositely bias the sample.  However, a comparison of the
submm-radio flux density ratio distribution for the radio-identified
sample of \citet{chapman05} and the similar ratio (corrected for
differences in observing frequency) constructed from SUMSS and SPT
measurements for our SUMSS-vetoed sources shows that these objects
emit a much larger fraction of their energy in the radio than even the
most extreme sources in \citet{chapman05} (see their Figure 7).
Likewise, sources that pass our SUMSS radio-veto are not biased
towards larger submm-radio flux density ratios than radio selected
samples from the literature due to the shallowness of the SUMSS
survey.  Therefore this veto should not preferentially exclude
low-redshift DSFGs, though optical spectroscopic measurements of the
excluded sources will be useful in determining which source classes
and which redshifts dominate the excluded objects.

%While radio emission in the radio-identified sources is presumed to arise from star formation, radio emission in 
%the SUMSS-vetoed SPT sources may arise in several ways, as noted in Section~\ref{Sect:lowzbias}. 
%This veto should not preferentially exclude low-redshift DSFGs, though optical spectroscopic measurements of the excluded 
%sources will be useful in determining which source classes and which redshifts dominate the excluded objects.

%---------------------------------------------
% Table 2 dn/dz
%---------------------------------------------

%%% OLD binning (z>1)
%\begin{deluxetable}{ccccccc}
%\centering
%\tablecaption{\label{Tab:dndz} Measured redshift distribution for SPT sources}
%\startdata
%\tableline
%\tableline
% $z$  & & N$^a$ &&d$n$/d$z$ & &$\pm$\\
%\tableline
%$1-2$ &\,\,\,\,\,\,\,&3&\,\,\,\,\,\,\,\,& 0.11 &\,\,\,\,\,\,\,\,& 0.06 \\
%$2-3$ &&7&& 0.25 && 0.09 \\
%$3-4$ &&8&& 0.29 && 0.10 \\
%$4-5$ &&8&& 0.29 && 0.10 \\
%$5-6$ &&2&& 0.07 && 0.05 \\
%\tableline
%\tablecomments{Reported redshifts are the most probable redshifts for 28 sources, 18 of which 
%have unambiguous spectroscopic redshifts (see \S\ref{Sect:lineIDs}).}
%\tablenotetext{a}{Number of sources per bin as listed in Table\,\ref{Tab:lineIDs} and including two
%SPT sources with previously known redshifts from \citet[][]{greve12}.}
%\end{deluxetable}

%NEW BINNING (z>1.5)
\begin{deluxetable}{ccccccc}
\centering
\tablecaption{\label{Tab:dndz} Measured redshift distribution for SPT sources}
\startdata
\tableline
\tableline
 $z$  & & N$^a$ &&d$n$/d$z$ & &$\pm$\\
\tableline
$1.5-2.5$ &\,\,\,\,\,\,\,&6&\,\,\,\,\,\,\,\,& 0.21 &\,\,\,\,\,\,\,\,& 0.09 \\
$2.5-3.5$ &&8&& 0.29 && 0.10 \\
$3.5-4.5$ &&9&& 0.32 && 0.11 \\
$4.5-5.5$ &&3&& 0.11 && 0.06 \\
$5.5-6.5$ &&2&& 0.07 && 0.05 \\
\tableline
\tablecomments{Reported redshifts are the most probable redshifts for 28 sources, 20 of which 
have unambiguous spectroscopic redshifts (see \S\ref{Sect:lineIDs}).}
\tablenotetext{a}{Number of sources per bin as listed in Table\,\ref{Tab:lineIDs} including two
SPT sources with previously known redshifts from \citet[][]{greve12}.}
\end{deluxetable}

%---------------------------------------------
% Fig 8  flux ratios
%---------------------------------------------

\begin{figure*}[ht]
\centering
\includegraphics[width=18.1cm,angle=0]{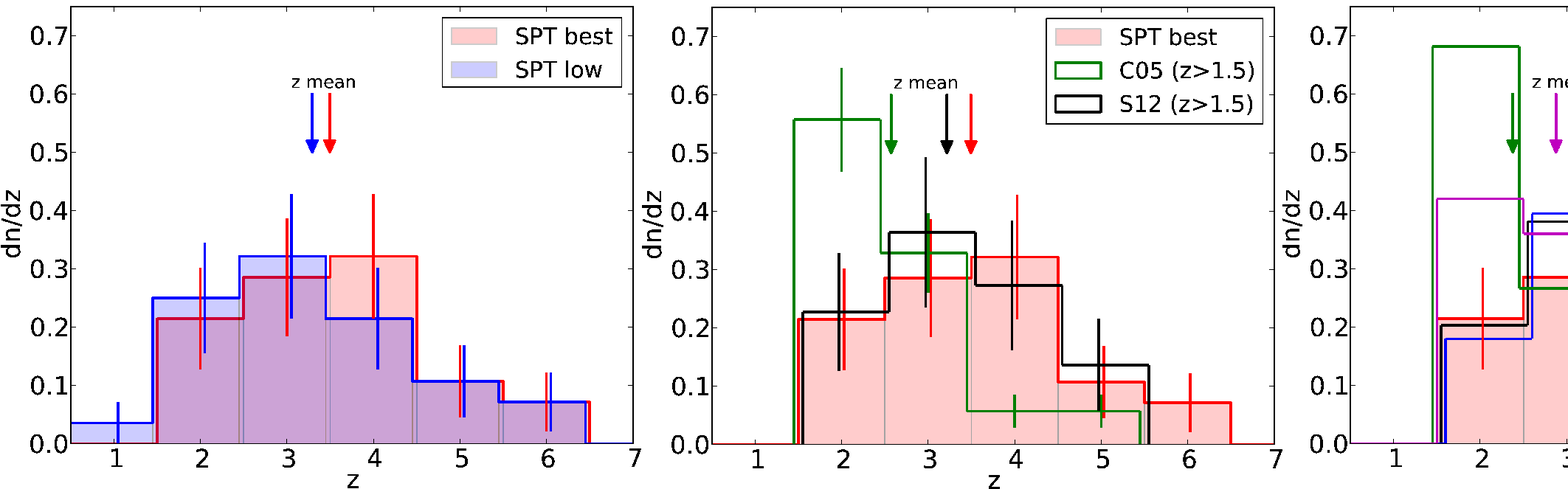}
\caption{{\it Left:} Redshift distribution of strongly lensed DSFGs derived 
from our molecular line survey. The red histogram shows the $z$-distributions for 
the SPT sources using the most likely redshift redshift identification for the sources 
with ambiguous redshifts; the blue histogram shows the same for the lowest redshift 
identification of these five sources (see Table\,\ref{Tab:lineIDs}). 
{\it Middle:} Redshift distribution of radio-identified DSFGs with spectroscopic 
redshifts at $z$$>$1.5 
(green, \citet{chapman05, capak08,coppin09,daddi09b, daddi09a, riechers10,banerji11,walter12}), 
mm-identified DSFGs with photometric redshifts at $z$$>$1.5 from  (black, \citet{smolcic12}), 
compared to the most likely SPT distribution (Red).  {\it Right:} Redshift distributions 
from the models of \citet{hayward12} (Black), \citet{bethermin12b} (Blue), \citet{benson12b} 
(Green) \citet{lacey10} (Purple) for $z$$>$1.5, compared to the most likely SPT distribution (Red). 
The arrows in all panels show the mean redshift of each distribution. In all panels the histograms 
are calculated over the same redshift bins but are plotted with slight shifts in $z$ for clarity.
} 
\label{Fig:NZ}
\end{figure*}

The determination of the shape of our redshift distribution is
currently hampered by the  eight ambiguous redshifts.  In
Figure\,\ref{Fig:NZ} ({\it left}) we compare two redshift distributions,
one using the lowest redshift option for all sources, and the other
assuming the most likely redshift.  In the first case, our redshift
distribution shows some evidence for a peak at $z\approx3$, consistent
with the findings of radio identified DSFGs, and then decreases out to
$z$$\sim$6. The decrease, however, is much shallower than suggested
from radio identified DSFGs. In the latter case our redshift
distribution rises up to $z\approx4$ and falls off at higher redshift. 
Within the errors both distributions are consistent with a flat redshift distribution 
between $z$=2--4. Note that to these distributions we have added two additional strongly
lensed SPT sources from \citet{greve12}.
%, at $z$=2.78 and $z$=2.73, that
%would have been in the ALMA sample except that redshifts were already
%known from VLT and Z-Spec/APEX spectroscopy \citep{greve12}.

We adopt the redshift distribution informed by our dust temperatures
and other data (``SPT best'' in Figure~\ref{Fig:NZ}) for the
discussion which follows, and report the values for d$n$/d$z$ in Table \ref{Tab:dndz}.

%---------------------------------------------
% Fig 9  redshift histograms
%---------------------------------------------

Figure\,\ref{Fig:NZ} ({\it center}) highlights the large difference
between our results and previous redshift surveys. Compared to
previous surveys with spectroscopic redshifts that rely on radio
counterpart identification \citep{chapman05, capak08,coppin09,daddi09b, 
daddi09a, riechers10,banerji11,walter12} we find a far greater fraction of high-redshift sources.
As discussed earlier, gravitational lensing may explain part of this
discrepancy if DSFGs are smaller at high redshifts, though extreme
evolution is required to explain the full difference.  Recent work
based on CO(1--0)-derived redshifts for a DSFG sample selected from
the H-ATLAS survey \citep[][not shown here]{harris12} implies a
redshift distribution in agreement with \citet{chapman05}. These
sources, however, were selected to peak in the SPIRE 350$\mu$m channel
to match the 2.1$\le$$z$$\le$ 3.5 redshift coverage of the instrument
used to measure redshifts \citep{harris12}. Despite this selection,
$>50$\% of their targeted sources remained undetected in CO, which may
imply that there are a significant number of sources at redshifts
larger than $z$=3.5 in this sample as well.

\citet{smolcic12} also find an increased fraction of DSFGs at $z$$>$3
through a combination of spectroscopic and photometric redshifts 
for a mixed sample of 1.1mm and 870\um\ selected sources in the COSMOS field. They
note that 50--70\% of their $z$$>$3 DSFGs have no radio counterpart
down to  $\sim$$10$\,$\mu$Jy at 1.4\,GHz, which supports the
prediction that including radio counterpart identification in the
process of surveying DSFG redshifts will suppress higher-$z$ sources,
as expected from SED templates.  The similarity in the redshift
distribution of unlensed sources compiled by \citet{smolcic12},
derived primarily from photometric redshifts, and our own
(Figure~\ref{Fig:NZ}, {\it center}) may be evidence that gravitational
lensing is not strongly affecting the underlying redshift
distribution.  However, greater numbers of molecular-line-derived
redshifts for both populations will likely be required to settle this
issue.

In the case of no size evolution in DSFGs, our study suggests that
previous spectroscopic DSFG redshift surveys, which are almost
exclusively based on radio identified sources, have missed $\ge$50\%
of the DSFG population as it resides at redshifts $z$$>$3 and the
putative high-redshift tail of DSFGs may in fact turn out to be a much
broader, flat-topped redshift distribution which could extend to
$z>4$.

\subsection{Comparison to models}

Redshift distributions (d$n$/d$z$) and number counts are the main
observational constraints to galaxy formation models.  Matching
available data for DSFGs with these models has been particularly
difficult \citep[e.g.,][]{baugh05}, requiring some ad hoc changes such
as top-heavy initial mass functions.  As argued above, our d$n$/d$z$
--- although currently based on only 28 sources --- appears
significantly different from the currently largest sample of
spectroscopic DSFG redshifts by \citet{chapman05}. With direct mm
identifications, a  71\% spectroscopic completeness, and likely
redshifts for an additional 18\%, our SPT DSFG d$n$/d$z$ represents an
important new observational constraint to these models.

We compare our measured d$n$/d$z$ with four recent models in
Figure~\ref{Fig:NZ} ({\it right}), removing sources at $z$$<$$1.5$ from
the models to mimic the strong lensing selection selection described in
Section~\ref{Sect:redshiftDist}. We discuss the individual models below and 
give the  $\chi^2$ for each model for the five redshift bins.
Despite the relatively small number of redshifts, our new SPT
d$n$/d$z$ already discriminates between galaxy formation models.

\citet{bethermin12b} present an empirical model starting from the
observed FIR number counts split into ``main-sequence'' and starburst
mode star-forming galaxies.  Their model includes the effects of
magnification by strong lensing, so it can directly predict the
d$n$/d$z$ for the SPT sample. For the comparison
with our data we use the predicted redshift distribution for sources with 
$S_{1.4{\rm mm}}$$>$15\,mJy, consistent with our source selection.
%The main effect of
%the lensing is to remove the low redshift infrared galaxy component at
%$z$$<$0.01 from the distribution, which is effectively equivalent to
%our IRAS veto. Of the sources at $z$$>$0.1, there is little
%redshift bias due to the lensing, although possible source size
%evolution with redshift is not included in the model.
This model
matches our redshift histogram very well, with a comparison to the 
five redshift bins giving a $\chi^2$ of 1.9 across five redshift bins.
%and will be described in more detail in a future publication.

 The \citet{lacey10} model is a semi-analytic model identical to
  that presented in \citet{baugh05}.  The model employs a top-heavy
  stellar initial mass function, which results in more luminosity and
  more dust produced per unit star formation rate, to better match the
  bright end of 850\,\um\ galaxy counts.  This model does not include
  the effects of strong lensing, and DSFG counts are based on a
  selection in S$_{1.4{\rm mm}}$ with $> 1$~mJy (C. Lacey, private
  communication).  The $\chi ^2$ between this model and our
  measurement across the five redshift bins is 10.7.

The \citet{benson12b} model is a semi-analytic model that
also expands upon the work of \citet{baugh05}.  Whereas the
  \citet{lacey10} model required a top-heavy stellar initial mass
function, the \citet{benson12b} model merely has enhanced dust
production in starbursts. This model does not include the effects of
strong lensing, and DSFG counts are based on a selection in
S$_{850\mu{\rm m}}$ ($> 5$~mJy). The predicted d$n$/d$z$ distribution
comes close to the \citet{chapman05} distribution, but clearly fails
to fit the SPT or \citet{smolcic12} measurements. Part of this
difference may be due to the 850\,\um\ instead of 1.4 mm source
selection, and a possible lensing bias.  The $\chi ^2$ between this
model and our measurement across the five redshift bins is 39.8.  Our
measurements are clearly at odds with this model.

The model by \citet{hayward12} combines a semi-empirical model
with 3D hydrodynamical simulations and a 3D dust radiative transfer.
Strong lensing is not included in the modeling and the model predicted
d$n$/d$z$ is determined using sources with $S_{1{\rm mm}}$$>$1\,mJy, consistent
with the expected intrinsic flux densities of our sample.
%The predicted d$n$/d$z$ is determined using $S_{1.1{\rm mm}}$$>$1 and
%$S_{1.1{\rm mm}}$$>$4\,mJy cuts, but strong lensing is not included in
%the count modeling. 
The distribution of the DSFGs in this model is close to the observed SPT 
d$n$/d$z$, with a  $\chi ^2$
of 2.8 between data and model.

%%%%%%%%%%%%%%%%%%%%%%%%%%%%%%%%%%%%%
% 5. Conclusion
\section{Summary and Conclusion}
\label{sect:summary}
%%%%%%%%%%%%%%%%%%%%%%%%%%%%%%%%%%%%%

We have used ALMA to measure or constrain the redshifts of 26 strongly lensed
DSFGs detected in the SPT-SZ survey data. The redshifts were derived
using molecular emission lines detected in frequency scans in the
3\,mm transmission window covering 84.2 to 114.9\,GHz.  As the
molecular emission lines can unambiguously be associated with the
thermal dust continuum emission at our selection wavelength of
1.4\,mm, this technique does not require any multi-wavelength
identification unlike other methods typically used to derive DSFG
redshifts.

In total we detect 44 spectral features in our survey which we
identify as redshifted emission lines of $^{12}$CO, $^{13}$CO, \ci,
H$_2$O, and H$_2$O$^+$. We find one or more lines in 23 sources,
yielding an unprecedented $\sim$90\% success rate of this survey.  In
12 sources we detect multiple lines. In 11 sources we robustly detect
a single line, and in one of those cases we can use that single line
to obtain an unambiguous redshift.  For an additional five galaxies,
in which we detect a single line with ALMA, we can determine the
redshift using additional spectral and
optical data yielding 18 unambiguous redshifts.  For five sources
with a single line detection we have used our excellent mm/submm
photometric coverage (3\,mm to 250\,$\mu$m) to narrow the line
identification and make a probabilistic estimate for the redshift
based on the FIR dust temperature derived from extensive broad band
photometric data.  In three sources we do not detect a line feature,
either because the lines are too weak, or because they are in the
redshift desert $z$=1.74--2.00.  Adding in two previously reported SPT
sources with spectroscopic redshifts from \citep{greve12}, we derive a
redshift distribution from 28 SPT sources.

We analyze the redshift biases inherent to our source selection and to
gravitational galaxy-galaxy lensing.  Our selection of bright 1.4~mm sources 
imposes a requirement that they be gravitationally lensed, effectively suppressing 
sources at $z\lesssim1.5$ due to the low probability of being lensed at these redshifts.
Beyond $z\sim2$, gravitational lensing does not significantly bias the redshift
distribution unless DSFGs undergo a systematic size evolution between
$z$=2--6 with decreasing source sizes for higher redshifts. An
analysis of the black body radii of unlensed DSFGs from the literature
does not support the existence of such an evolution, but it also
cannot be excluded conclusively at this point. 

Our sample mean redshift is $\bar z$=3.5. This finding is in
contrast to the redshift distribution of radio identified DSFGs which
have a significantly lower mean redshift of $\bar z$=2.3, and for
which only 10-15\% of the population is expected to be at $z$$>$3
\citep[e.g.,][]{chapman05}.  The redshift distribution of our sample
appears almost flat between $z$=2--4.  Our study suggests that
previous spectroscopic redshift surveys of DSFGs based on radio
identified sources are likely biased towards lower redshift and have
missed a large fraction ($\ge50$\%) of the DSFG population at
redshifts $z$$>$3.

With a 90\% detection rate, our ALMA+SPT CO redshift survey is the
most complete DSFG survey to date.  It demonstrates the power of ALMA,
with its broad-band receivers and large collecting area, to provide
the critical galaxy redshift information needed to measure the cosmic
history of obscured star formation, particularly at the highest
redshifts where other techniques falter.  The magnification of the SPT
sources by intervening mass \citep[factors of $\sim$10 or
more,][]{hezaveh13} has allowed us to obtain these results in the
early science phase of ALMA, with only 16, of the eventual array of
54, 12-meter antennas.  With the full array, such studies will be
possible on unlensed sources, highlighting the enormous scientific
impact ALMA will have in the coming decades.  With spectroscopic
redshifts for a large number of DSFGs, it is now possible to study the
conditions of the interstellar medium at high redshift in great detail
through spatially resolved spectroscopy of FIR molecular and atomic
lines.  The SPT sources presented here represent less than 25\% of the
entire sample of high-redshift, strongly lensed DSFGs.  Obtaining
redshifts for the remaining sources will enable us to definitively
constrain the redshift evolution of DSFGs.

\acknowledgments 

The authors would like to thank A. Blain and N. Scoville for many
useful discussions related to this work, and A.  Benson, C. Baugh, C.
Hayward, and C. Lacey for providing us with the predicted redshift
distributions from their models and useful discussions regarding their
implications.  This paper makes use of the following ALMA data:
ADS/JAO.ALMA\#2011.0.00957.S. ALMA is a partnership of ESO
(representing its member states), NSF (USA) and NINS (Japan), together
with NRC (Canada) and NSC and ASIAA (Taiwan), in cooperation with the
Republic of Chile.  The Joint ALMA Observatory is operated by ESO,
AUI/NRAO and NAOJ.  Based on observations taken with European Southern
Observatory Very Large Telescope, Paranal, Chile, with program ID
088.A-0902(C), and with the Atacama Pathfinder Experiment under
program IDs 086.A-0793(A), 086.A-1002(A), 087.A-0815(A) and
087.A-0968(A). APEX is a collaboration between the Max-Planck-Institut
f\"ur Radioastronomie, the European Southern Observatory, and the
Onsala Space Observatory.  This research has made use of the NASA/IPAC
Extragalactic Database (NED) which is operated by the Jet Propulsion
Laboratory, California Institute of Technology, under contract with
the National Aeronautics and Space Administration.  The SPT is
supported by the National Science Foundation through grant
ANT-0638937, with partial support through PHY-1125897, the Kavli
Foundation and the Gordon and Betty Moore Foundation.  The National
Radio Astronomy Observatory is a facility of the National Science
Foundation operated under cooperative agreement by Associated
Universities, Inc.  Partial support for this work was provided by NASA
through grant HST-GO-12659 from the Space Telescope Science Institute.
This work is based in part on observations made with {\it Herschel}, a
European Space Agency Cornerstone Mission with significant
participation by NASA, and supported through an award issued by
JPL/Caltech for OT2\_jvieira\_5.  TRG gratefully acknowledges support
from a STFC Advanced Fellowship.

\bibliographystyle{apj}

\bibliography{spt_smg}

\begin{thebibliography}{84}
\expandafter\ifx\csname natexlab\endcsname\relax\def\natexlab#1{#1}\fi

\bibitem[{{Appenzeller} {et~al.}(1998){Appenzeller}, {Fricke}, {F{\"u}rtig},
  {G{\"a}ssler}, {H{\"a}fner}, {Harke}, {Hess}, {Hummel}, {J{\"u}rgens},
  {Kudritzki}, {Mantel}, {Meisl}, {Muschielok}, {Nicklas}, {Rupprecht},
  {Seifert}, {Stahl}, {Szeifert}, \& {Tarantik}}]{appenzeller98}
{Appenzeller}, I., {Fricke}, K., {F{\"u}rtig}, W., {et~al.} 1998, The
  Messenger, 94, 1

\bibitem[{{Ashby} {et~al.}(2006){Ashby}, {Dye}, {Huang}, {Eales}, {Willner},
  {Webb}, {Barmby}, {Rigopoulou}, {Egami}, {McCracken}, {Lilly}, {Miyazaki},
  {Brodwin}, {Blaylock}, {Cadien}, \& {Fazio}}]{ashby06}
{Ashby}, M.~L.~N., {Dye}, S., {Huang}, J.-S., {et~al.} 2006, \apj, 644, 778

\bibitem[{{Austermann} {et~al.}(2009){Austermann}, {Aretxaga}, {Hughes},
  {Kang}, {Kim}, {Lowenthal}, {Perera}, {Sanders}, {Scott}, {Scoville},
  {Wilson}, \& {Yun}}]{austermann09}
{Austermann}, J.~E., {Aretxaga}, I., {Hughes}, D.~H., {et~al.} 2009, \mnras,
  393, 1573

\bibitem[{{Banerji} {et~al.}(2011){Banerji}, {Chapman}, {Smail},
  {Alaghband-Zadeh}, {Swinbank}, {Dunlop}, {Ivison}, \& {Blain}}]{banerji11}
{Banerji}, M., {Chapman}, S.~C., {Smail}, I., {et~al.} 2011, \mnras, 418, 1071

\bibitem[{{Baugh} {et~al.}(2005){Baugh}, {Lacey}, {Frenk}, {Granato}, {Silva},
  {Bressan}, {Benson}, \& {Cole}}]{baugh05}
{Baugh}, C.~M., {Lacey}, C.~G., {Frenk}, C.~S., {et~al.} 2005, \mnras, 356,
  1191

\bibitem[{{Benford} {et~al.}(1999){Benford}, {Cox}, {Omont}, {Phillips}, \&
  {McMahon}}]{benford99}
{Benford}, D.~J., {Cox}, P., {Omont}, A., {Phillips}, T.~G., \& {McMahon},
  R.~G. 1999, \apjl, 518, L65

\bibitem[{{Benson}(2012)}]{benson12b}
{Benson}, A.~J. 2012, \na, 17, 175

\bibitem[{{B{\'e}thermin} {et~al.}(2012){B{\'e}thermin}, {Daddi}, {Magdis},
  {Sargent}, {Hezaveh}, {Elbaz}, {Le Borgne}, {Mullaney}, {Pannella}, {Buat},
  {Charmandaris}, {Lagache}, \& {Scott}}]{bethermin12b}
{B{\'e}thermin}, M., {Daddi}, E., {Magdis}, G., {et~al.} 2012, \apjl, 757, L23

\bibitem[{{Biggs} {et~al.}(2011){Biggs}, {Ivison}, {Ibar}, {Wardlow},
  {Dannerbauer}, {Smail}, {Walter}, {Wei{\ss}}, {Chapman}, {Coppin}, {De
  Breuck}, {Dickinson}, {Knudsen}, {Mainieri}, {Menten}, \&
  {Papovich}}]{biggs11}
{Biggs}, A.~D., {Ivison}, R.~J., {Ibar}, E., {et~al.} 2011, \mnras, 413, 2314

\bibitem[{{Blain} \& {Longair}(1993)}]{blain93}
{Blain}, A.~W., \& {Longair}, M.~S. 1993, \mnras, 264, 509

\bibitem[{{Blain} {et~al.}(1999){Blain}, {Smail}, {Ivison}, \&
  {Kneib}}]{blain99b}
{Blain}, A.~W., {Smail}, I., {Ivison}, R.~J., \& {Kneib}, J.-P. 1999, \mnras,
  302, 632

\bibitem[{{Bock} {et~al.}(1998){Bock}, {Turtle}, \& {Green}}]{bock98}
{Bock}, D.~C.-J., {Turtle}, A.~J., \& {Green}, A.~J. 1998, \aj, 116, 1886

\bibitem[{{Borys} {et~al.}(2003){Borys}, {Chapman}, {Halpern}, \&
  {Scott}}]{borys03}
{Borys}, C., {Chapman}, S., {Halpern}, M., \& {Scott}, D. 2003, \mnras, 344,
  385

\bibitem[{{Bouwens} {et~al.}(2010){Bouwens}, {Illingworth}, {Oesch},
  {Stiavelli}, {van Dokkum}, {Trenti}, {Magee}, {Labb{\'e}}, {Franx},
  {Carollo}, \& {Gonzalez}}]{bouwens10}
{Bouwens}, R.~J., {Illingworth}, G.~D., {Oesch}, P.~A., {et~al.} 2010, \apjl,
  709, L133

\bibitem[{{Bouwens} {et~al.}(2011){Bouwens}, {Illingworth}, {Labbe}, {Oesch},
  {Trenti}, {Carollo}, {van Dokkum}, {Franx}, {Stiavelli}, {Gonz{\'a}lez},
  {Magee}, \& {Bradley}}]{bouwens11}
{Bouwens}, R.~J., {Illingworth}, G.~D., {Labbe}, I., {et~al.} 2011, \nat, 469,
  504

\bibitem[{{Bradford} {et~al.}(2004){Bradford}, {Ade}, {Aguirre}, {Bock},
  {Dragovan}, {Duband}, {Earle}, {Glenn}, {Matsuhara}, {Naylor}, {Nguyen},
  {Yun}, \& {Zmuidzinas}}]{bradford04}
{Bradford}, C.~M., {Ade}, P.~A.~R., {Aguirre}, J.~E., {et~al.} 2004, in Society
  of Photo-Optical Instrumentation Engineers (SPIE) Conference Series, Vol.
  5498, Society of Photo-Optical Instrumentation Engineers (SPIE) Conference
  Series, ed. C.~M. {Bradford}, P.~A.~R. {Ade}, J.~E. {Aguirre}, J.~J. {Bock},
  M.~{Dragovan}, L.~{Duband}, L.~{Earle}, J.~{Glenn}, H.~{Matsuhara}, B.~J.
  {Naylor}, H.~T. {Nguyen}, M.~{Yun}, \& J.~{Zmuidzinas}, 257

\bibitem[{{Capak} {et~al.}(2008){Capak}, {Carilli}, {Lee}, {Aldcroft},
  {Aussel}, {Schinnerer}, {Wilson}, {Yun}, {Blain}, {Giavalisco}, {Ilbert},
  {Kartaltepe}, {Lee}, {McCracken}, {Mobasher}, {Salvato}, {Sasaki}, {Scott},
  {Sheth}, {Shioya}, {Thompson}, {Elvis}, {Sanders}, {Scoville}, \&
  {Tanaguchi}}]{capak08}
{Capak}, P., {Carilli}, C.~L., {Lee}, N., {et~al.} 2008, \apjl, 681, L53

\bibitem[{{Carilli} {et~al.}(2011){Carilli}, {Hodge}, {Walter}, {Riechers},
  {Daddi}, {Dannerbauer}, \& {Morrison}}]{carilli11}
{Carilli}, C.~L., {Hodge}, J., {Walter}, F., {et~al.} 2011, \apjl, 739, L33

\bibitem[{{Carlstrom} {et~al.}(2011){Carlstrom}, {Ade}, {Aird}, {Benson},
  {Bleem}, {Busetti}, {Chang}, {Chauvin}, {Cho}, {Crawford}, {Crites}, {Dobbs},
  {Halverson}, {Heimsath}, {Holzapfel}, {Hrubes}, {Joy}, {Keisler}, {Lanting},
  {Lee}, {Leitch}, {Leong}, {Lu}, {Lueker}, {Luong-van}, {McMahon}, {Mehl},
  {Meyer}, {Mohr}, {Montroy}, {Padin}, {Plagge}, {Pryke}, {Ruhl}, {Schaffer},
  {Schwan}, {Shirokoff}, {Spieler}, {Staniszewski}, {Stark}, {Tucker},
  {Vanderlinde}, {Vieira}, \& {Williamson}}]{carlstrom11}
{Carlstrom}, J.~E., {Ade}, P.~A.~R., {Aird}, K.~A., {et~al.} 2011, \pasp, 123,
  568

\bibitem[{{Chapman} {et~al.}(2003){Chapman}, {Blain}, {Ivison}, \&
  {Smail}}]{chapman03}
{Chapman}, S.~C., {Blain}, A.~W., {Ivison}, R.~J., \& {Smail}, I.~R. 2003,
  \nat, 422, 695

\bibitem[{{Chapman} {et~al.}(2005){Chapman}, {Blain}, {Smail}, \&
  {Ivison}}]{chapman05}
{Chapman}, S.~C., {Blain}, A.~W., {Smail}, I., \& {Ivison}, R.~J. 2005, \apj,
  622, 772

\bibitem[{{Coppin} {et~al.}(2006){Coppin}, {Chapin}, {Mortier}, {Scott},
  {Borys}, {Dunlop}, {Halpern}, {Hughes}, {Pope}, {Scott}, {Serjeant}, {Wagg},
  {Alexander}, {Almaini}, {Aretxaga}, {Babbedge}, {Best}, {Blain}, {Chapman},
  {Clements}, {Crawford}, {Dunne}, {Eales}, {Edge}, {Farrah}, {Gazta{\~n}aga},
  {Gear}, {Granato}, {Greve}, {Fox}, {Ivison}, {Jarvis}, {Jenness}, {Lacey},
  {Lepage}, {Mann}, {Marsden}, {Martinez-Sansigre}, {Oliver}, {Page},
  {Peacock}, {Pearson}, {Percival}, {Priddey}, {Rawlings}, {Rowan-Robinson},
  {Savage}, {Seigar}, {Sekiguchi}, {Silva}, {Simpson}, {Smail}, {Stevens},
  {Takagi}, {Vaccari}, {van Kampen}, \& {Willott}}]{coppin06}
{Coppin}, K., {Chapin}, E.~L., {Mortier}, A.~M.~J., {et~al.} 2006, \mnras, 372,
  1621

\bibitem[{{Coppin} {et~al.}(2009){Coppin}, {Smail}, {Alexander}, {Weiss},
  {Walter}, {Swinbank}, {Greve}, {Kovacs}, {De Breuck}, {Dickinson}, {Ibar},
  {Ivison}, {Reddy}, {Spinrad}, {Stern}, {Brandt}, {Chapman}, {Dannerbauer},
  {van Dokkum}, {Dunlop}, {Frayer}, {Gawiser}, {Geach}, {Huynh}, {Knudsen},
  {Koekemoer}, {Lehmer}, {Menten}, {Papovich}, {Rix}, {Schinnerer}, {Wardlow},
  \& {van der Werf}}]{coppin09}
{Coppin}, K.~E.~K., {Smail}, I., {Alexander}, D.~M., {et~al.} 2009, \mnras,
  395, 1905

\bibitem[{{Cox} {et~al.}(2011){Cox}, {Krips}, {Neri}, {Omont}, {G{\"u}sten},
  {Menten}, {Wyrowski}, {Wei{\ss}}, {Beelen}, {Gurwell}, {Dannerbauer},
  {Ivison}, {Negrello}, {Aretxaga}, {Hughes}, {Auld}, {Baes}, {Blundell},
  {Buttiglione}, {Cava}, {Cooray}, {Dariush}, {Dunne}, {Dye}, {Eales},
  {Frayer}, {Fritz}, {Gavazzi}, {Hopwood}, {Ibar}, {Jarvis}, {Maddox},
  {Micha{\l}owski}, {Pascale}, {Pohlen}, {Rigby}, {Smith}, {Swinbank}, {Temi},
  {Valtchanov}, {van der Werf}, \& {de Zotti}}]{cox11}
{Cox}, P., {Krips}, M., {Neri}, R., {et~al.} 2011, \apj, 740, 63

\bibitem[{{Crawford} {et~al.}(2010){Crawford}, {Switzer}, {Holzapfel},
  {Reichardt}, {Marrone}, \& {Vieira}}]{crawford10}
{Crawford}, T.~M., {Switzer}, E.~R., {Holzapfel}, W.~L., {et~al.} 2010, \apj,
  718, 513

\bibitem[{{Daddi} {et~al.}(2009{\natexlab{a}}){Daddi}, {Dannerbauer}, {Krips},
  {Walter}, {Dickinson}, {Elbaz}, \& {Morrison}}]{daddi09b}
{Daddi}, E., {Dannerbauer}, H., {Krips}, M., {et~al.} 2009{\natexlab{a}},
  \apjl, 695, L176

\bibitem[{{Daddi} {et~al.}(2009{\natexlab{b}}){Daddi}, {Dannerbauer}, {Stern},
  {Dickinson}, {Morrison}, {Elbaz}, {Giavalisco}, {Mancini}, {Pope}, \&
  {Spinrad}}]{daddi09a}
{Daddi}, E., {Dannerbauer}, H., {Stern}, D., {et~al.} 2009{\natexlab{b}}, \apj,
  694, 1517

\bibitem[{{Eales} {et~al.}(2010){Eales}, {Dunne}, {Clements}, {Cooray}, {de
  Zotti}, {Dye}, {Ivison}, {Jarvis}, {Lagache}, {Maddox}, {Negrello},
  {Serjeant}, {Thompson}, {Kampen}, {Amblard}, {Andreani}, {Baes}, {Beelen},
  {Bendo}, {Benford}, {Bertoldi}, {Bock}, {Bonfield}, {Boselli}, {Bridge},
  {Buat}, {Burgarella}, {Carlberg}, {Cava}, {Chanial}, {Charlot},
  {Christopher}, {Coles}, {Cortese}, {Dariush}, {da Cunha}, {Dalton}, {Danese},
  {Dannerbauer}, {Driver}, {Dunlop}, {Fan}, {Farrah}, {Frayer}, {Frenk},
  {Geach}, {Gardner}, {Gomez}, {Gonz{\'a}lez-Nuevo}, {Gonz{\'a}lez-Solares},
  {Griffin}, {Hardcastle}, {Hatziminaoglou}, {Herranz}, {Hughes}, {Ibar},
  {Jeong}, {Lacey}, {Lapi}, {Lawrence}, {Lee}, {Leeuw}, {Liske},
  {L{\'o}pez-Caniego}, {M{\"u}ller}, {Nandra}, {Panuzzo}, {Papageorgiou},
  {Patanchon}, {Peacock}, {Pearson}, {Phillipps}, {Pohlen}, {Popescu},
  {Rawlings}, {Rigby}, {Rigopoulou}, {Robotham}, {Rodighiero}, {Sansom},
  {Schulz}, {Scott}, {Smith}, {Sibthorpe}, {Smail}, {Stevens}, {Sutherland},
  {Takeuchi}, {Tedds}, {Temi}, {Tuffs}, {Trichas}, {Vaccari}, {Valtchanov},
  {van der Werf}, {Verma}, {Vieria}, {Vlahakis}, \& {White}}]{eales10}
{Eales}, S., {Dunne}, L., {Clements}, D., {et~al.} 2010, \pasp, 122, 499

\bibitem[{{Engel} {et~al.}(2010){Engel}, {Tacconi}, {Davies}, {Neri}, {Smail},
  {Chapman}, {Genzel}, {Cox}, {Greve}, {Ivison}, {Blain}, {Bertoldi}, \&
  {Omont}}]{engel10}
{Engel}, H., {Tacconi}, L.~J., {Davies}, R.~I., {et~al.} 2010, \apj, 724, 233

\bibitem[{{Fathi} {et~al.}(2012){Fathi}, {Gatchell}, {Hatziminaoglou}, \&
  {Epinat}}]{fathi12}
{Fathi}, K., {Gatchell}, M., {Hatziminaoglou}, E., \& {Epinat}, B. 2012,
  \mnras, 423, L112

\bibitem[{{Frayer} {et~al.}(2011){Frayer}, {Harris}, {Baker}, {Ivison},
  {Smail}, {Negrello}, {Maddalena}, {Aretxaga}, {Baes}, {Birkinshaw},
  {Bonfield}, {Burgarella}, {Buttiglione}, {Cava}, {Clements}, {Cooray},
  {Dannerbauer}, {Dariush}, {De Zotti}, {Dunlop}, {Dunne}, {Dye}, {Eales},
  {Fritz}, {Gonzalez-Nuevo}, {Herranz}, {Hopwood}, {Hughes}, {Ibar}, {Jarvis},
  {Lagache}, {Leeuw}, {Lopez-Caniego}, {Maddox}, {Micha{\l}owski}, {Omont},
  {Pohlen}, {Rigby}, {Rodighiero}, {Scott}, {Serjeant}, {Smith}, {Swinbank},
  {Temi}, {Thompson}, {Valtchanov}, {van der Werf}, \& {Verma}}]{frayer11}
{Frayer}, D.~T., {Harris}, A.~I., {Baker}, A.~J., {et~al.} 2011, \apjl, 726,
  L22

\bibitem[{{Genzel} {et~al.}(2003){Genzel}, {Baker}, {Tacconi}, {Lutz}, {Cox},
  {Guilloteau}, \& {Omont}}]{genzel03}
{Genzel}, R., {Baker}, A.~J., {Tacconi}, L.~J., {et~al.} 2003, \apj, 584, 633

\bibitem[{{Greve} {et~al.}(2008){Greve}, {Pope}, {Scott}, {Ivison}, {Borys},
  {Conselice}, \& {Bertoldi}}]{greve08}
{Greve}, T.~R., {Pope}, A., {Scott}, D., {et~al.} 2008, \mnras, 389, 1489

\bibitem[{{Greve} {et~al.}(2012){Greve}, {Vieira}, {Wei{\ss}}, {Aguirre},
  {Aird}, {Ashby}, {Benson}, {Bleem}, {Bradford}, {Brodwin}, {Carlstrom},
  {Chang}, {Chapman}, {Crawford}, {de Breuck}, {de Haan}, {Dobbs}, {Downes},
  {Fassnacht}, {Fazio}, {George}, {Gladders}, {Gonzalez}, {Halverson},
  {Hezaveh}, {High}, {Holder}, {Holzapfel}, {Hoover}, {Hrubes}, {Johnson},
  {Keisler}, {Knox}, {Lee}, {Leitch}, {Lueker}, {Luong-Van}, {Malkan},
  {Marrone}, {McIntyre}, {McMahon}, {Mehl}, {Menten}, {Meyer}, {Montroy},
  {Murphy}, {Natoli}, {Padin}, {Plagge}, {Pryke}, {Reichardt}, {Rest},
  {Rosenman}, {Ruel}, {Ruhl}, {Schaffer}, {Sharon}, {Shaw}, {Shirokoff},
  {Stalder}, {Stanford}, {Staniszewski}, {Stark}, {Story}, {Vanderlinde},
  {Walsh}, {Welikala}, \& {Williamson}}]{greve12}
{Greve}, T.~R., {Vieira}, J.~D., {Wei{\ss}}, A., {et~al.} 2012, \apj, 756, 101

\bibitem[{{Harris} {et~al.}(2012){Harris}, {Baker}, {Frayer}, {Smail},
  {Swinbank}, {Riechers}, {van der Werf}, {Auld}, {Baes}, {Bussmann},
  {Buttiglione}, {Cava}, {Clements}, {Cooray}, {Dannerbauer}, {Dariush},
  {DeZotti}, {Dunne}, {Dye}, {Eales}, {Fritz}, {Gonzalez-Nuevo}, {Hopwood},
  {Ibar}, {Ivison}, {Jarvis}, {Maddox}, {Negrello}, {Rigby}, {Smith}, {Temi},
  \& {Wardlow}}]{harris12}
{Harris}, A.~I., {Baker}, A.~J., {Frayer}, D.~T., {et~al.} 2012, ArXiv e-prints

\bibitem[{{Hayward} {et~al.}(2012){Hayward}, {Narayanan}, {Kere{\v s}},
  {Jonsson}, {Hopkins}, {Cox}, \& {Hernquist}}]{hayward12}
{Hayward}, C.~C., {Narayanan}, D., {Kere{\v s}}, D., {et~al.} 2012, ArXiv
  e-prints

\bibitem[{{Hezaveh} \& {Holder}(2011)}]{hezaveh11}
{Hezaveh}, Y.~D., \& {Holder}, G.~P. 2011, \apj, 734, 52

\bibitem[{{Hezaveh} {et~al.}(2012){Hezaveh}, {Marrone}, {Fassnacht}, {Spilker},
  {Vieira}, {et~al.}}]{hezaveh12b}
{Hezaveh}, Y.~D., {Marrone}, D.~P., {Fassnacht}, C.~D., {et~al.} 2012,
  submitted to ApJ

\bibitem[{{Hezaveh} {et~al.}(2013){Hezaveh}, {Marrone}, {Fassnacht}, {Spilker},
  \& {Vieira}}]{hezaveh13}
{Hezaveh}, Y.~D., {Marrone}, D.~P., {Fassnacht}, C.~D., {Spilker}, J.~S., \&
  {Vieira}, J.~D. e.~a. 2013, submitted to ApJ

\bibitem[{{Hopkins} \& {Beacom}(2006)}]{hopkins06}
{Hopkins}, A.~M., \& {Beacom}, J.~F. 2006, \apj, 651, 142

\bibitem[{{Hughes} {et~al.}(1998){Hughes}, {Serjeant}, {Dunlop},
  {Rowan-Robinson}, {Blain}, {Mann}, {Ivison}, {Peacock}, {Efstathiou}, {Gear},
  {Oliver}, {Lawrence}, {Longair}, {Goldschmidt}, \& {Jenness}}]{hughes98}
{Hughes}, D.~H., {Serjeant}, S., {Dunlop}, J., {et~al.} 1998, \nat, 394, 241

\bibitem[{{Ivison} {et~al.}(2011){Ivison}, {Papadopoulos}, {Smail}, {Greve},
  {Thomson}, {Xilouris}, \& {Chapman}}]{ivison11}
{Ivison}, R.~J., {Papadopoulos}, P.~P., {Smail}, I., {et~al.} 2011, \mnras,
  412, 1913

\bibitem[{{Ivison} {et~al.}(2010){Ivison}, {Smail}, {Papadopoulos}, {Wold},
  {Richard}, {Swinbank}, {Kneib}, \& {Owen}}]{ivison10b}
{Ivison}, R.~J., {Smail}, I., {Papadopoulos}, P.~P., {et~al.} 2010, \mnras,
  404, 198

\bibitem[{{Ivison} {et~al.}(2002){Ivison}, {Greve}, {Smail}, {Dunlop}, {Roche},
  {Scott}, {Page}, {Stevens}, {Almaini}, {Blain}, {Willott}, {Fox}, {Gilbank},
  {Serjeant}, \& {Hughes}}]{ivison02}
{Ivison}, R.~J., {Greve}, T.~R., {Smail}, I., {et~al.} 2002, \mnras, 337, 1

\bibitem[{{Ivison} {et~al.}(2007){Ivison}, {Greve}, {Dunlop}, {Peacock},
  {Egami}, {Smail}, {Ibar}, {van Kampen}, {Aretxaga}, {Babbedge}, {Biggs},
  {Blain}, {Chapman}, {Clements}, {Coppin}, {Farrah}, {Halpern}, {Hughes},
  {Jarvis}, {Jenness}, {Jones}, {Mortier}, {Oliver}, {Papovich},
  {P{\'e}rez-Gonz{\'a}lez}, {Pope}, {Rawlings}, {Rieke}, {Rowan-Robinson},
  {Savage}, {Scott}, {Seigar}, {Serjeant}, {Simpson}, {Stevens}, {Vaccari},
  {Wagg}, \& {Willott}}]{ivison07}
{Ivison}, R.~J., {Greve}, T.~R., {Dunlop}, J.~S., {et~al.} 2007, \mnras, 380,
  199

\bibitem[{{Karim} {et~al.}(2012){Karim}, {Swinbank}, {Hodge}, {Smail},
  {Walter}, {Biggs}, {Simpson}, {Danielson}, {Alexander}, {Bertoldi},
  {Chapman}, {Coppin}, {Dannerbauer}, {Edge}, {Greve}, {Ivison}, {Knudsen},
  {Menten}, {Schinnerer}, {Wardlow}, {Wei{\ss}}, \& {van der Werf}}]{karim12}
{Karim}, A., {Swinbank}, M., {Hodge}, J., {et~al.} 2012, ArXiv e-prints

\bibitem[{{Komatsu} {et~al.}(2011){Komatsu}, {Smith}, {Dunkley}, {Bennett},
  {Gold}, {Hinshaw}, {Jarosik}, {Larson}, {Nolta}, {Page}, {Spergel},
  {Halpern}, {Hill}, {Kogut}, {Limon}, {Meyer}, {Odegard}, {Tucker}, {Weiland},
  {Wollack}, \& {Wright}}]{komatsu11}
{Komatsu}, E., {Smith}, K.~M., {Dunkley}, J., {et~al.} 2011, \apjs, 192, 18

\bibitem[{{Lacey} {et~al.}(2010){Lacey}, {Baugh}, {Frenk}, {Benson}, {Orsi},
  {Silva}, {Granato}, \& {Bressan}}]{lacey10}
{Lacey}, C.~G., {Baugh}, C.~M., {Frenk}, C.~S., {et~al.} 2010, \mnras, 405, 2

\bibitem[{{Lilly} {et~al.}(1996){Lilly}, {Le Fevre}, {Hammer}, \&
  {Crampton}}]{lilly96}
{Lilly}, S.~J., {Le Fevre}, O., {Hammer}, F., \& {Crampton}, D. 1996, \apjl,
  460, L1

\bibitem[{{Lupu} {et~al.}(2012){Lupu}, {Scott}, {Aguirre}, {Aretxaga}, {Auld},
  {Barton}, {Beelen}, {Bertoldi}, {Bock}, {Bonfield}, {Bradford},
  {Buttiglione}, {Cava}, {Clements}, {Cooke}, {Cooray}, {Dannerbauer},
  {Dariush}, {De Zotti}, {Dunne}, {Dye}, {Eales}, {Frayer}, {Fritz}, {Glenn},
  {Hughes}, {Ibar}, {Ivison}, {Jarvis}, {Kamenetzky}, {Kim}, {Lagache},
  {Leeuw}, {Maddox}, {Maloney}, {Matsuhara}, {Murphy}, {Naylor}, {Negrello},
  {Nguyen}, {Omont}, {Pascale}, {Pohlen}, {Rigby}, {Rodighiero}, {Serjeant},
  {Smith}, {Temi}, {Thompson}, {Valtchanov}, {Verma}, {Vieira}, \&
  {Zmuidzinas}}]{lupu12}
{Lupu}, R.~E., {Scott}, K.~S., {Aguirre}, J.~E., {et~al.} 2012, \apj, 757, 135

\bibitem[{{Madau} {et~al.}(1996){Madau}, {Ferguson}, {Dickinson}, {Giavalisco},
  {Steidel}, \& {Fruchter}}]{madau96}
{Madau}, P., {Ferguson}, H.~C., {Dickinson}, M.~E., {et~al.} 1996, \mnras, 283,
  1388

\bibitem[{{McMullin} {et~al.}(2007){McMullin}, {Waters}, {Schiebel}, {Young},
  \& {Golap}}]{mcmullin07}
{McMullin}, J.~P., {Waters}, B., {Schiebel}, D., {Young}, W., \& {Golap}, K.
  2007, in Astronomical Society of the Pacific Conference Series, Vol. 376,
  Astronomical Data Analysis Software and Systems XVI, ed. R.~A. {Shaw},
  F.~{Hill}, \& D.~J. {Bell}, 127

\bibitem[{{Moshir} {et~al.}(1992){Moshir}, {Kopman}, \& {Conrow}}]{moshir92}
{Moshir}, M., {Kopman}, G., \& {Conrow}, T.~A.~O., eds. 1992, {IRAS Faint
  Source Survey, Explanatory supplement version 2} (Pasadena: Infrared
  Processing and Analysis Center, California Institute of Technology, 1992,
  edited by Moshir, M.; Kopman, G.; Conrow, T. a.o.)

\bibitem[{{Negrello} {et~al.}(2010){Negrello}, {Hopwood}, {De Zotti}, {Cooray},
  {Verma}, {Bock}, {Frayer}, {Gurwell}, {Omont}, {Neri}, {Dannerbauer},
  {Leeuw}, {Barton}, {Cooke}, {Kim}, {da Cunha}, {Rodighiero}, {Cox},
  {Bonfield}, {Jarvis}, {Serjeant}, {Ivison}, {Dye}, {Aretxaga}, {Hughes},
  {Ibar}, {Bertoldi}, {Valtchanov}, {Eales}, {Dunne}, {Driver}, {Auld},
  {Buttiglione}, {Cava}, {Grady}, {Clements}, {Dariush}, {Fritz}, {Hill},
  {Hornbeck}, {Kelvin}, {Lagache}, {Lopez-Caniego}, {Gonzalez-Nuevo}, {Maddox},
  {Pascale}, {Pohlen}, {Rigby}, {Robotham}, {Simpson}, {Smith}, {Temi},
  {Thompson}, {Woodgate}, {York}, {Aguirre}, {Beelen}, {Blain}, {Baker},
  {Birkinshaw}, {Blundell}, {Bradford}, {Burgarella}, {Danese}, {Dunlop},
  {Fleuren}, {Glenn}, {Harris}, {Kamenetzky}, {Lupu}, {Maddalena}, {Madore},
  {Maloney}, {Matsuhara}, {Michaowski}, {Murphy}, {Naylor}, {Nguyen},
  {Popescu}, {Rawlings}, {Rigopoulou}, {Scott}, {Scott}, {Seibert}, {Smail},
  {Tuffs}, {Vieira}, {van der Werf}, \& {Zmuidzinas}}]{negrello10}
{Negrello}, M., {Hopwood}, R., {De Zotti}, G., {et~al.} 2010, Science, 330, 800

\bibitem[{{Oliver} {et~al.}(2010){Oliver}, {Wang}, {Smith}, {Altieri},
  {Amblard}, {Arumugam}, {Auld}, {Aussel}, {Babbedge}, {Blain}, {Bock},
  {Boselli}, {Buat}, {Burgarella}, {Castro-Rodr{\'{\i}}guez}, {Cava},
  {Chanial}, {Clements}, {Conley}, {Conversi}, {Cooray}, {Dowell}, {Dwek},
  {Eales}, {Elbaz}, {Fox}, {Franceschini}, {Gear}, {Glenn}, {Griffin},
  {Halpern}, {Hatziminaoglou}, {Ibar}, {Isaak}, {Ivison}, {Lagache},
  {Levenson}, {Lu}, {Madden}, {Maffei}, {Mainetti}, {Marchetti},
  {Mitchell-Wynne}, {Mortier}, {Nguyen}, {O'Halloran}, {Omont}, {Page},
  {Panuzzo}, {Papageorgiou}, {Pearson}, {P{\'e}rez-Fournon}, {Pohlen},
  {Rawlings}, {Raymond}, {Rigopoulou}, {Rizzo}, {Roseboom}, {Rowan-Robinson},
  {S{\'a}nchez Portal}, {Savage}, {Schulz}, {Scott}, {Seymour}, {Shupe},
  {Stevens}, {Symeonidis}, {Trichas}, {Tugwell}, {Vaccari}, {Valiante},
  {Valtchanov}, {Vieira}, {Vigroux}, {Ward}, {Wright}, {Xu}, \&
  {Zemcov}}]{oliver10}
{Oliver}, S.~J., {Wang}, L., {Smith}, A.~J., {et~al.} 2010, \aap, 518, L21+

\bibitem[{{Petry} {et~al.}(2012)}]{petry12}
{Petry}, D., {et~al.} 2012, ArXiv e-prints

\bibitem[{{Pope} {et~al.}(2005){Pope}, {Borys}, {Scott}, {Conselice},
  {Dickinson}, \& {Mobasher}}]{pope05}
{Pope}, A., {Borys}, C., {Scott}, D., {et~al.} 2005, \mnras, 358, 149

\bibitem[{{Pope} {et~al.}(2006){Pope}, {Scott}, {Dickinson}, {Chary},
  {Morrison}, {Borys}, {Sajina}, {Alexander}, {Daddi}, {Frayer}, {MacDonald},
  \& {Stern}}]{pope06}
{Pope}, A., {Scott}, D., {Dickinson}, M., {et~al.} 2006, \mnras, 370, 1185

\bibitem[{{Riechers} {et~al.}(2011{\natexlab{a}}){Riechers}, {Hodge}, {Walter},
  {Carilli}, \& {Bertoldi}}]{riechers11b}
{Riechers}, D.~A., {Hodge}, J., {Walter}, F., {Carilli}, C.~L., \& {Bertoldi},
  F. 2011{\natexlab{a}}, \apjl, 739, L31

\bibitem[{{Riechers} {et~al.}(2010){Riechers}, {Capak}, {Carilli}, {Cox},
  {Neri}, {Scoville}, {Schinnerer}, {Bertoldi}, \& {Yan}}]{riechers10}
{Riechers}, D.~A., {Capak}, P.~L., {Carilli}, C.~L., {et~al.} 2010, \apjl, 720,
  L131

\bibitem[{{Riechers} {et~al.}(2011{\natexlab{b}}){Riechers}, {Cooray}, {Omont},
  {Neri}, {Harris}, {Baker}, {Cox}, {Frayer}, {Carpenter}, {Auld}, {Aussel},
  {Beelen}, {Blundell}, {Bock}, {Brisbin}, {Burgarella}, {Chanial}, {Chapman},
  {Clements}, {Conley}, {Dowell}, {Eales}, {Farrah}, {Franceschini}, {Gavazzi},
  {Glenn}, {Griffin}, {Gurwell}, {Ivison}, {Kim}, {Krips}, {Mortier}, {Oliver},
  {Page}, {Papageorgiou}, {Pearson}, {P{\'e}rez-Fournon}, {Pohlen}, {Rawlings},
  {Raymond}, {Rodighiero}, {Roseboom}, {Rowan-Robinson}, {Scott}, {Seymour},
  {Smith}, {Symeonidis}, {Tugwell}, {Vaccari}, {Vieira}, {Vigroux}, {Wang},
  {Wardlow}, \& {Wright}}]{riechers11}
{Riechers}, D.~A., {Cooray}, A., {Omont}, A., {et~al.} 2011{\natexlab{b}},
  \apjl, 733, L12

\bibitem[{{Rujopakarn} {et~al.}(2011){Rujopakarn}, {Rieke}, {Eisenstein}, \&
  {Juneau}}]{rujopakarn11}
{Rujopakarn}, W., {Rieke}, G.~H., {Eisenstein}, D.~J., \& {Juneau}, S. 2011,
  \apj, 726, 93

\bibitem[{{Smail} {et~al.}(1997){Smail}, {Ivison}, \& {Blain}}]{smail97}
{Smail}, I., {Ivison}, R.~J., \& {Blain}, A.~W. 1997, \apjl, 490, L5+

\bibitem[{{Smolcic} {et~al.}(2012){Smolcic}, {Aravena}, {Navarrete},
  {Schinnerer}, {Riechers}, {Bertoldi}, {Feruglio}, {Finoguenov}, {Salvato},
  {Sargent}, {McCracken}, {Albrecht}, {Karim}, {Capak}, {Carilli},
  {Cappelluti}, {Elvis}, {Ilbert}, {Kartaltepe}, {Lilly}, {Sanders}, {Sheth},
  {Scoville}, \& {Taniguchi}}]{smolcic12}
{Smolcic}, V., {Aravena}, M., {Navarrete}, F., {et~al.} 2012, ArXiv e-prints

\bibitem[{{Stern} {et~al.}(2006){Stern}, {Chary}, {Eisenhardt}, \&
  {Moustakas}}]{stern06}
{Stern}, D., {Chary}, R.-R., {Eisenhardt}, P.~R.~M., \& {Moustakas}, L.~A.
  2006, \aj, 132, 1405

\bibitem[{{Story} {et~al.}(2012){Story}, {Reichardt}, {Hou}, {Keisler}, {Aird},
  {Benson}, {Bleem}, {Carlstrom}, {Chang}, {Cho}, {Crawford}, {Crites}, {de
  Haan}, {Dobbs}, {Dudley}, {Follin}, {George}, {Halverson}, {Holder},
  {Holzapfel}, {Hoover}, {Hrubes}, {Joy}, {Knox}, {Lee}, {Leitch}, {Lueker},
  {Luong-Van}, {McMahon}, {Mehl}, {Meyer}, {Millea}, {Mohr}, {Montroy},
  {Padin}, {Plagge}, {Pryke}, {Ruhl}, {Sayre}, {Schaffer}, {Shaw}, {Shirokoff},
  {Spieler}, {Staniszewski}, {Stark}, {van Engelen}, {Vanderlinde}, {Vieira},
  {Williamson}, \& {Zahn}}]{story12}
{Story}, K.~T., {Reichardt}, C.~L., {Hou}, Z., {et~al.} 2012, ArXiv e-prints

\bibitem[{{Swinbank} {et~al.}(2010){Swinbank}, {Smail}, {Longmore}, {Harris},
  {Baker}, {De Breuck}, {Richard}, {Edge}, {Ivison}, {Blundell}, {Coppin},
  {Cox}, {Gurwell}, {Hainline}, {Krips}, {Lundgren}, {Neri}, {Siana},
  {Siringo}, {Stark}, {Wilner}, \& {Younger}}]{swinbank10}
{Swinbank}, A.~M., {Smail}, I., {Longmore}, S., {et~al.} 2010, \nat, 464, 733

\bibitem[{{Tacconi} {et~al.}(2006){Tacconi}, {Neri}, {Chapman}, {Genzel},
  {Smail}, {Ivison}, {Bertoldi}, {Blain}, {Cox}, {Greve}, \&
  {Omont}}]{tacconi06}
{Tacconi}, L.~J., {Neri}, R., {Chapman}, S.~C., {et~al.} 2006, \apj, 640, 228

\bibitem[{{Tacconi} {et~al.}(2008){Tacconi}, {Genzel}, {Smail}, {Neri},
  {Chapman}, {Ivison}, {Blain}, {Cox}, {Omont}, {Bertoldi}, {Greve},
  {F{\"o}rster Schreiber}, {Genel}, {Lutz}, {Swinbank}, {Shapley}, {Erb},
  {Cimatti}, {Daddi}, \& {Baker}}]{tacconi08}
{Tacconi}, L.~J., {Genzel}, R., {Smail}, I., {et~al.} 2008, \apj, 680, 246

\bibitem[{{van der Werf} {et~al.}(2010){van der Werf}, {Isaak}, {Meijerink},
  {Spaans}, {Rykala}, {Fulton}, {Loenen}, {Walter}, {Wei{\ss}}, {Armus},
  {Fischer}, {Israel}, {Harris}, {Veilleux}, {Henkel}, {Savini}, {Lord},
  {Smith}, {Gonz{\'a}lez-Alfonso}, {Naylor}, {Aalto}, {Charmandaris}, {Dasyra},
  {Evans}, {Gao}, {Greve}, {G{\"u}sten}, {Kramer}, {Mart{\'{\i}}n-Pintado},
  {Mazzarella}, {Papadopoulos}, {Sanders}, {Spinoglio}, {Stacey}, {Vlahakis},
  {Wiedner}, \& {Xilouris}}]{vanderWerf10}
{van der Werf}, P.~P., {Isaak}, K.~G., {Meijerink}, R., {et~al.} 2010, \aap,
  518, L42

\bibitem[{{Vernet} {et~al.}(2011){Vernet}, {Dekker}, {D'Odorico}, {Kaper},
  {Kjaergaard}, {Hammer}, {Randich}, {Zerbi}, {Groot}, {Hjorth}, {Guinouard},
  {Navarro}, {Adolfse}, {Albers}, {Amans}, {Andersen}, {Andersen}, {Binetruy},
  {Bristow}, {Castillo}, {Chemla}, {Christensen}, {Conconi}, {Conzelmann},
  {Dam}, {de Caprio}, {de Ugarte Postigo}, {Delabre}, {di Marcantonio},
  {Downing}, {Elswijk}, {Finger}, {Fischer}, {Flores}, {Fran{\c c}ois},
  {Goldoni}, {Guglielmi}, {Haigron}, {Hanenburg}, {Hendriks}, {Horrobin},
  {Horville}, {Jessen}, {Kerber}, {Kern}, {Kiekebusch}, {Kleszcz}, {Klougart},
  {Kragt}, {Larsen}, {Lizon}, {Lucuix}, {Mainieri}, {Manuputy}, {Martayan},
  {Mason}, {Mazzoleni}, {Michaelsen}, {Modigliani}, {Moehler}, {M{\o}ller},
  {Norup S{\o}rensen}, {N{\o}rregaard}, {P{\'e}roux}, {Patat}, {Pena}, {Pragt},
  {Reinero}, {Rigal}, {Riva}, {Roelfsema}, {Royer}, {Sacco}, {Santin},
  {Schoenmaker}, {Spano}, {Sweers}, {Ter Horst}, {Tintori}, {Tromp}, {van
  Dael}, {van der Vliet}, {Venema}, {Vidali}, {Vinther}, {Vola}, {Winters},
  {Wistisen}, {Wulterkens}, \& {Zacchei}}]{Vernet11}
{Vernet}, J., {Dekker}, H., {D'Odorico}, S., {et~al.} 2011, \aap, 536, A105

\bibitem[{{Vieira} {et~al.}(2010){Vieira}, {Crawford}, {Switzer}, {Ade},
  {Aird}, {Ashby}, {Benson}, {Bleem}, {Brodwin}, {Carlstrom}, {Chang}, {Cho},
  {Crites}, {de Haan}, {Dobbs}, {Everett}, {George}, {Gladders}, {Hall},
  {Halverson}, {High}, {Holder}, {Holzapfel}, {Hrubes}, {Joy}, {Keisler},
  {Knox}, {Lee}, {Leitch}, {Lueker}, {Marrone}, {McIntyre}, {McMahon}, {Mehl},
  {Meyer}, {Mohr}, {Montroy}, {Padin}, {Plagge}, {Pryke}, {Reichardt}, {Ruhl},
  {Schaffer}, {Shaw}, {Shirokoff}, {Spieler}, {Stalder}, {Staniszewski},
  {Stark}, {Vanderlinde}, {Walsh}, {Williamson}, {Yang}, {Zahn}, \&
  {Zenteno}}]{vieira10}
{Vieira}, J.~D., {Crawford}, T.~M., {Switzer}, E.~R., {et~al.} 2010, \apj, 719,
  763

\bibitem[{{Vieira} {et~al.}(2013)}]{vieira13}
{Vieira}, J.~D., {et~al.} 2013, \nat, submitted

\bibitem[{{Walter} {et~al.}(2004){Walter}, {Carilli}, {Bertoldi}, {Menten},
  {Cox}, {Lo}, {Fan}, \& {Strauss}}]{walter04}
{Walter}, F., {Carilli}, C., {Bertoldi}, F., {et~al.} 2004, \apjl, 615, L17

\bibitem[{{Walter} {et~al.}(2009){Walter}, {Riechers}, {Cox}, {Neri},
  {Carilli}, {Bertoldi}, {Weiss}, \& {Maiolino}}]{walter09}
{Walter}, F., {Riechers}, D., {Cox}, P., {et~al.} 2009, \nat, 457, 699

\bibitem[{{Walter} {et~al.}(2012){Walter}, {Decarli}, {Carilli}, {Bertoldi},
  {Cox}, {da Cunha}, {Daddi}, {Dickinson}, {Downes}, {Elbaz}, {Ellis}, {Hodge},
  {Neri}, {Riechers}, {Weiss}, {Bell}, {Dannerbauer}, {Krips}, {Krumholz},
  {Lentati}, {Maiolino}, {Menten}, {Rix}, {Robertson}, {Spinrad}, {Stark}, \&
  {Stern}}]{walter12}
{Walter}, F., {Decarli}, R., {Carilli}, C., {et~al.} 2012, \nat, 486, 233

\bibitem[{{Wardlow} {et~al.}(2011){Wardlow}, {Smail}, {Coppin}, {Alexander},
  {Brandt}, {Danielson}, {Luo}, {Swinbank}, {Walter}, {Wei{\ss}}, {Xue},
  {Zibetti}, {Bertoldi}, {Biggs}, {Chapman}, {Dannerbauer}, {Dunlop},
  {Gawiser}, {Ivison}, {Knudsen}, {Kov{\'a}cs}, {Lacey}, {Menten}, {Padilla},
  {Rix}, \& {van der Werf}}]{wardlow11}
{Wardlow}, J.~L., {Smail}, I., {Coppin}, K.~E.~K., {et~al.} 2011, \mnras, 415,
  1479

\bibitem[{{Wei{\ss}} {et~al.}(2007){Wei{\ss}}, {Downes}, {Neri}, {Walter},
  {Henkel}, {Wilner}, {Wagg}, \& {Wiklind}}]{weiss07}
{Wei{\ss}}, A., {Downes}, D., {Neri}, R., {et~al.} 2007, \aap, 467, 955

\bibitem[{{Wei{\ss}} {et~al.}(2009{\natexlab{a}}){Wei{\ss}}, {Ivison},
  {Downes}, {Walter}, {Cirasuolo}, \& {Menten}}]{weiss09a}
{Wei{\ss}}, A., {Ivison}, R.~J., {Downes}, D., {et~al.} 2009{\natexlab{a}},
  \apjl, 705, L45

\bibitem[{{Wei{\ss}} {et~al.}(2009{\natexlab{b}}){Wei{\ss}}, {Kov{\'a}cs},
  {Coppin}, {Greve}, {Walter}, {Smail}, {Dunlop}, {Knudsen}, {Alexander},
  {Bertoldi}, {Brandt}, {Chapman}, {Cox}, {Dannerbauer}, {De Breuck},
  {Gawiser}, {Ivison}, {Lutz}, {Menten}, {Koekemoer}, {Kreysa}, {Kurczynski},
  {Rix}, {Schinnerer}, \& {van der Werf}}]{weiss09}
{Wei{\ss}}, A., {Kov{\'a}cs}, A., {Coppin}, K., {et~al.} 2009{\natexlab{b}},
  \apj, 707, 1201

\bibitem[{{Williamson} {et~al.}(2011){Williamson}, {Benson}, {High},
  {Vanderlinde}, {Ade}, {Aird}, {Andersson}, {Armstrong}, {Ashby}, {Bautz},
  {Bazin}, {Bertin}, {Bleem}, {Bonamente}, {Brodwin}, {Carlstrom}, {Chang},
  {Chapman}, {Clocchiatti}, {Crawford}, {Crites}, {de Haan}, {Desai}, {Dobbs},
  {Dudley}, {Fazio}, {Foley}, {Forman}, {Garmire}, {George}, {Gladders},
  {Gonzalez}, {Halverson}, {Holder}, {Holzapfel}, {Hoover}, {Hrubes}, {Jones},
  {Joy}, {Keisler}, {Knox}, {Lee}, {Leitch}, {Lueker}, {Luong-Van}, {Marrone},
  {McMahon}, {Mehl}, {Meyer}, {Mohr}, {Montroy}, {Murray}, {Padin}, {Plagge},
  {Pryke}, {Reichardt}, {Rest}, {Ruel}, {Ruhl}, {Saliwanchik}, {Saro},
  {Schaffer}, {Shaw}, {Shirokoff}, {Song}, {Spieler}, {Stalder}, {Stanford},
  {Staniszewski}, {Stark}, {Story}, {Stubbs}, {Vieira}, {Vikhlinin}, \&
  {Zenteno}}]{williamson11}
{Williamson}, R., {Benson}, B.~A., {High}, F.~W., {et~al.} 2011, \apj, 738, 139

\bibitem[{{Younger} {et~al.}(2007){Younger}, {Fazio}, {Huang}, {Yun}, {Wilson},
  {Ashby}, {Gurwell}, {Lai}, {Peck}, {Petitpas}, {Wilner}, {Iono}, {Kohno},
  {Kawabe}, {Hughes}, {Aretxaga}, {Webb}, {Mart{\'{\i}}nez-Sansigre}, {Kim},
  {Scott}, {Austermann}, {Perera}, {Lowenthal}, {Schinnerer}, \& {Smol{\v
  c}i{\'c}}}]{younger07}
{Younger}, J.~D., {Fazio}, G.~G., {Huang}, J.-S., {et~al.} 2007, \apj, 671,
  1531

\bibitem[{{Younger} {et~al.}(2010){Younger}, {Fazio}, {Ashby}, {Civano},
  {Gurwell}, {Huang}, {Iono}, {Peck}, {Petitpas}, {Scott}, {Wilner}, {Wilson},
  \& {Yun}}]{younger10}
{Younger}, J.~D., {Fazio}, G.~G., {Ashby}, M.~L.~N., {et~al.} 2010, \mnras,
  407, 1268

\bibitem[{{Yun} {et~al.}(2012){Yun}, {Scott}, {Guo}, {Aretxaga}, {Giavalisco},
  {Austermann}, {Capak}, {Chen}, {Ezawa}, {Hatsukade}, {Hughes}, {Iono},
  {Johnson}, {Kawabe}, {Kohno}, {Lowenthal}, {Miller}, {Morrison}, {Oshima},
  {Perera}, {Salvato}, {Silverman}, {Tamura}, {Williams}, \& {Wilson}}]{yun12}
{Yun}, M.~S., {Scott}, K.~S., {Guo}, Y., {et~al.} 2012, \mnras, 420, 957

\end{thebibliography}
\appendix
\section{Supplementary redshift information}
\label{sect:supplementary_redshift_info}

In this appendix, we show the supplementary observations that resolve redshift 
ambiguities in our ALMA observations:\\

\begin{figure*}[htb]
\centering
\includegraphics[width=7cm,angle=0]{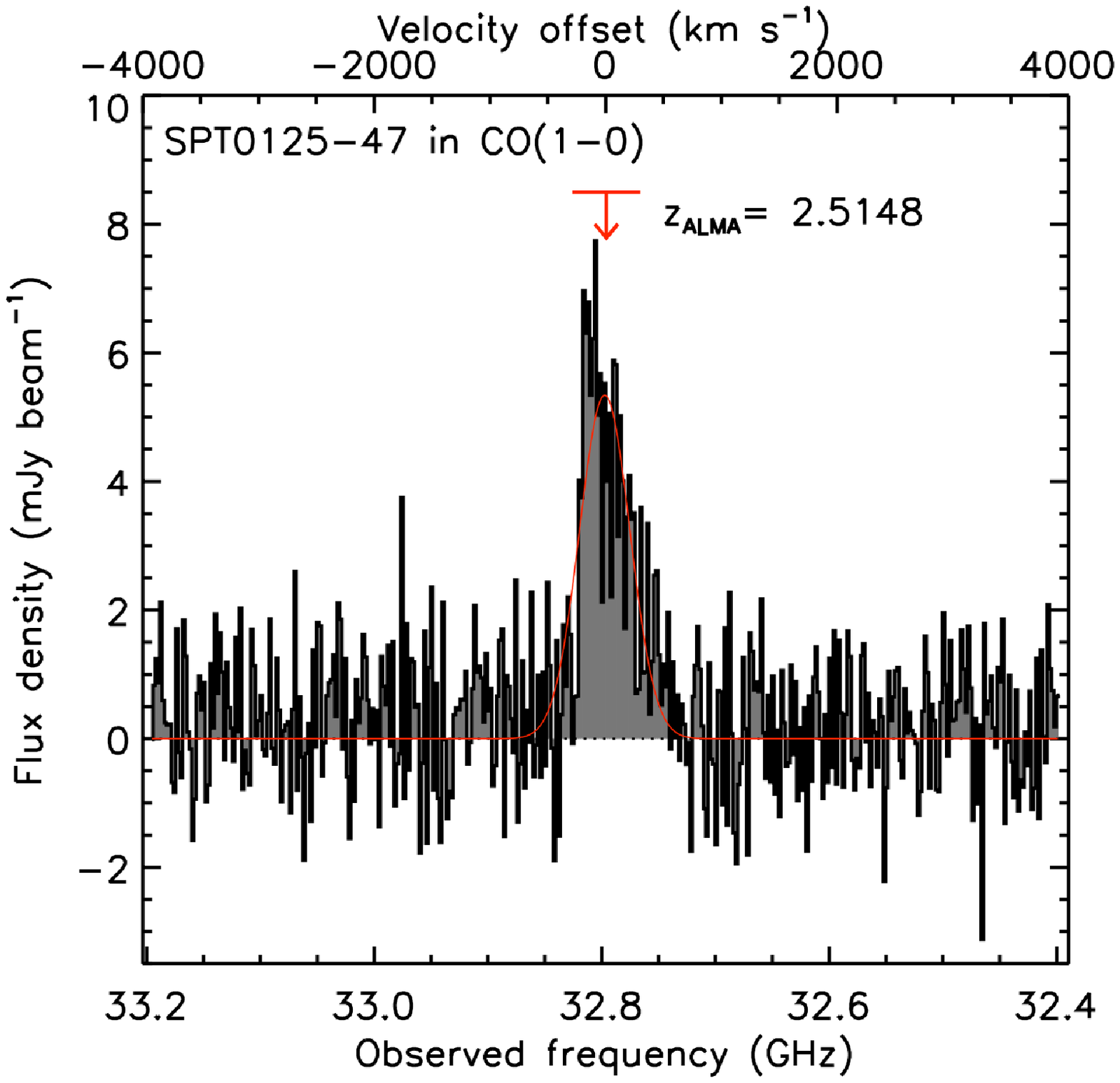}
\caption{Australia Telescope Compact Array spectrum of SPT~0125-47 showing the CO(1--0) line confirming the 
single ALMA line as CO(3--2) at $z$=2.5148.}
\label{Fig:SPT0125ATCA}
\end{figure*}

\emph{SPT0125-47}:
The identification of the 98~GHz line as CO(3--2) is confirmed with a
CO(1--0) detection using the Australia Telescope Compact Array
(Figure~\ref{Fig:SPT0125ATCA}).\\

\begin{figure*}[htb]
\centering
\includegraphics[width=7cm,angle=0]{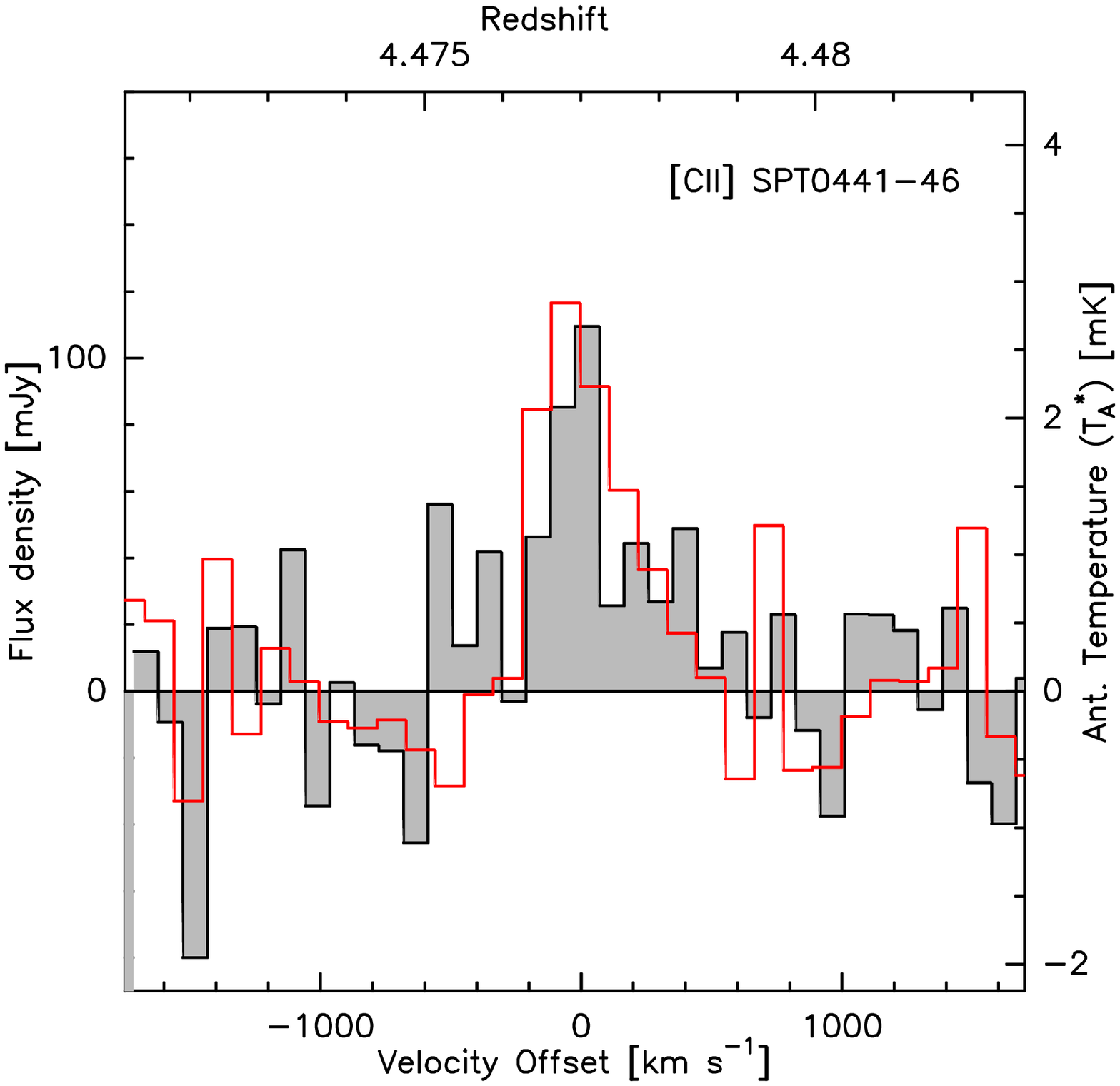}
\caption{APEX/FLASH spectrum of SPT~0441-46 showing the [CII]~$\lambda$158$\mu$m 
line (filled histogram) confirming the single ALMA line as CO(5-4) (red line, 
scaled to allow for a comparison between the line profiles) at $z$=4.4771.}
\label{Fig:SPT0441FLASH}
\end{figure*}

\emph{SPT0441-46}:
The identification of the 105~GHz line as CO(5-4) is confirmed with a
[CII]\,158$\mu$m detection with the First Light APEX Submillimetre
Heterodyne receiver (FLASH) on APEX (Figure~\ref{Fig:SPT0441FLASH}).
The low S/N [CI](1-0) detection with ALMA further strengthens this redshift 
identification.\\

\begin{figure*}[htb]
\centering
\includegraphics[width=7cm,angle=0]{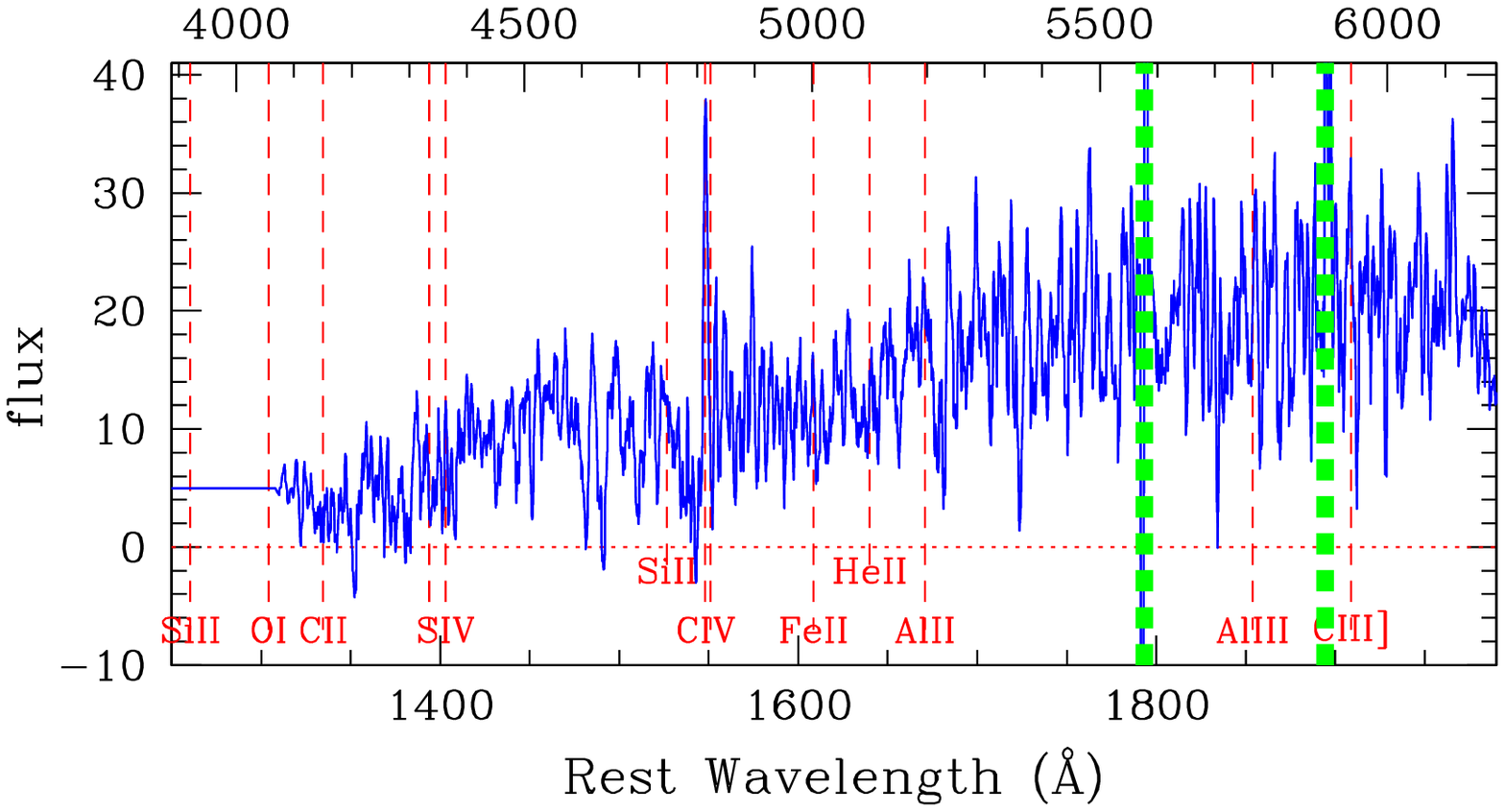}
\caption{VLT/FORS2 spectrum of SPT~0551-50 showing the 
C\ts {\scriptsize IV}~$\lambda$1549\AA\ line confirming the single ALMA 
line as CO(3-2) at $z$=2.123. Thin red dashed lines indicate the wavelengths 
of expected spectroscopic features, while thick green dotted lines mark areas 
dominated by skylines.
} 
\label{Fig:SPT0551VLT}
\end{figure*}

\emph{SPT0551-50}:
A strong emission line is visible at $\sim$4800\,\AA\ using the VLT
FOcal Reducer and Spectrograph \citep[FORS2;][]{appenzeller98}, which
is consistent with the 3~mm CO(3--2) line if we ascribe it to C{\small IV}\,${\rm 1550\,\AA}$. 
See Figure~\ref{Fig:SPT0551VLT}. \\

\begin{figure*}[htb]
\centering
\includegraphics[width=7cm,angle=0]{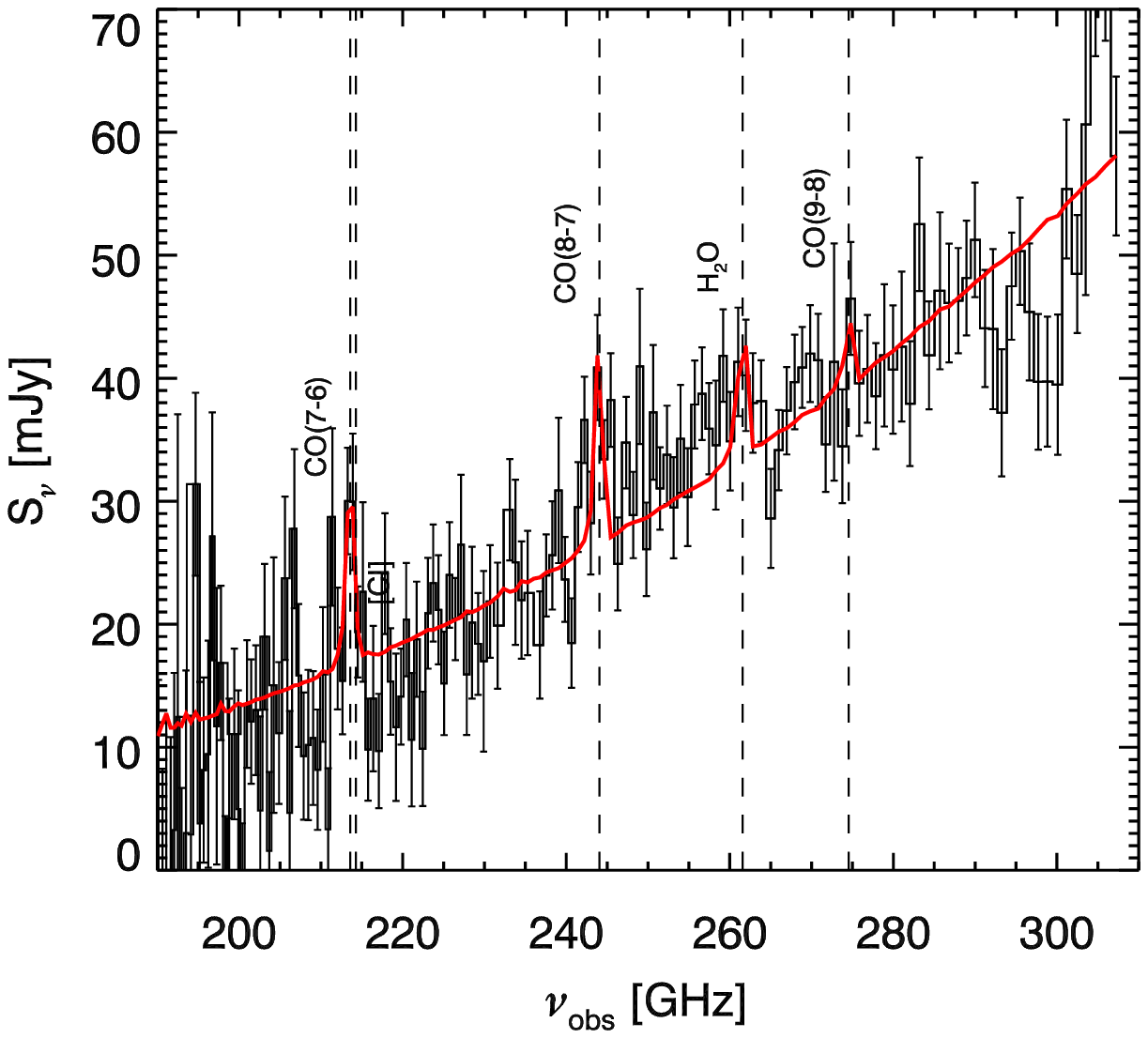}
\caption{APEX/Z-spec spectrum of SPT~2134-50 showing 2--3$\sigma$ detections 
of the CO(7--6) and CO(8-7) lines confirming the single ALMA line as CO(3--2) at $z$=2.779. 
Dashed lines mark the expected frequencies of CO and H$_2$O features. The combined significance of 
the lines detections is 5.6$\sigma$. 
} 
\label{Fig:SPT2134Zspec}
\end{figure*}

\begin{figure*}[htb]
\centering
\includegraphics[width=7cm,angle=0]{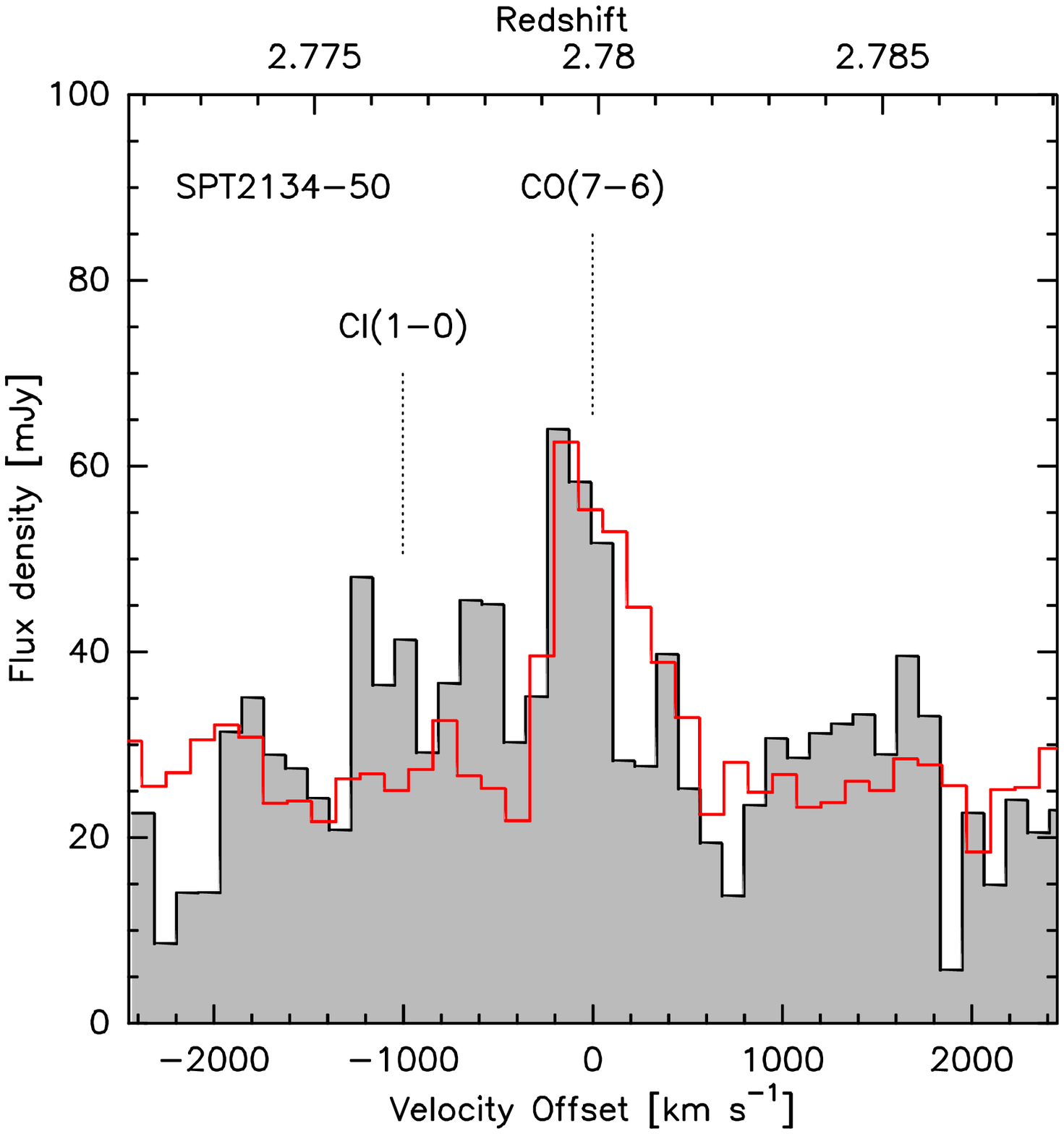}
\caption{SMA spectrum (filled histogram) of SPT~2134-50 showing CO(7--6) and evidence for [CI](2--1) 
confirming the single ALMA line as CO(3--2) (red line, scaled to allow for a comparison between the 
line profiles) at $z$=2.779.
} 
\label{Fig:SPT2134SMA}
\end{figure*}

\emph{SPT2134-50}:
The CO(7--6) and CO(8--7) lines are detected in a 190--310\,GHz
spectrum (Figure~\ref{Fig:SPT2134Zspec}) obtained with Z-Spec/APEX
\citep[][]{bradford04}, and subsequently confirmed through
Submillimeter Array (SMA) observations of CO(7--6) and [CI](2-1) (See
Figure~\ref{Fig:SPT2134SMA}). The ALMA data, released later, agree with
this identification, with ALMA detecting the CO(3--2) line at
91.5~GHz.\\

\begin{figure*}[htb]
\centering
\includegraphics[width=7cm,angle=0]{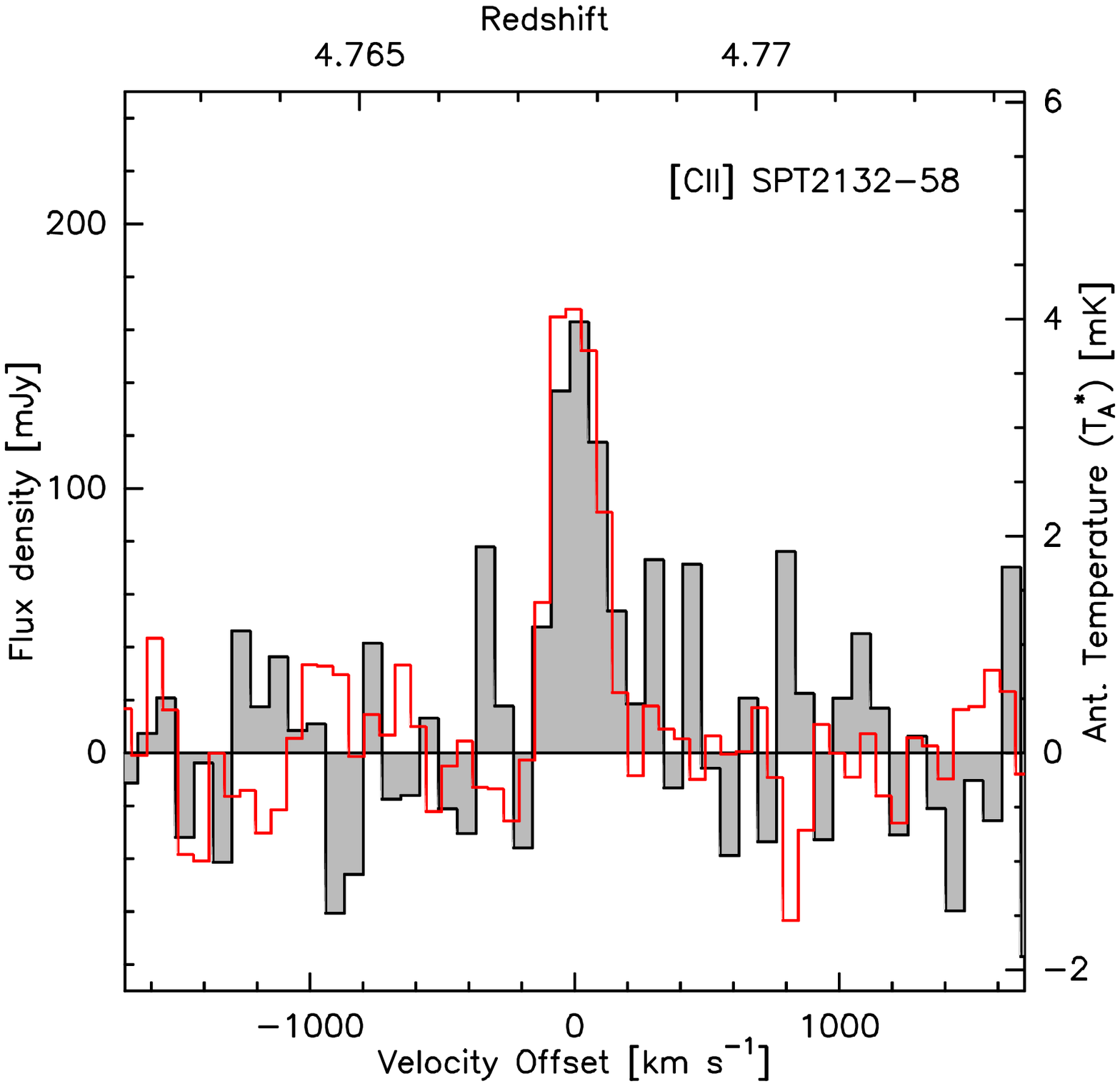}
\caption{APEX/FLASH spectrum of SPT~2132-58 showing the [CII]~$\lambda$158$\mu$m 
line (filled histogram) confirming the single ALMA line as CO(5-4) (red line, 
scaled to allow for a comparison between the line profiles) at $z$=4.7677.
} 
\label{Fig:SPT2132FLASH}
\end{figure*}

\emph{SPT2132-58}:
The identification of the 100~GHz line as CO(5-4) is confirmed with a
[CII]\,158$\mu$m detection with the First Light APEX Submillimetre
Heterodyne receiver (FLASH) on APEX (Figure~\ref{Fig:SPT2132FLASH}).\\



\section{Supplementary information for sources with a no or single line detections}
\label{sect:supplementary_singleline}
Below, we discuss the 9 individual cases which have zero or one CO line detected with ALMA 
and no additional spectroscopic observations.\\

%\emph{SPT0125-47}: 
%Despite detecting a bright line at 98.34\,GHz, we
%find no other lines in the ALMA spectrum.  CO(4--3) can be excluded as
%\cone\ should have been detected given the bright CO line.  In case of
%CO(5--4) at $z$=4.857 the \cone\ line is not covered by our spectrum,
%but we exclude this identification based on the high implied dust
%temperature (\tdust=84\,K). CO(3--2) at $z$=2.514, implying 
%\tdust=41\,K, is our preferred identification, but CO(2--1) at $z$=1.343 
%with \tdust=24\,K remains an alternative solution.\\

\emph{SPT0125-50}: 
In this galaxy we detect a second tentative line feature at 99.20\,GHz
which is consistent with the expected frequency for \cone\ if the
93.03\,GHz line is CO(4--3). This is our preferred identification,
giving $z$=3.959. In case the weak 99.20\,GHz feature is not real,
CO(5--4) as identification for the bright line can be excluded as
CO(6--5) should have been detected too. For CO(2--1) at $z$=1.343, the
implied dust temperature would be 17\,K, lower than any we observe.
An additional plausible identification is CO(3--2) at $z$=2.717 (\tdust=30\,K).\\

\emph{SPT0128-51}:
No line is detected in this spectrum. If it is in the $z$=1.74--2.00
redshift desert, the dust temperature is a low $\tdust \approx 19$\,K.
Alternatively, at higher redshift the line-to-continuum ratio should
be smaller and could go undetected. If SPT0128-51 has the same \tdust\
as the median temperature of the unambiguously identified
population, 37\,K, its corresponding photometric redshift would be $z=4.3$.\\

\emph{SPT0300-46}:
This source is similar to SPT0125-50 and has a clear CO detection at 100.30\,GHz
and a tentative \cone\ line at 107.08\,GHz which implies CO(4--3) at
$z$=3.594. If the latter feature is not real, CO(3--2) at
$z$=2.446 and \tdust=27\,K is an alternative interpretation. CO(2--1) at
$z$=1.298 would imply \tdust=17\,K, which we consider unlikely. CO(5--4) can be
ruled out as CO(6--5) would have also been detected.\\

\emph{SPT0319-47}:
%There is a tentative line feature at 104.38\,GHz. The implied line
%width would be large ($\Delta v$=1800\,kms) but not without precedent
%for DSFGs. Given, however, the low significance of the line we regard
%it as undetected. 
No line is detected in this spectrum. The dust temperature would be
$\approx 20$\,K if the source is in the $z$=1.74-2.00 redshift desert.  As with
SPT0128-51, a higher redshift with weak lines cannot be ruled out.
Matching this source to the median temperature of the known sample
yields a photometric redshift of $z=4.0$.\\

\emph{SPT0452-50 \label{SPT0452-50}}: There is a clear line detection
at the very edge of the band (114.87\,GHz).  CO(4--3) and CO(5--4) can
be excluded as a second CO line would be detected in the band.
CO(2--1) at $z$=1.007 can be excluded as it would
imply \tdust=13\,K. This identifies the line as CO(3--2) at $z$=2.010.\\

\emph{SPT0457-49}:
There is no line detected in the spectrum. The dust temperature would
be $\approx 22$\,K if the source is in the $z$=1.74-2.00 redshift
desert.  As with SPT0128-51, a higher redshift with weak lines cannot
be ruled out. This source would lie at $z$=3.3 were its \tdust\ the
same as the median of the unambiguous sample.\\

\emph{SPT0459-58}:
A single CO line is detected at 98.40\,GHz. If the line is identified
as CO(4--3) at $z$=3.685, the \cone\ transition is in the band as well
at 105.12\,GHz . In this case the \cone/CO(4--3) flux density ratio
limit is $<$0.15 (3$\sigma$), comparable to the limit we observe for
SPT0345-47.  Therefore CO(4--3) cannot be excluded but would require
an unusually low (but not unprecedented) \ci/CO line ratio. CO(2--1)
at $z$=1.343 can be excluded based on the dust temperature
(\tdust=14\,K). CO(3--2) at $z$=2.514 implies \tdust=22\,K.  The
most plausible identification is CO(5--4) at $z$=4.856 with \tdust=41\,K.\\

\emph{SPT0512-59}:
A single CO line is detected at 106.94\,GHz. CO(4--3) and CO(5--4) can
be excluded as \cone\ should have been detected given the bright CO
line.  CO(2--1) at $z$=1.156 is unlikely as it implies \tdust=20\,K,
but cannot be ruled out. Our preferred identification is CO(3--2) at
$z=2.234$ with \tdust=33\,K. \\

\emph{SPT0550-53}:
A single bright CO line is identified at 111.67\,GHz. CO(2--1) at
$z$=1.064 is excluded (\tdust=14\,K); for CO(5--4) at $z$=4.160
CO(4--3) should have been detected. CO(3--2) at $z$=2.096 and CO(4--3)
at $z$=3.128 are both plausible identifications with \tdust=22 and
31\,K, respectively.\\

%\emph{SPT2132-58}:
%A single CO line identified at 99.91\,GHz. CO(2--1) at $z$=1.307 is
%excluded based on the dust temperature (\tdust=13\,K). CO(3--2) at
%$z$=2.461, CO(4--3) at $z$=3.614, and CO(5--4) at $z$=4.498 are all
%possible identifications with \tdust = 21, 29, and 37\,K, respectively.
%A \cone\ detection is not expected given the SNR of the CO line.\\

\section{Supplementary Far-Infrared Photometry}
\label{sect:supplementary_photometry}

In this appendix, we show the supplementary FIR through mm photometric
measurements used to determine dust temperatures, assign
probabilistic redshift estimates to the sources with single-line
detections and to show the representativeness of the dust colors of
this subsample for the larger sample of 1.4~mm selected SPT sources meeting
the same selection criteria.

We used the LABOCA instruments at APEX to obtain 870 imaging. The observations 
took place during ESO and MPIfR observing time between 2010 September and 2012 May. The
observing strategy and data processing are described in
\citep{greve12}.

\textit{Herschel}-SPIRE maps at 250, 350, and 500\,\um\ were observed as
part of program OT2\_jvieira\_5. The SPIRE data consists of a triple
repetition map, with coverage complete to a radius of 5 arcmin from
the nominal SPT position. The maps were produced via the standard
reduction pipeline HIPE v9.0, the SPIRE Photometer Interactive
Analysis (SPIA) package v1.7, and the calibration product v8.1.
Photometry was extracted by fitting a gaussian profile to the 
SPIRE counterpart of the SPT detection and the noise was estimated 
by taking the RMS in the central 5 arcmin of the map which is then added in 
quadrature tothe absolute calibration uncertainty.

For SED fits, we have added in quadrature an absolute calibration uncertainty of $10\%$ for SPIRE,  
$15\%$ for LABOCA,  $10\%$ for SPT, and  $10\%$ for ALMA.

\begin{table*}
\centering
\caption{\label{tab:photometry}Far-Infrared and mm Photometry}
\begin{tabular}{cccccccccc}
\hline
\hline
 & SPIRE & SPIRE & SPIRE & LABOCA & SPT & SPT & ALMA  \\
& $250$ \um & $350$ \um & $500$ \um &   $870$ \um & $1.4$ mm  & $2.0$ mm  & $3.0$ mm\\
ID &S$_{\nu}$ [mJy] &S$_{\nu}$ [mJy] &S$_{\nu}$ [mJy] & S$_{\nu}$ [mJy] & S$_{\nu}$ [mJy]& S$_{\nu}$ [mJy]& S$_{\nu}$ [mJy]\\
\hline
SPT0103-45 & $ 121 \pm 15 $ & $ 210 \pm 23 $ & $ 222 \pm 24 $ & $ 132 \pm 22 $ & $ 36.4 \pm 6.8 $ & $ 8.4 \pm 1.6 $ & $ 1.46 \pm 0.23 $ \\
SPT0113-46 & $ 22 \pm 8 $ & $ 54 \pm 10 $ & $ 82 \pm 11 $ & $ 71 \pm 15 $ & $ 29.3 \pm 6.7 $ & $ 9.0 \pm 1.8 $ & $ 1.28 \pm 0.20 $ \\
SPT0125-47 & $ 785 \pm 79 $ & $ 722 \pm 73 $ & $ 488 \pm 50 $ & $ 138 \pm 24 $ & $ 41.3 \pm 7.0 $ & $ 8.9 \pm 1.6 $ & $ 1.88 \pm 0.29 $ \\
SPT0125-50 & $ 156 \pm 18 $ & $ 183 \pm 20 $ & $ 156 \pm 18 $ & $ 122 \pm 23 $ & $ 36.0 \pm 6.7 $ & $ 8.1 \pm 1.6 $ & $ 1.51 \pm 0.24 $ \\
SPT0128-51 & $ 40 \pm 9 $ & $ 38 \pm 9 $ & $ 38 \pm 9 $ & $ 29 \pm 8 $ & $ 19.3 \pm 5.5 $ & $ 4.3 \pm 1.5 $ & $ 0.41 \pm 0.09 $ \\
SPT0243-49 & $ 18 \pm 8 $ & $ 26 \pm 8 $ & $ 59 \pm 11 $ & $ 73 \pm 12 $ & $ 35.5 \pm 6.6 $ & $ 11.0 \pm 1.8 $ & $ 3.16 \pm 0.48 $ \\
SPT0300-46 & $ 78 \pm 11 $ & $ 124 \pm 15 $ & $ 136 \pm 16 $ & $ 50 \pm 10 $ & $ 20.0 \pm 5.5 $ & $ 4.9 \pm 1.7 $ & $ 1.01 \pm 0.16 $ \\
SPT0319-47 & $ 71 \pm 11 $ & $ 105 \pm 13 $ & $ 102 \pm 13 $ & $ 74 \pm 14 $ & $ 24.6 \pm 5.8 $ & $ 5.6 \pm 1.5 $ & $ 1.20 \pm 0.20 $ \\
SPT0345-47 & $ 242 \pm 25 $ & $ 279 \pm 29 $ & $ 215 \pm 23 $ & $ 89 \pm 16 $ & $ 26.3 \pm 6.0 $ & $ 5.3 \pm 1.3 $ & $ 1.48 \pm 0.24 $ \\
SPT0346-52 & $ 136 \pm 16 $ & $ 202 \pm 22 $ & $ 194 \pm 21 $ & $ 138 \pm 24 $ & $ 43.7 \pm 7.1 $ & $ 11.2 \pm 1.6 $ & $ 2.82 \pm 0.43 $ \\
SPT0418-47 & $ 115 \pm 14 $ & $ 189 \pm 20 $ & $ 187 \pm 20 $ & $ 100 \pm 20 $ & $ 33.5 \pm 6.4 $ & $ 7.2 \pm 1.5 $ & $ 0.79 \pm 0.13 $ \\
SPT0441-46 & $ 62 \pm 10 $ & $ 98 \pm 12 $ & $ 105 \pm 13 $ & $ 79 \pm 17 $ & $ 28.2 \pm 6.2 $ & $ 6.8 \pm 1.5 $ & $ 1.26 \pm 0.20 $ \\
SPT0452-50 & $ 38 \pm 9 $ & $ 79 \pm 11 $ & $ 84 \pm 12 $ & $ 54 \pm 10 $ & $ 17.5 \pm 5.2 $ & $ 4.0 \pm 0.9 $ & $ 0.67 \pm 0.11 $ \\
SPT0457-49 & $ 38 \pm 8 $ & $ 60 \pm 9 $ & $ 67 \pm 10 $ & $ 25 \pm 6 $ & $ 16.3 \pm 5.4 $ & $ 3.8 \pm 0.9 $ & $ 0.28 \pm 0.07 $ \\
SPT0459-58 & $ 47 \pm 9 $ & $ 62 \pm 9 $ & $ 79 \pm 11 $ & $ 47 \pm 10 $ & $ 22.4 \pm 4.9 $ & $ 4.5 \pm 1.1 $ & $ 0.96 \pm 0.16 $ \\
SPT0459-59 & $ 35 \pm 10 $ & $ 54 \pm 10 $ & $ 61 \pm 11 $ & $ 67 \pm 13 $ & $ 20.9 \pm 4.5 $ & $ 7.3 \pm 1.5 $ & $ 1.19 \pm 0.19 $ \\
SPT0512-59 & $ 322 \pm 33 $ & $ 368 \pm 38 $ & $ 264 \pm 28 $ & $ 102 \pm 18 $ & $ 22.7 \pm 4.5 $ & $ 5.5 \pm 1.3 $ & $ 0.98 \pm 0.16 $ \\
SPT0529-54 & $ 74 \pm 13 $ & $ 137 \pm 17 $ & $ 162 \pm 19 $ & $ 122 \pm 20 $ & $ 35.4 \pm 5.9 $ & $ 9.2 \pm 1.6 $ & $ 1.51 \pm 0.23 $ \\
SPT0532-50 & $ 214 \pm 23 $ & $ 269 \pm 28 $ & $ 256 \pm 27 $ & $ 125 \pm 21 $ & $ 40.8 \pm 6.6 $ & $ 13.4 \pm 1.9 $ & $ 3.04 \pm 0.47 $ \\
SPT0550-53 & $ 65 \pm 18 $ & $ 78 \pm 16 $ & $ 79 \pm 15 $ & $ 71 \pm 15 $ & $ 17.3 \pm 4.6 $ & $ 3.9 \pm 1.1 $ & $ 0.61 \pm 0.12 $ \\
SPT0551-50 & $ 150 \pm 17 $ & $ 191 \pm 21 $ & $ 189 \pm 21 $ & $ 72 \pm 13 $ & $ 26.7 \pm 5.0 $ & $ 5.0 \pm 1.0 $ & $ 1.04 \pm 0.17 $ \\
SPT2103-60 & $ 43 \pm 10 $ & $ 72 \pm 11 $ & $ 108 \pm 15 $ & $ 70 \pm 13 $ & $ 28.5 \pm 5.4 $ & $ 8.1 \pm 1.4 $ & $ 0.99 \pm 0.16 $ \\
SPT2132-58 & $ 55 \pm 11 $ & $ 75 \pm 12 $ & $ 78 \pm 12 $ & $ 56 \pm 10 $ & $ 28.7 \pm 5.5 $ & $ 5.7 \pm 1.2 $ & $ 1.42 \pm 0.23 $ \\
SPT2134-50 & $ 346 \pm 36 $ & $ 339 \pm 35 $ & $ 257 \pm 28 $ & $ 100 \pm 17 $ & $ 24.5 \pm 5.8 $ & $ 5.5 \pm 1.5 $ & $ 1.13 \pm 0.18 $ \\
SPT2146-55 & $ 58 \pm 12 $ & $ 79 \pm 14 $ & $ 82 \pm 14 $ & $ 55 \pm 9 $ & $ 21.8 \pm 5.1 $ & $ 4.7 \pm 1.4 $ & $ 1.18 \pm 0.19 $ \\
SPT2147-50 & $ 73 \pm 12 $ & $ 114 \pm 14 $ & $ 116 \pm 15 $ & $ 50 \pm 9 $ & $ 21.7 \pm 5.2 $ & $ 4.8 \pm 1.5 $ & $ 0.76 \pm 0.12 $ \\
\hline
\end{tabular}
\tablecomments{
 Fluxes are given in units of mJy and include absolute calibration uncertainties. 2\,mm \& 1.4\,mm fluxes have 
been deboosted. All other flux densities are photometric measurements at the ALMA position of the 1.4mm source.
We note that source blending is typically not a problem for the photometry as the 
strong galaxy-galaxy lensing implies that the FIR light is dominated by a single lensed background 
object. Contamination by the lensing foreground galaxy can be ruled out by our ALMA high angular resolution
 870\um\ imaging \citep[][]{vieira13}.}
\end{table*}

%
%---------------------------------------------
% Fig 0 Sample flux-flux plots
%---------------------------------------------
\begin{figure*}[htb]
\centering
\includegraphics[width=16.5cm,angle=0]{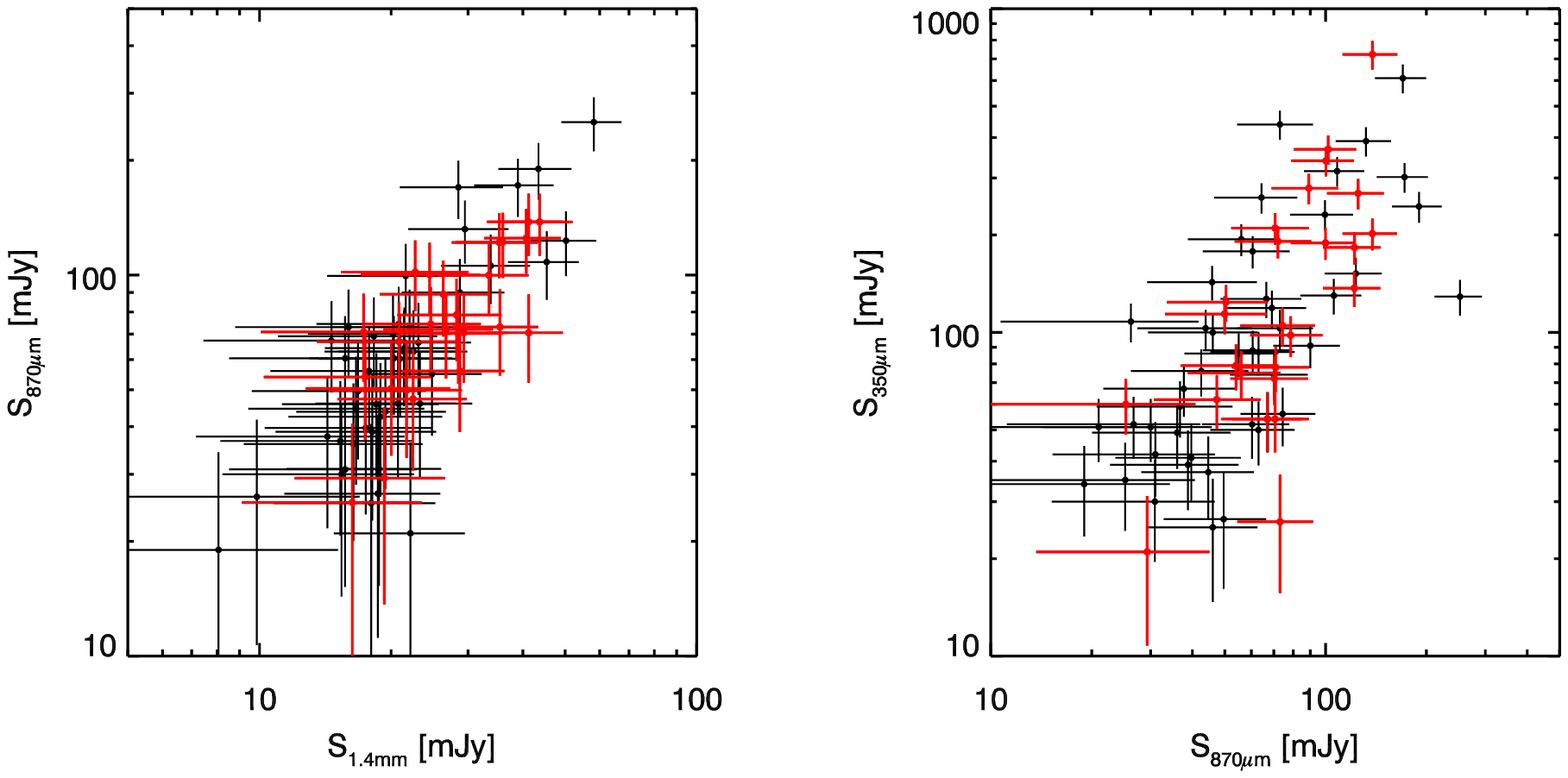}
\caption{{\it Left:} 870\micron\ LABOCA flux density as a function of 1.4\,mm SPT flux density for
the 26 sources discussed in this paper (red) compared to the full sample of SPT sources which
have been observed with LABOCA and {\it Herschel}-SPIRE (black). {\it Right}: Same as to the left but
with 350\micron\ {\it Herschel}-SPIRE flux density as a function of 870\micron\ LABOCA flux density.}
\label{Fig:sample}
\end{figure*}

\end{document}